%% file: main.tex
\documentclass[twocolumn]{aastex631}
\input{preamble.tex}

\begin{document}

\title{Observational Signatures of Disk Winds in Protoplanetary Disks: Differentiating Magnetized and Photoevaporative Outflows With Fully Coupled Thermochemistry}

\author[0000-0003-3201-4549]{Xiao Hu (\cntextsc{胡晓})}\thanks{E-mail: xiao.hu.astro@gmail.com}
\affiliation{Department of Astronomy, University of Florida, Gainesville, FL 32608, USA}

\author[0000-0001-7258-770X]{Jaehan Bae}
\affiliation{Department of Astronomy, University of Florida, Gainesville, FL 32608, USA}

\author[0000-0003-3616-6822]{Zhaohuan Zhu (\cntextsc{朱照寰})}
\affiliation{Department of Physics and Astronomy, University of
  Nevada, Las Vegas, 4505 S. Maryland Parkway, Las Vegas,
  NV 89154, USA}
\affiliation{Nevada Center for Astrophysics, University of Nevada, Las Vegas, 4505 South Maryland Parkway, Las Vegas, NV 89154, USA}

\author[0000-0002-6540-7042]{Lile Wang (\cntextsc{王力乐})}\thanks{E-mail: lilew@pku.edu.cn}
\affiliation{The Kavli Institute for Astronomy and Astrophysics, 
Peking University, Beijing 100084, China}

\begin{abstract}
Magnetized winds and photoevaporative winds are critical in shaping protoplanetary disk evolution. Using 2D axisymmetric (magneto-)hydrodynamic simulations with fully coupled thermochemistry, we investigate the signatures of the two winds in CO and [C~I] ALMA observations, and examine the potential to distinguish the origins. Our simulations reveal fundamental differences between the two winds: magnetized winds are colder and denser, exhibiting super-Keplerian rotation with small poloidal velocities of $\lesssim 1~{\rm km~s}^{-1}$ in the atmosphere ($z/R\gtrsim0.45$), while photoevaporative winds are hotter and less dense, exhibiting sub-Keplerian rotation with higher poloidal velocities of several ${\rm km~s}^{-1}$. In addition to previously identified factors like thermal pressure gradient and disk's self-gravity, we demonstrate that magnetic force and advection significantly influence rotational velocities of the gas in the wind, which lead to emission patterns that are distinct from Keplerian rotation in synthetic ALMA observations. Magnetized winds are visible in CO channel maps when wind loss rates are $\gtrsim10^{-8}~M_\odot~{\rm yr}^{-1}$. When wind loss rates are lower, magnetized winds produce subtle perturbations that resemble those produced by protoplanets. While strong XUV radiation photodissociates CO in photoevaporative winds, they can create observable ring-like substructures at disk surfaces. [C~I] emission is optically thin and could be most effective at detecting both winds in disks with high gas mass and/or high [C~I] abundance. Due to the spatially extended nature of the winds, using a large beam ($\simeq0\farcs4$ for disks in nearby star-forming regions) will be helpful regardless of the tracer used.

\end{abstract}



\section{Introduction} 
\label{sec:intro}

After decades of studying both solar and exoplanetary systems, we are now closer than ever to observing planet formation in action. Protoplanetary disks (PPDs) play a crucial role in this process, serving as reservoirs of gas and dust necessary for planetary development. Recent observations with the Atacama Large Millimeter/submillimeter Array \citep[ALMA,][]{2015ApJ...808L...3A} have spatially resolved dust substructures, showing that bright rings and dark gaps are common features in PPDs~\citep[e.g.,][]{2018ApJ...869L..42H,2020ARA&A..58..483A}. High-resolution observations of molecular line emissions have revealed even more intricate gas structures within these disks \citep[e.g.,][]{2019Natur.574..378T,2020ApJ...890L...9P}. Additionally, kinematic patterns observed in gas tracers provide insights into disk dynamics and have gained popularity for identifying embedded forming planets, especially when the substructures' azimuthal velocities align with perturbations caused by planets \citep{2018ApJ...860L..13P}.

On the other hand, the disk itself is a very dynamic environment, hosting various hydrodynamic, magnetic, thermodynamic, and gravitational instabilities that could form substructures without planets. These processes include vertical shearing instability~\citep[VSI, e.g.,][]{nelson2013,2017ApJ...850..131F}, snow lines of various volatiles \citep{2015ApJ...806L...7Z, 2021ApJ...913..133H}, transition at the dead zone boundary \citep[e.g.,][]{2015A&A...574A..68F},
secular gravitational instability \citep[e.g.,][]{2016AJ....152..184T}, 
zonal flows \citep[e.g.,][]{2009ApJ...697.1269J,2018ApJ...865..105K}, 
and magnetic disk winds with magnetic diffusivities \citep[e.g.,][]{2017MNRAS.468.3850S, 2020A&A...639A..95R,2021MNRAS.507.1106C, 2022MNRAS.516.2006H}. In particular, disk wind launched by the magnetohydrodynamic (MHD) process has been considered the main mechanism that drives disk accretion, while the magnetorotational instability (MRI) which supports viscous accretion, is likely to be suppressed by Ohmic resistivity in the cold, weakly ionized disk midplane \citep{2011ApJ...739...50B}. The MHD wind extracts mass and angular momentum from the disk surface, inducing radial mass flow within the disk~\citep[e.g.,][]{2015ApJ...801...84G, 2017ApJ...845...75B, 2019ApJ...874...90W, 2023ASPC..534..465L}.

Photoevaporation (PE), i.e., outflow driven by high-energy radiation is another important dispersal mechanism for protoplanetary disks. Ultraviolet (UV) and X-ray radiation heats the gas in the upper layers of the disk, causing it to reach escape velocities and flow outward in the form of a wind \citep{2006MNRAS.369..216A, 2006MNRAS.369..229A, 2010MNRAS.401.1415O}. This thermally driven outflow is most effective in the outer regions of the disk, where the gravitational pull from the star is weaker and the disk is optically thinner, which allows radiation to penetrate deeper into the gas \citep{2014prpl.conf..475A}. At the late stage of disk evolution, the low density makes the PE wind more dominant over other disk dispersal mechanisms. This would set a time limit for planet growth and effectively shut off further planet formation by cutting off the material supply. As PE wind only carries its own share of angular momentum, it removes mass without altering disk accretion. 
This could modify the disk density profile over long-term evolution, complicating the distinction between viscous and wind-driven disks \citep{2024MNRAS.527.7588C}.

Observational evidence for winds in Class II sources is inferred mainly from the kinematics of spatially unresolved, blueshifted forbidden emission of atomic lines, using optical spectroscopic measurements to identify high- and low-velocity components \citep[e.g.,][]{2019ApJ...870...76B, 2020ApJ...903...78P, 2023ApJ...956...25C}. The origin of these outflows is usually at the sub-au scale, indicating they are magnetic-driven. High-resolution [O I] line spectral mapping of TW Hya directly confines 80\% of the emission to within 1 au radially from the star~\citep{2023NatAs...7..905F}. For molecular lines like CO, the molecular outflow of HH30 demonstrates multiple shells and can be explained by magnetocentrifugal disk winds with launching radii $<$~4 au~\citep{2024ApJ...962...28L}. Recently,
the large program Molecules with ALMA at Planet-forming Scales \citep[MAPS;][]{2021ApJS..257....1O} targeted five protoplanetary disks (MWC 480, IM Lup, GM Aur, HD 163296, and AS 209) in several molecular lines, unveiling abundant velocity structures and even candidates for embedded planets \citep{2022ApJ...934L..20B}. Some of the kinematic structures have been proposed as potential indicators of underlying wind-driven processes \citep{2023ApJ...950..147G,izquierdo2023}, yet the clear, unambiguous identification of disk winds launched from outer disk remains elusive \citep{2021ApJS..257...16B}.

The absence of disk wind signatures in current molecular line observations towards Class II sources could imply that the wind density or temperature lies below the current detection threshold, or that the tracer molecule like CO is sufficiently depleted from the relevant wind regions \citep{2021ApJS..257....5Z}. Even if direct detection proves challenging, we ask whether indirect evidence of disk winds might still be inferred from current observations. These high-resolution kinematics data already indirectly constrain some important disk properties. Models assuming vertical hydrostatic equilibrium can derive properties like disk mass and thermal stratification from a more precise rotation curve \citep{2024A&A...686A...9M,2024ApJ...970..153A}, but a fully (magneto-)hydrodynamics model is necessary for studying disk winds. A model that consistently includes both MHD wind and PE wind is essential to determine the origin of the outflow.

Proper treatment of thermochemistry with (magneto-)hydrodynamics is crucial to wind launching since thermal pressure is the driving force for the PE wind, and disk ionization controls the coupling between magnetic fields and the gas in the MHD wind. However, global protoplanetary disk (PPD) simulations typically employ a $\beta$ cooling scheme that relaxes temperature to the initial profile \citep[e.g.,][]{2017ApJ...836...46B,2022MNRAS.516.2006H}. 
This simplified approach 
fails to provide accurate temperatures for the highly dynamic disk surface, which is believed to be the source of $^{12}$CO line emissions \citep[e.g.,][]{2021ApJS..257....4L}. In the context of MHD winds, this transitional zone marks the shift from a toroidal field-dominated disk to a poloidal field-dominated atmosphere, where angular momentum is extracted. For photoevaporative winds, the disk surface represents the transition from an optically opaque disk to a region that is transparent to some of the central star's major heating energy bands. This is also where a significant jump in ionization level occurs, which is crucial for non-ideal MHD diffusion processes. These transitions involve complex heating, dissociation, or ionization of the gas through a variety of mechanisms. Properly modeling the disk surface or wind base requires a (magneto-)hydrodynamics setup that includes an on-the-fly consistent thermochemistry scheme with an appropriate treatment of radiation \citep{2017ApJ...847...11W, 2019ApJ...874...90W}.

In this study, we aim to provide a framework for interpreting disk wind observations by conducting two-dimensional axisymmetric (magneto-)hydrodynamic simulations with fully coupled thermochemistry, post processed with radiative transfer to explore the observational signatures of both magnetically driven and photoevaporative disk winds. \S\ref{sec:setup} details our simulation setup, including the thermochemical model of radiation, CO freeze-out, and magnetic diffusion. In \S\ref{sec:results} we present the key results directly from our simulations, analyzing physical and dynamic properties of both winds. We built simple parametric disk models to produce line-of-sight (LOS) velocity maps that would help interpret channel maps of disk winds in \S\ref{sec:LOS}. We then make synthetic CO and [C I] line emission observations in \S\ref{sec:obs}, with a focus on distinguishing between wind types. \S\ref{sec:velocity} analyzes kinematic properties to be measured at different emission layers and force balance analysis of disk rotation, shedding light on the importance of advection terms and magnetic forces. We discuss the cases of disks with 1/10 of our fiducial mass and the indications for future observations in \S\ref{sec:discuss}. Finally, we summarize our conclusions in key points in \S\ref{sec:summary}. This study aims to provide a framework for interpreting disk wind observations.

\section{Numerical Setup} 
\label{sec:setup}
\subsection{disk}
\label{sec:disk}

We performed 2D axisymmetric (magneto-)hydrodynamic simulations using \texttt{Athena++} \citep{2020ApJS..249....4S,2019ApJ...874...90W}, ray-tracing radiative transfer for high-energy photons, and consistent thermochemistry. For each (M)HD timestep, the non-equilibrium thermochemistry is co-evolved in each zone throughout the simulation domain with a semi-implicit method. The standard resolution is 240 radial by 64 latitudinal. The radial zones are spaced logarithmically. The latitudinal zones have grid spacing decreasing in geometric progression from pole to midplane so that $\delta\theta$ at the midplane is 1/4 as large as near the pole. As the midplane scale height at R=100~au is roughly $h_{\rm mid}\sim$ 0.10R, this grid geometry gives more than 8 latitudinal zones per $h_{\rm mid}$.  

The initial midplane temperature and density profiles are:
\begin{equation}
    T_{\rm mid} = T_{\rm mid,0}\left(\frac{R}{R_0}\right)^{q}
    \label{eq:Tmid}
\end{equation}
\begin{equation}
 \rho_{\rm mid} = \rho_{\rm mid,0}\left(\frac{R}{R_0}\right)^{p}\exp\left[-\left(\frac{R}{R_0}\right)^n\right]
 \label{eq:rhomid}
\end{equation}
We adopt q=-0.59, p=-2.21, and n=1 with $T_{\rm mid,0}=28.5K$ and $\rho_{\rm mid,0}=2.1\times10^{10}m_{\rm p}~{\rm cm}^{-3}=3.5\times10^{-14}~{\rm g~cm}^{-3}$ at $R_0=80$~au. The surface density structure is then:
\begin{equation}
    \Sigma = 9.86 \left(\frac{R}{R_0}\right)^{-1}\exp\left[-\left(\frac{R}{R_0}\right)^1\right]~{\rm g~cm}^{-2}
\end{equation}

\begin{table}
  \caption{Properties of Disk Model (\S\ref{sec:disk}) 
    \label{table:fiducial-model}
  }
  \centering
  \scriptsize
\begin{tabular}{l r}
\hline
  Item & Value\\
  \hline
  Radial domain & $10~\au \le r \le \ 500~\au$\\
  Latitudinal domain & $0.06~{\rm rad}\le\theta\le \pi/2~{\rm rad}$ \\
  Resolution & $N_{\log r} = 240$, $N_\theta= 64$ \\
  \\
  Stellar mass & $1.0~M_\odot$ \\
  \\[2pt]
  Initial mid-plane density & see Eq.~\ref{eq:rhomid}\\
  Initial mid-plane plasma $\beta$ & $10^4,10^5,\infty$ \\
  Initial mid-plane temperature &
  $25(R/100\au)^{-0.59}~\K$ \\
  Dust mid-plane temperature &
  $25(R/100\au)^{-0.59}~\K$ \\
  \\
  Luminosities [photon~$\s^{-1}$] & \\[5pt]
  $7~\eV$ (``soft'' FUV)  & $4.5\times 10^{42}$ \\
  $12~\eV$ (LW) & $1.6\times 10^{40}$ \\
  $300~\eV$ (XUV, optional) & $1.8\times 10^{40}$ \\ 
  $3~\keV$ (X-ray) & $1.0\times 10^{38}$ \\
  \\
  Initial abundances [$n_{\chem{X}}/n_{\chem{H}}$] & \\[5pt]
  \chem{H_2} & 0.5\\
  He & 0.1\\
  \chem{H_2O} & $1.8 \times 10^{-4}$\\
  CO & $1.4 \times 10^{-4}$\\
  S  & $2.8 \times 10^{-5}$\\
  SiO & $1.7 \times 10^{-6}$\\
  \chem{N_2}  & $1 \times 10^{-5}$\\
  \\
  Dust/PAH properties & \\
  $a_\Gr$ & $5$ \AA \\
  $\sigma_\Gr/\chem{H}$ & $7.8\times 10^{-21}~{\rm cm^2}$ \\
  \hline
\end{tabular}
\end{table}

The exponentially tapered power-law surface density profile is for a self-similarly viscous disk~\citep[e.g.,][]{1974MNRAS.168..603L, 1998ApJ...495..385H, 2011ApJ...732...42A}. The choice of $R_0$ and power-law index is adopted from the AS209 disk \citep{2021ApJS..257....5Z} with an enhanced surface density. The total gas disk mass is 0.02 $M_\odot$, close to GM Aur's mass, when estimated by the CO emission and local CO abundance \citep{2021ApJS..257....5Z}. The temperature profile also follows AS209, as $T_{\rm mid}=25~K$ at 100 au from \citet{2021ApJS..257....4L}. Note we did not use the midplane temperature profile from the 2D fit directly, as $q_{mid}=-0.18$ is too flat compared to a typical PPD model used in simulations. Instead, we choose $q=-0.59$ which is the ``surface'' temperature slope. Both the $^{12}$CO and $^{13}$CO layer in AS209 have a steeper temperature slope ($\sim0.8$). This slope is closer to the power-law of the $^{13}$CO layer in MWC480, and the $^{12}$CO layer IM lup and GM Aur.

For the magnetized cases, the disk is initially threaded by a large-scale poloidal magnetic field, with a midplane plasma~$\beta$ (defined as the ratio between thermal pressure and magnetic pressure $\beta\equiv2P_{gas}/B^2$) of $10^4$ or $10^5$. The corresponding initial vector potential is adopted from \citet{2007A&A...469..811Z}:
\begin{equation}
A_\phi(r, \theta) = \frac{2B_{z0}R_0}{4+p+q}\left(\frac{r\sin\theta}{r_0}\right)^{\frac{p+q}{2}+1}\
[1+(m\tan\theta)^{-2}]^{-\frac{5}{8}}
\end{equation}
where $p$, $q$ from Eq.\ref{eq:Tmid},\ref{eq:rhomid}, $r_0$=$R_0$ and $m$ is a parameter that specifies the degree that poloidal fields bend, with $m\rightarrow\infty$ giving a pure vertical field. We chose $m=0.5$ the same as \citet{2017ApJ...836...46B}. 

In addition to the equation of continuity, the momentum and energy equations are especially important for wind-launching:
\begin{eqnarray}
    \frac{\partial\rho}{\partial t} + \nabla \cdot (\rho \v) = 0\ ;
    \\
   \frac{\partial(\rho \v)}{\partial t} + \nabla \cdot \left(\rho \v \v -
      \dfrac{\B\B}{4\pi} + P_\tot \mathbf{I} \right) = -
    \nabla \Phi \ ;
    \\
    \frac{\partial\epsilon}{\partial t} + \nabla \cdot \left[
      \left(\epsilon + P_\tot \right) \v -
      \dfrac{(\B\cdot\v)\B}{4\pi} + {\bm S}' \right]
    = 0\ , 
\end{eqnarray}
where $\rho$, $\v$ and $p$ are the gas density, velocity and
gas thermal pressure, $\B$ is the magnetic
field, $P_\tot\equiv p + B^2/(8\pi)$ is the total pressure,
$\epsilon \equiv p / (\gamma - 1) + \rho (v^2/2 + \Phi) +
B^2/(8\pi)$ is the total energy density with $\gamma$ as the
adiabatic index, $\Phi$ is the gravitational potential, and
$\mathbf{I}$ is the identity tensor, $\mathbf{S}'$ is the Poynting flux associated with co-moving electric field $\E'$ due to nonideal MHD effects, which reads
${\bm S}' = c\E'\times\B / (4\pi)$. No explicit viscosity is used to model any turbulence on a subgrid level. Non-adiabatic processes that affect gas energy are calculated separately
via operator-splitting, i.e., the gas's internal energy is updated with the extra contribution from thermochemical processes. Governing the evolution of magnetic fields,
the non-ideal induction equation is:
\begin{eqnarray}
\frac{\partial {\bm B}}{\partial t}=\nabla \times \left({\bm v}\times {\bm B}\right)
-\frac{4\pi}{c}\nabla \times ( \eta_\mathrm{O} {\bm J} +
\eta_\mathrm{A} {\bm J}_{\bot}),
\label{eq:induction}
\end{eqnarray}
where ${\bm b} =
{\bm B}/|B|$ is the unit vector representing field line direction.  
$\bm J$ is current density vector, and $\bm J_{\bot}$ is the current component perpendicular to the magnetic field. Note that the second term on the right-hand side is $c\E'$, and the ambipolar diffusion and Ohmic resistivity are included in our non-ideal MHD simulations but not the Hall effect. The Hall effect is neglected for several reasons: Ambipolar diffusion and Ohmic resistivity are sufficient to maintain wind-driven accretion, as shown by numerous studies \citep[e.g.,][]{2013ApJ...769...76B, 2015ApJ...801...84G, 2019ApJ...874...90W}. Also, recent theoretical advancements suggest that the Hall effect may be less significant than previously thought \citep{2024arXiv240506026H}. We calculate the diffusion profiles based on the local densities of charged species, including charged grains. The general framework is from \citet{2007Ap&SS.311...35W}. For the details of diffusivity calculations the chemical reaction network that produces the ionization structure, and their thermal effects we refer the readers to our previous works \citep{2019ApJ...874...90W, 2021ApJ...913..133H, 2023MNRAS.523.4883H}.  For both $\eta_\mathrm{O}$ and $\eta_\mathrm{A}$, we capped the diffusivities as $\eta_\mathrm{cap}=[10c_{s,\mathrm{mid}}h_\mathrm{mid}]_{r=r_\mathrm{min}}$, so the diffusion timestep is not significantly smaller than the hydro timestep of the innermost region. Here $c_{s,\mathrm{mid}}$ denotes the adiabatic sound speed at the midplane, $h_\mathrm{mid}$ is the midplane scale height, and $r_\mathrm{min}$ is the radial inner boundary. The $\eta_\mathrm{A}$ cap is sometimes reached in a limited area around the inner disk region, while the $\eta_\mathrm{O}$ is never reached. This cap does not harm our simulations as we focus on more ionized regions like the disk surfaces and atmosphere where the plasma is mostly in the ideal MHD regime. For regions with much weaker ionization, the magnetic diffusivity is dominated by ohmic resistivity. 

The boundary conditions are as follows. On the outer radial boundary, we use the outflow condition for the hydro variables, with inflow set to zero to prevent collapsing. The magnetic fields are extrapolated from the last active zone with different power-laws for three components, as $B_r\propto r^{-2}$, $B_\phi\propto r^{-1}$, and $B_\theta$ is directly copied. For the inner boundary, the magnetic treatment is the same. The density and velocity are fixed from the initial values, while the gas temperature and relative abundances of chemical species copy the values in the innermost active zone. Reflective conditions are used at both $\theta$ boundaries, with special treatment on magnetic fields to maintain zero divergences, i.e., $B_r$ and $B_\phi$ are flipped, while $B_\theta$ maintains mirror symmetry, i.e., at the exact boundary, $B_r=B_\phi=\partial_\theta B_\theta=0.$. Since this is a 2.5D setup, periodic conditions are applied at $\phi$ boundaries.

\subsection{Chemistry}
\label{sec:chem}

We used a reduced chemical network consisting of 33 species that are most relevant to the disk's thermal and ionization structure, including charged grains. We effectively remove gas phase CO from the network when t$<$20 K to account for CO freeze-out. The dust grains are both charge carriers and heat reservoirs.
They are crucial in the ionization balance of
disks, especially near the midplane. In general, they assist recombination by adsorbing charged particles. A high abundance of dust grains helps maintain the disk temperature from efficient line cooling. In order to have a reasonable CO freeze-out radius, we use an enhanced dust abundance $n_{\rm gr}/n_{\rm H}=10^{-6}$ compared to $n_{\rm gr}/n_{\rm H}=10^{-7}$ in \citet{2017ApJ...847...11W}.

For two-body reactions, e.g., between CO molecules and the grain, the reaction rate corresponds to
the de Kooij-Arrhenius (KA), or modified Arrhenius, formula~\citep{2013A&A...550A..36M}: 
\begin{equation}
    k=\alpha_A\left(\frac{T}{300}\right)^{\beta_A} \exp(\frac{-\gamma_A}{T})~~ {\rm cm^3s^{-1}}
\end{equation}

We added three reactions to account for the CO capture (``freeze-out'') on the grain surfaces: $\ce{CO + \widetilde{Gr} -> CO^* + \widetilde{Gr}}$. Here ${\rm CO^*}$ is the solid state CO, and $\widetilde{\rm Gr}$ represents all three types of grains: \Gr, $\Gr^+$, and $\Gr^-$, i.e., the reaction is independent of grain's charge status. The capture process should have a weak temperature dependence, except for the collision frequency that scales to $T^{0.5}$. Thus we have $\beta_A=0.5$ and $\gamma_A=0$.  The pre-exponential factor is relatively arbitrary as long as the CO capture is faster than most other reactions in the network. We use $\alpha_A=8.8\times10^{-9}$ which is 10 times faster than the CO to $\ce{HCO^+}$ reaction $\ce{CO + N2H+ -> HCO+ +N2}$. For the three reactions of CO release from grain surface, $\ce{CO^* + \widetilde{Gr} -> CO + \widetilde{Gr}}$, the CO molecule needs to break a potential barrier of 0.1 eV, which gives $\gamma_A=1160$ \citep{2014ApJ...790...97F}. The detailed balance at $T=20$ K requires $\alpha_A=1.36\times10^{17}$, and we keep $\beta_A=0.5$. Note that the grain is unchanged in all reactions, which means a single grain can ``catalyze'' the capture and release of multiple CO molecules. Thus, we make sure that all CO would be turned into a solid state efficiently when $T<20~{\rm K}$. This differs from the grain-electron reactions, where all charged grains can only capture/lose one electron. 

The central star radiates at three energy bands: 7 eV (FUV), 12 eV (Lyman-Werner band), and 3000 eV (X-ray). For the non-magnetized setup, we add an extra energy band of 300 eV to mimic the combined effect of EUV and soft X-ray. In \citet{2017ApJ...847...11W}, 25 eV was used as EUV photons for the inner disk region (R~$<$~100 au). 25 eV energy bin can only scarcely penetrate an intermediate layer (0.3$<$z/R$<$0.6) and excite disk wind above. This penetration is inadequate to reach and launch wind at outer ($>$100~au) disks. Since the absorption cross-section is inversely proportional to the cube of photon energy, 300 eV radiation effectively represents the energy band with intermediate penetration. The 300 eV photons have a penetration potential that is in between the EUV and X-ray band, so they could reach the surface at a larger radius while still depositing most of their energy at the surface. Also, X-rays have shown a relatively low efficiency on the thermal-to-mechanical conversion ratio, i.e., the ability to transfer thermal energy to the kinematic energy of the wind~\citep{2017ApJ...847...11W}.

\begin{table*}[!t]
    \centering
    \begin{tabular}{l|l|c|c|c}
    Model No. & Setup   &  Initial Field (100 au) & Evolved Field (100 au) & wind $\dot{m}$~[$M_\odot~yr^{-1}$]\\ 
    \hline
    1 (\texttt{strongB})& $\beta=10^4$  & $\beta=2.9\times10^3$,0.33 mG & 3 mG & $4.0\times10^{-8}$\\
    2 (\texttt{strongB-XUV}) & $\beta=10^4$, 300 eV  & $\beta=2.9\times10^3$, 0.33 mG & 3 mG& $4.0\times10^{-8}$\\
    3 (\texttt{weakB})&  $\beta=10^5$  & $\beta=2.9\times10^4$, 0.1 mG & 1 mG & $4.5\times10^{-9}$\\
    4 (\texttt{weakB-XUV})& $\beta=10^5$, 300 eV  & $\beta=2.9\times10^4$, 0.1 mG & 1 mG & $1.8\times10^{-8}$\\
    5 (\texttt{XUV}) &  $\beta=\inf$, 300 eV & 0 mG & 0 mG   & $9.2\times10^{-9}$\\
    6 no XUV & $\beta=\inf$  & 0 mG & 0 mG & $1.0\times10^{-10}$\\
    \end{tabular}
    \caption{List of simulations,  their initial and evolved field strength, and wind loss rates.}
    \label{tab:list}
\end{table*}

We conducted six simulations, combining three magnetic field strengths ($\beta = 10^4$, $10^5$, and $\infty$, i.e., no magnetic field) with two radiation energy distributions (with and without a 300 eV XUV energy bin), as listed in Table~\ref{tab:list}.  Note that because of the extra exponential taper in the initial density profile, the $beta$ value is only accurate in the innermost region. Thus we also list the initial $\beta$s at 100 au, with field strength listed in milligauss (mG). The field strength at the current snapshot of t=$6\times10^4$ yr is also listed as ``Evolved Field''.  To measure the wind mass-loss rate, we sample along the polar angle from the pole to the midplane at each radius. The wind base is defined as the first layer where the poloidal gas velocity becomes subsonic relative to its midplane value. The sphere-integrated radial mass flux increases rapidly with radius and flattens beyond approximately 300 au. Although we do not perform time averaging, comparisons between snapshots around $t=6\times10^{4}~{\rm yr}$ show only minor temporal fluctuations. Both strongly magnetized models, \texttt{strongB} and \texttt{strongB-XUV}, exhibit the highest wind loss rate of $4 \times 10^8~M_\odot~\text{yr}^{-1}$. The MHD wind loss rate is approximately inversely proportional to plasma $\beta$, reducing to $4.5 \times 10^9~M_\odot~\text{yr}^{-1}$ in \texttt{weakB}. When photoevaporation is included, the loss rate in \texttt{weakB-XUV} increases to $1.8 \times 10^8~M_\odot~\text{yr}^{-1}$. The pure photoevaporative (PE) wind model, \texttt{XUV}, doubles $\dot{m}$ compared to \texttt{weakB}. The final model, with no XUV or magnetic field, produces a negligible disk wind. Our analysis will focus on the first five models only. We also have a set of simulations of less massive disks, which will be discussed separately in \S\ref{sec:lowmass}.


\section{Simulation Results}

\label{sec:results}
\subsection{Physical properties of winds}
\label{sec:2dquant}
\begin{figure*}
    \centering
    \includegraphics[width=0.99\textwidth]{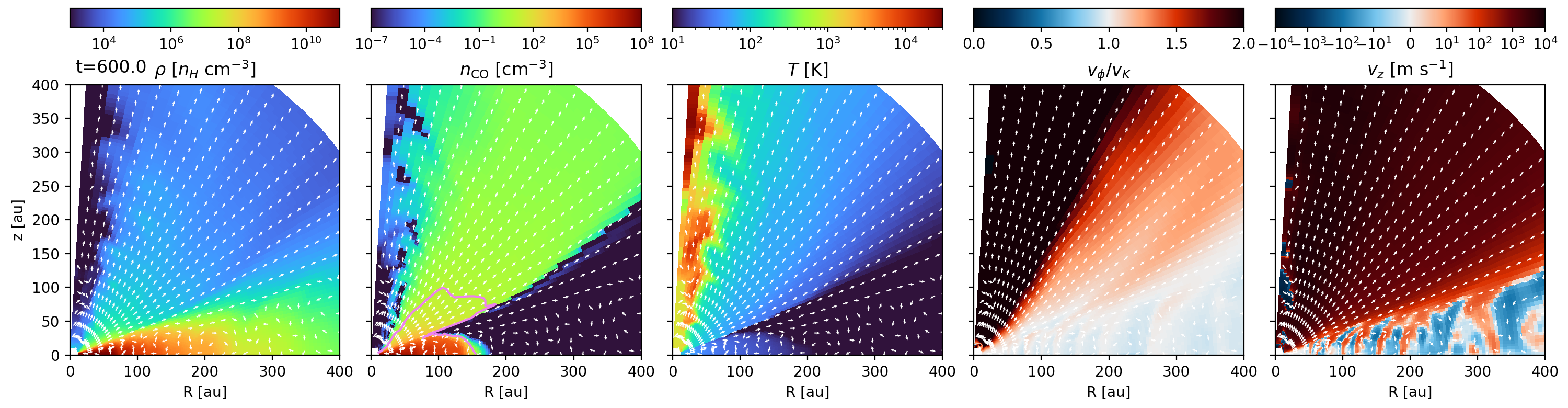}
    \includegraphics[width=0.99\textwidth]{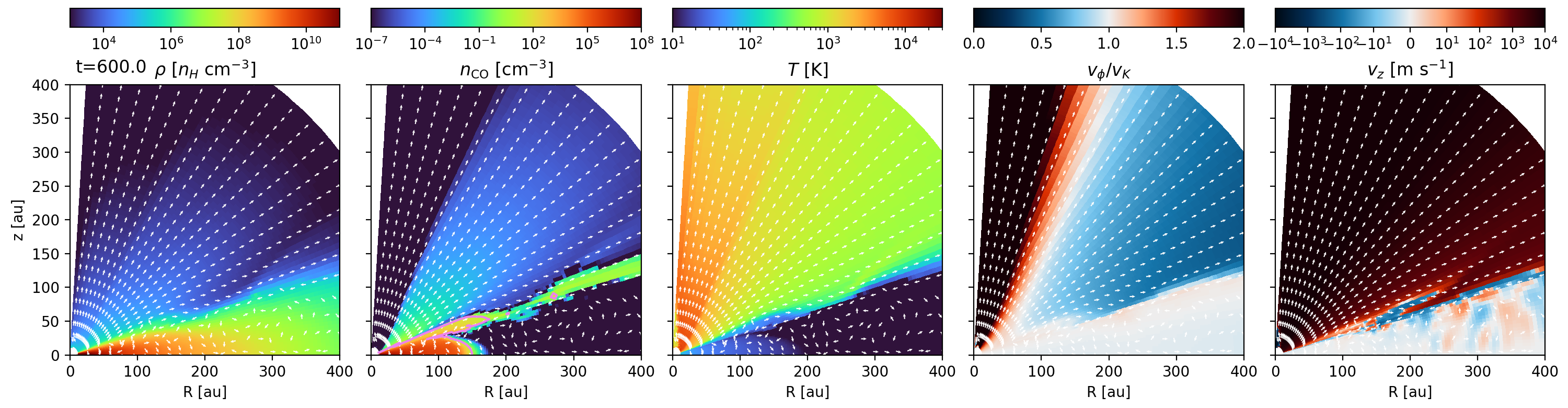}
    \caption{A comparison of vertical structure between model \texttt{strongB} (top panels) and \texttt{XUV} (bottom panels). All of the variables have been taken at the snapshot of t=$6\times10^4$ yr. From left to right are gas density in terms of the number density of hydrogen nuclei, the number density of molecular CO, gas temperature, azimuthal velocity scaled by local Keplerian, and vertical velocity in m/s. The violet contours in the CO number density panels mark $n_{CO}=10~{\rm cm^{-3}}.$ }
    \label{fig:Trhovzco}
\end{figure*}

We focus on the two most distinctive cases, MHD wind Model \texttt{strongB} with a midplane plasma $\beta=10^4$, and Model \texttt{XUV}, with photoevaporative wind driven by 300 eV XUV photons. The overall results are presented in Figure~\ref{fig:Trhovzco}. We choose the quantities that are more relevant to molecular line observations: gas density $\rho$, molecular CO density, gas temperature, azimuthal velocity, and vertical velocity. The MHD wind can be categorized into two types \citep{2017ApJ...845...75B}: (1) magnetocentrifugal winds, where the poloidal magnetic fields are strong enough to enforce corotation near the wind base, driving an outflow via centrifugal forces; and (2) magneto-thermal winds, which are driven by the gradient of total pressure, particularly the energy density in the toroidal magnetic fields. 1D global modelling have suggested the latter exists in protoplanetary disks \citep{2016ApJ...818..152B} and is confirmed by more complicated numerical simulations \citep{2019ApJ...874...90W}. Though the wind is only partially launched by centrifugal force, the poloidal magnetic field lines still couple the disk atmosphere with the rotating disk. The MHD wind could then extract angular momentum from the disk and rotate faster in the upper atmosphere. In our case, model \texttt{strongB} has a denser and colder (tens to couple hundred K) wind, and $v_\phi$ above the disk is faster than the local Keplerian velocity. Throughout this paper, we use the vertically supported Keplerian velocity $v_{K}=\sqrt{GM_*R^2/r^3}$ that is widely used in gas kinematics of PPDs. It represents the rotational velocity of a gas parcel in a steadily rotating disk, where the gravitational force in the cylindrical radial direction (R) is balanced by centrifugal force, and the vertical component of gravity is counteracted by the vertical pressure gradient. It reduces to the standard Keplerian velocity used in \citet{2019ApJ...874...90W} at the disk midplane.

Photoevaporative (PE) wind is launched when the local thermal velocity dispersion can break the gravity potential \citep{1994ApJ...428..654H}, and the high energy photons can provide enough ``fuel'' to overcome the cooling from adiabatic expansion in the outward movement. In fact, any significant heating is diminished after wind launching because of a lack of neutrals in the highly photoionized atmosphere \citep{2024arXiv240115419L}. The gas patch that leaves the disk surface is effectively detached from the rotating disk, thus exerting no torque on the disk, and the angular momentum of the gas patch in the wind is conversed. The wind-filled atmosphere is mainly pressure-supported and rotates slowly. The gas density in model \texttt{XUV}'s wind region is an order of magnitude lower than that the MHD wind, while being much hotter ($\sim10^3$~K). The wind region also rotates slower than Keplerian.

The CO distributions below the disk surfaces are nearly identical, with significant freeze-outs beyond 150 au. The additional 300 eV photons cannot penetrate the dense disk region, leading to similar thermal structures below the wind base. However, the differences are pronounced from the disk surface to the wind. \texttt{strongB} exhibits a dense CO wind, with CO maintaining a uniform abundance in the wind, similar to the rest of the warm ($>20$K) disk. The exception is the polar region, where the gas density is so low that thermal dissociation can destroy molecules.  Low density and strong magnetization also increase MHD heating rate. In contrast, model \texttt{XUV} features a concentrated CO layer at the disk surface and a very thin CO wind (CO abundance $<10^{-6}$).  This difference comes from the warmer ($\sim$500K) atmosphere in model \texttt{XUV} maintained by 300 eV photons. In the PE wind region, CO also lacks the column density (radial column to the central star) \citep[e.g., $10^{17} {\rm cm^{-2}}$ for interstellar radiation field,][]{2017A&A...602A.105H} required for effective self-shielding.

In addition to the differences in CO distribution and the magnitudes of poloidal velocity, the radiation-driven photoevaporative (PE) wind exhibits velocity perturbations at the disk surface. This results from the pressure balance between the warm disk surface and the warmer wind. For $R < 150$ AU, the disk surface is dense enough to support the warmer disk wind due to the ``puffed-up'' layer caused by incident radiation heating. Between 150 and 270 AU, the wind region expands downward to the surface layer as the density is insufficient to counterbalance the wind pressure. This downward motion is indicated by the slender blue area in the $v_z$ panel. The ``collapsing'' flow is subsequently bounced back by the high-density disk, creating a thin positive $v_z$ region beyond 200 AU, just below the negative $v_z$ slice.  The warm, expanding wind serves as the primary heating source for the thin CO layer at z/R = 0.3. An additional contributing factor is the deeper penetration of 7, 12, and 300 eV radiation, enabled by the reduced density at the wind base of photoevaporation.

\begin{figure}
    \centering
    \includegraphics[width=0.45\textwidth]{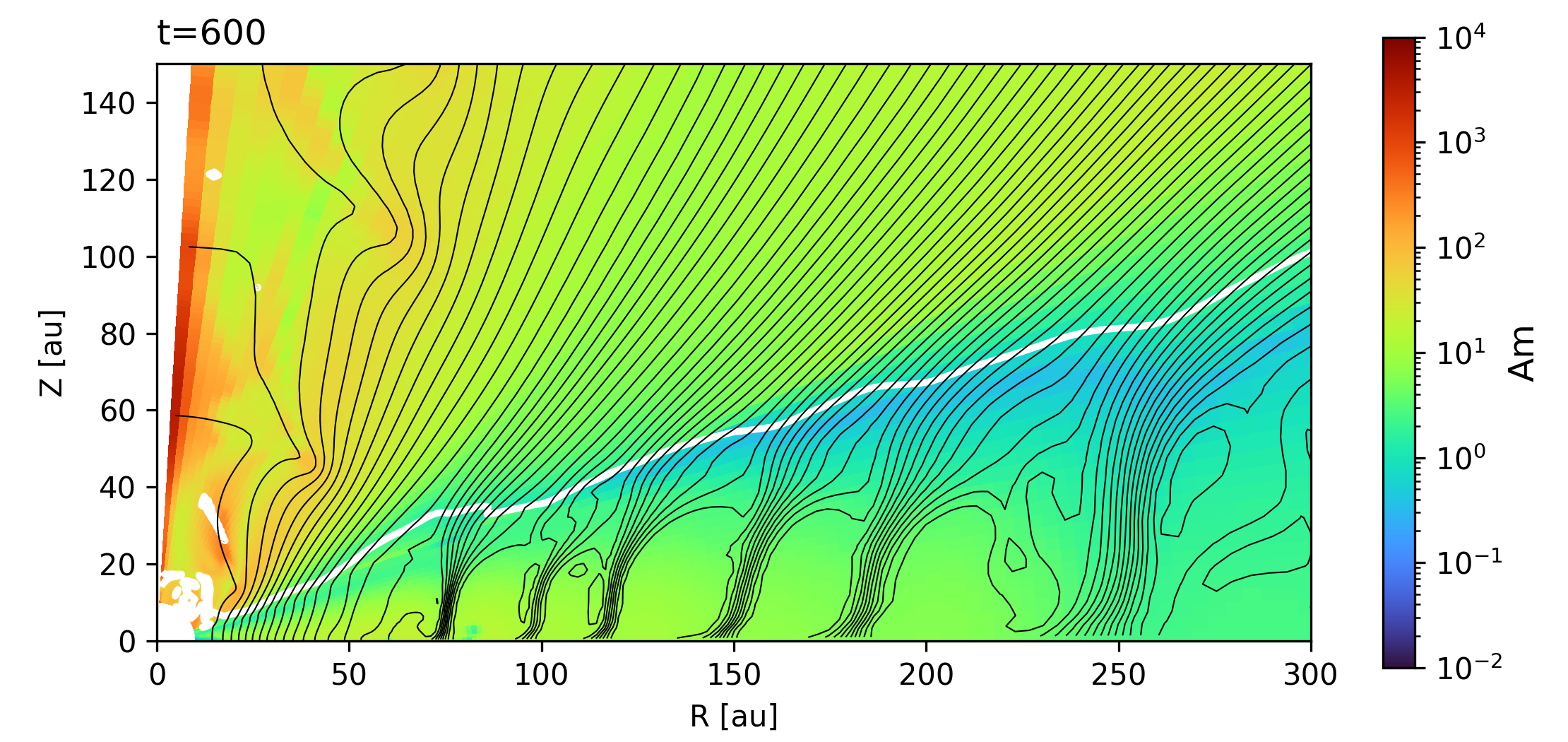} 
    \includegraphics[width=0.45\textwidth]{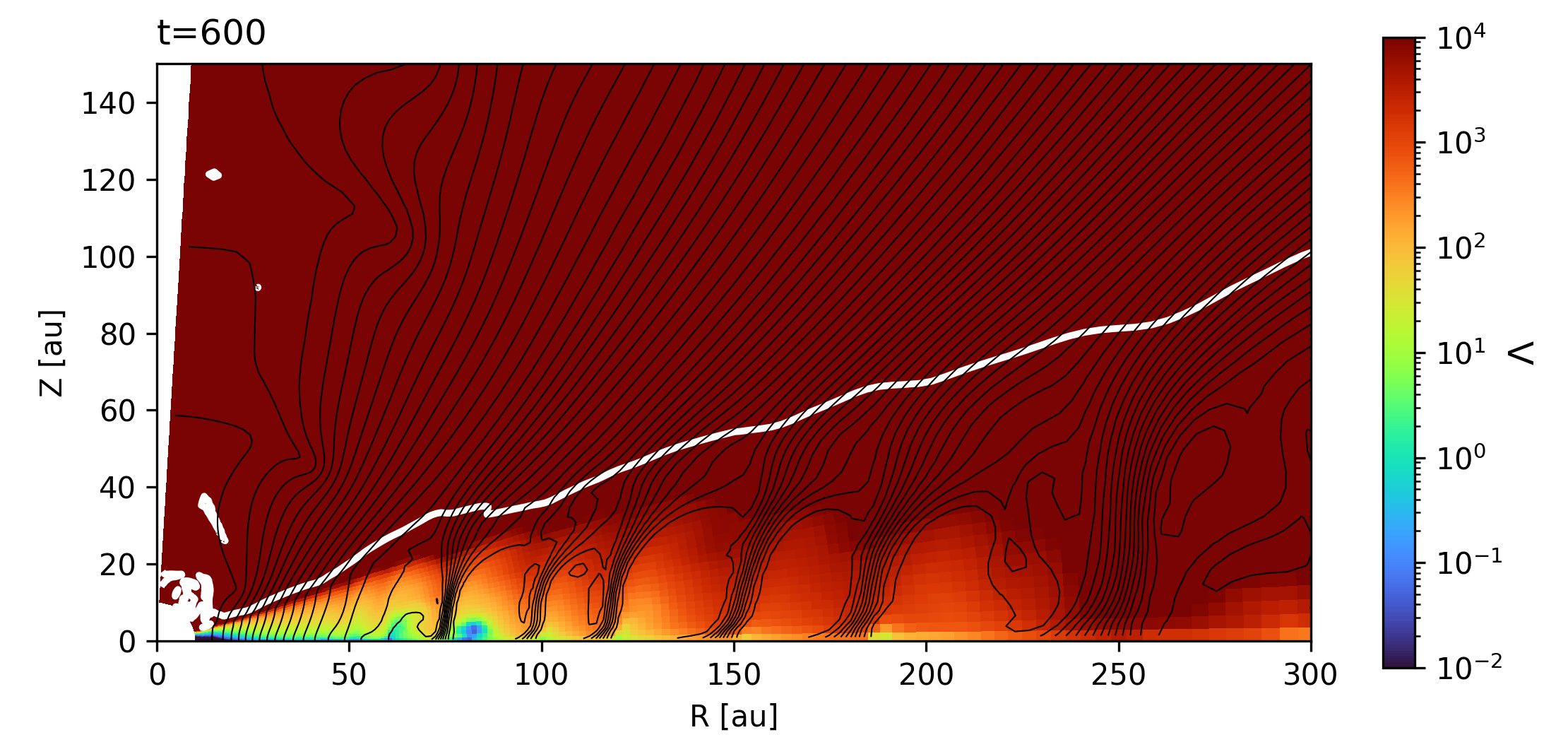} 
    \caption{Magnetic diffusivities of the \texttt{strongB} model. The upper panel is the ambipolar diffusion Elsasser number $Am$ and the lower panel is the Ohmic resistivity Elsasser number $\Lambda$. The poloidal magnetic fields are illustrated as black solid lines. The white solid contours marking $v_{pol}/c_{s,{\rm mid}=1}$ are representative of the wind base.}
    \label{fig:am}
\end{figure}

\begin{figure}
    \centering
    \includegraphics[width=0.45\textwidth]{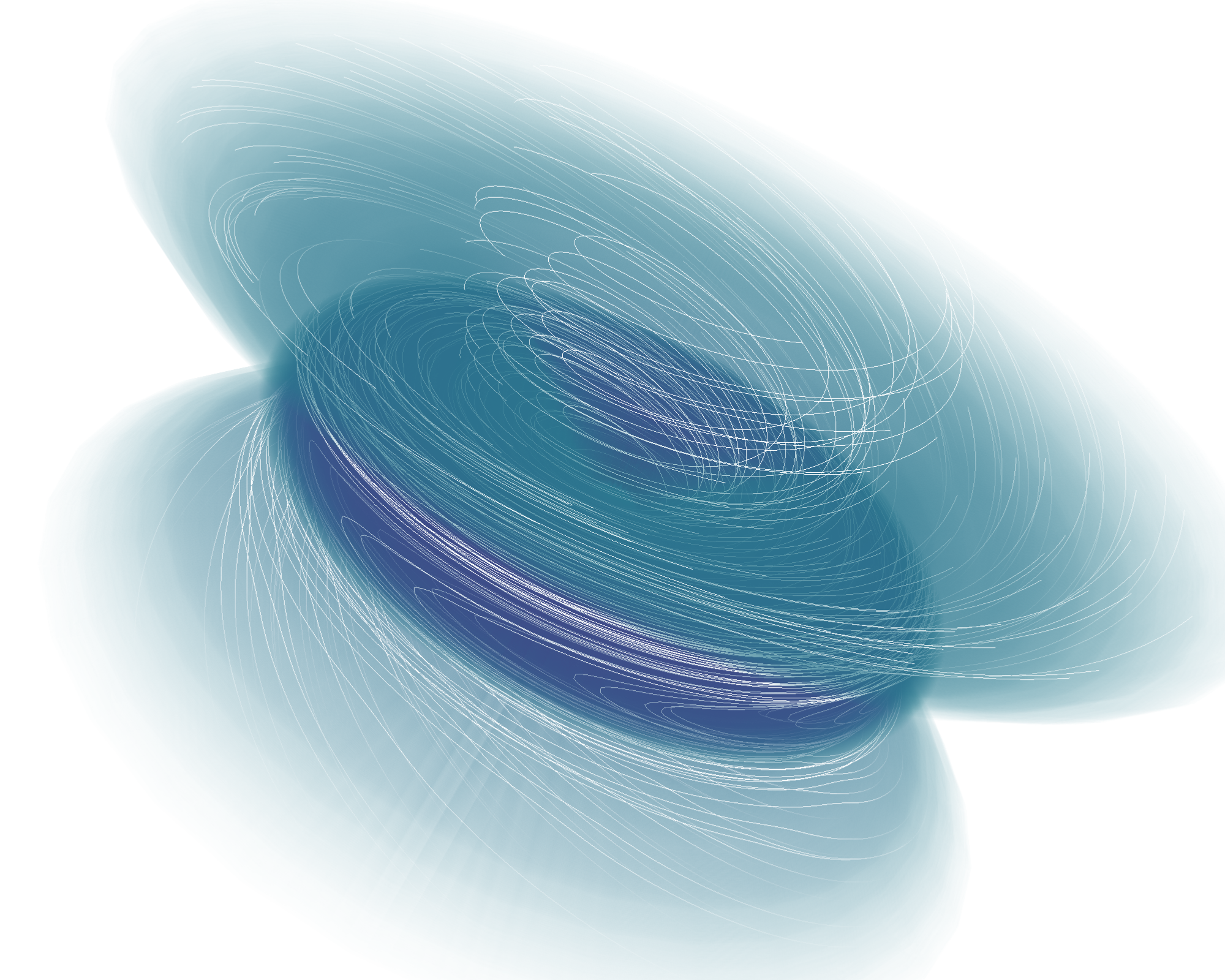} 
    \caption{A volume rendering of CO distribution in model \texttt{weakB}, and the white lines illustrate the highly twisted magnetic fields.}
    \label{fig:render}
\end{figure}
We also note that the outer disk has weaker magnetic diffusivities compared to previous work using the same on-the-fly thermochemistry calculation \citep{2019ApJ...874...90W, 2021ApJ...913..133H, 2023MNRAS.523.4883H} that focus on inner ($<$~50 au) disk. The outer disk's low density makes the ionization by diffused radiation much more effective. The attenuation column density of down-scattered X-rays are $1.5\times10^{21}~{\rm cm^{-2}}$ for atomic H and $7.5\times10^{23}~{\rm cm^{-2}}$ for molecular H$_2$ \citep{1999ApJ...518..848I, 2019ApJ...874...90W}. The column density of H$_2$ at 100 au in our setup is $\sim6\times10^{23}~{\rm cm^{-2}}$, below the attenuation limit. The lower density also means less frequent charge-neutral interaction, significantly reducing Ohmic resistivity. 
The ambipolar diffusion is also reduced, giving an Elsasser number $1\sim10$ compared to $0.01\sim0.1$ of the inner disk setups. This is similar to the value used in studies of spontaneous substructure formation \citep[e.g.,][]{2018MNRAS.477.1239S, 2021MNRAS.507.1106C, 2022MNRAS.516.2006H}. The outer disk also presents magnetic flux concentration, distorting local $v_\phi$ and surface density. With $Am>1$ and $\Lambda>100$, the MRI won't be suppressed. A higher resolution ($>$ 32 cells/h) 3D simulation is required to resolve MRI as suggested by previous works \citep[e.g.,][]{2022MNRAS.516.4660C}. As we focus on wind properties, this is beyond the scope of this work. Even though the inner region has a similar $Am$, the magnetic flux redistribution stops within 50 au. This is likely due to the strong Ohmic resistivity, as the Ohmic Elasser number $\Lambda$ drops well below unity at this portion of the disk, as shown in the bottom panel of Figure~\ref{fig:am}. 

We also generated a volume rendering of the CO distribution of the MHD wind in Figure~\ref{fig:render}. We can clearly see the dense disk and a diffusive wind that extends beyond the CO disk's radial boundary. There is also a cavity in the pole region as CO molecules are being photodissociated. Note for illustrative purposes, the data used for this rendering is not exactly the same as model \texttt{weakB} that only includes the upper hemisphere, instead, it has both hemispheres, i.e., a ``full disk'' setup.

\subsection{Dynamic properties of winds}
\label{sec:wind-dyn}

\begin{figure*}[t]
    \centering
    \includegraphics[width=1\textwidth]{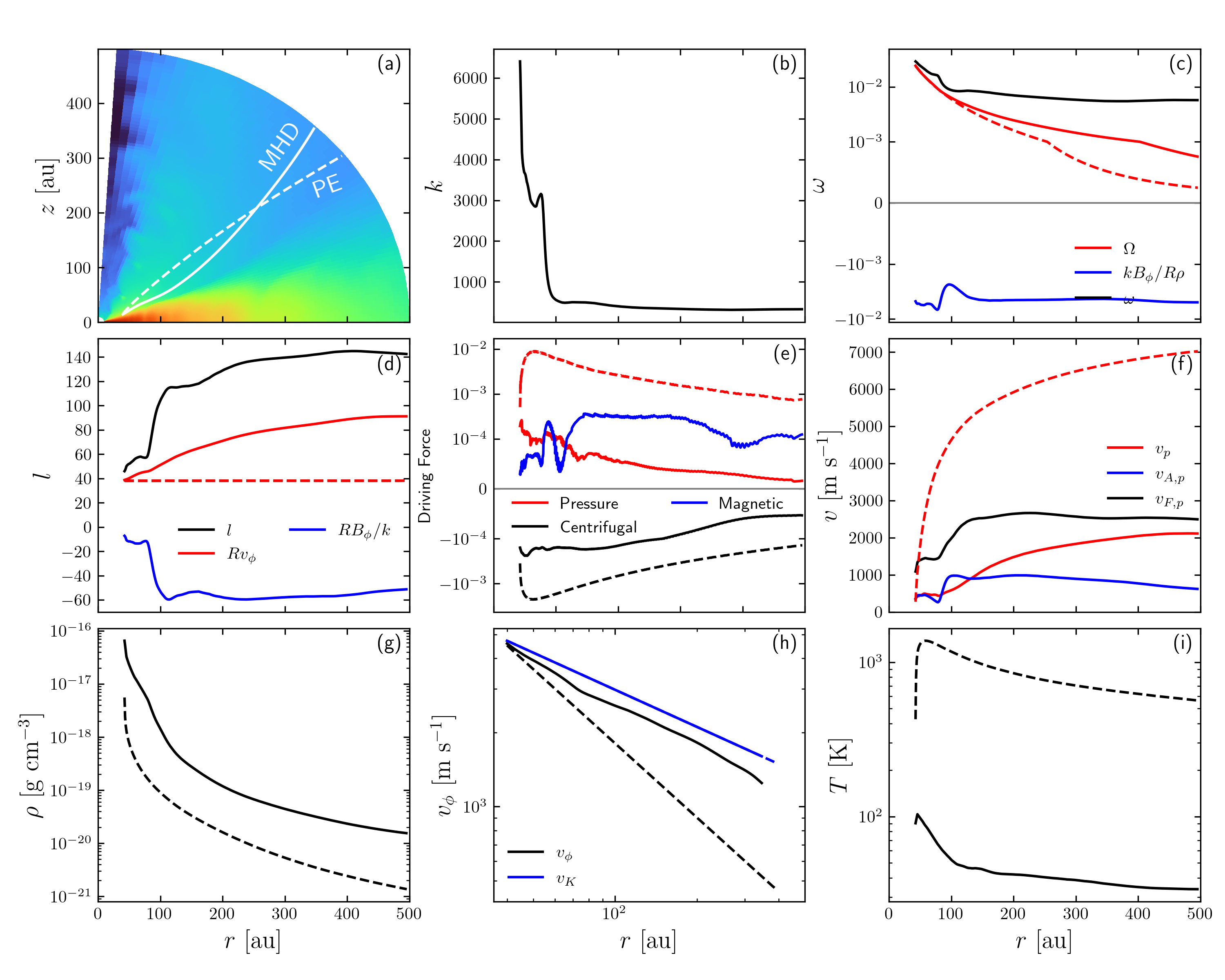}
    \caption{Various conserved quantities (k, l, $\omega$), the wind driving force, characteristic speeds (poloidal speed, Alfv\'{e}n speed, and fast magnetosonic speed) gas density, azimuthal velocity, and temperature along a fluid streamline shown as the white curve in the upper-left panel. The solid lines are from model \texttt{strongB} and the dashed lines are from \texttt{XUV}. The gas density background in panel (a) is from \texttt{strongB}. All of the variables have been at the snapshot of t=$6\times10^4$~yr. ``Symlog'' y-axis scale is used in panel (c) and (e), with a linear threshold set at $10^{-3}$ and $10^{-4}$, respectively.}
    \label{fig:lkwe}
\end{figure*}

In this section, we analyze the dynamic properties of the two different disk winds: photoevaporative (PE) wind and magnetohydrodynamic (MHD) wind. PE wind is launched when the local thermal velocity dispersion can break the gravity potential \citep{1994ApJ...428..654H}, and the high energy photons can provide enough ``fuel'' to overcome the cooling from adiabatic expansion in the outward movement. In fact, any significant heating is diminished after wind launching because of a lack of neutrals in the highly photoionized atmosphere \citep{2024arXiv240115419L}. The gas patch that leaves the disk surface is effectively detached from the rotating disk, thus exerting no torque on the disk, and the angular momentum of the gas patch in the wind is conversed. The wind-filled atmosphere is mainly pressure-supported and rotates slowly. 

To better compare the wind dynamics, in Figure~\ref{fig:lkwe} we set the anchor point at R=40 au and z=14.5 au above the midplane for both Model \texttt{strongB} and Model \texttt{XUV}, at the t=$6\times10^4$ year snapshot. In panel (a), The MHD wind streamline (solid) has an overall convex shape while the PE wind streamline (dashed) follows a slightly concave function. At the PE wind launch point, the gas pressure gradient is dominated by the vertical component because of the denser disk, so the initial velocity is more vertical in the meridional plane. It quickly turns into radial motion as the effect from the disk is minimized in the upper atmosphere. 

In a steady laminar disk wind, the conservation of mass, angular momentum, and energy along poloidal streamlines is beneficial for diagnosing the mechanism of wind launching and acceleration. In the MHD model, the poloidal streamlines in the wind are well aligned with the magnetic field lines. The analysis will be performed on streamlines starting at the same wind launch point in both cases.

By separating {\bf v} and {\bf B} into the poloidal and the toroidal components ($\mathbf{v}=\mathbf{v_p}+\Omega R \hat{\phi}$, $\mathbf{B}=\mathbf{B_p}+\mathbf{B_\phi}$), and using mass conservation equation, we have the mass loading constant \citep{1967ApJ...148..217W, 1982MNRAS.199..883B,1996ASIC..477..249S,1999MNRAS.309..233O}: 
\begin{equation}
    k=\frac{\rho v_p}{B_p}.
    \label{eq:k}
\end{equation}
It describes the mass flux density per unit of poloidal magnetic flux, is conserved along a field line, so each field line has its own mass flux. Using the same equations in the azimuthal direction, the second constant is
\begin{equation}
    \omega=\Omega-\frac{k B_\phi}{\rho R}
\end{equation}
It can be loosely interpreted as the rotation rate of the field line. This constant describes how the rotation is ``lagging behind'' over a large distance \citep{1996ASIC..477..249S}. 
With the angular momentum equation, we have the third constant
\begin{equation}
    l=R(v_\phi-\frac{B_\phi}{k})
\label{eq:l}
\end{equation}
which is the specific angular momentum of the wind. 

These conserved quantities (k, $\omega$, l) are plotted in panels (b), (c), and (d), of Figure~\ref{fig:lkwe}. For the MHD disk wind (solid lines), these quantities only start to follow the conservation laws beyond r$>$140 au. As shown by the characteristic speeds in Figure~\ref{fig:lkwe}f, the Alfv\'{e}n point, where the poloidal velocity $v_p=\sqrt{v_r^2+v_\theta^2}$ equals the poloidal Alfv\'{e}n speed $v_{A,p}=\sqrt{B_p^2/\rho}$ is also $\sim$ 140 au. The poloidal fast magnetosonic velocity $v_{F,p}^2 =(v_A^2+c_s^2)/2+\sqrt{(v_A^2+c_s^2)^2-4c_s^2v_{A,p}^2}$, where the $v_{A}=\sqrt{(B_p^2+B_\phi^2)/\rho}$ is the Alfv\'{e}n speed, is always faster than the poloidal velocity in our simulation domain. The fast magnetosonic point, where $v_p=v_{F,p}$, is usually at very large distances \citep{2016ApJ...818..152B}. Since the Alfv\'{e}n point is quickly stabilized, containing fast magnetosonic point is not crucial to wind kinematics \citep{2017ApJ...845...75B}.

The wind acceleration process is best understood by decomposing the poloidal forces \citep{2017ApJ...845...75B}:
\begin{equation}
\frac{\d v_p}{\d t}=-\frac{1}{\rho}\frac{\d p}{\d s} + \left( \frac{v_\phi^2}{R}\frac{\d R}{\d s} - \frac{\d\Phi}{\d s}\right) - \frac{B_\phi}{\rho R}\frac{\d(RB_\phi)}{\d s}  
\end{equation}
where the three terms on the right-hand side correspond to the thermal pressure
gradient, net centrifugal force, and Lorentz force from the
toroidal magnetic pressure gradient. We define the net centrifugal force as the difference between the centrifugal force and gravitational acceleration. As seen in Figure \ref{fig:lkwe}e this force is consistently negative, indicating that corotation is not maintained to drive centrifugal ejection, as in the traditional magnetocentrifugal wind model. Instead, the acceleration is dominated by the magnetic pressure gradient. This occurs because the poloidal fields in protoplanetary disk winds are too weak to enforce corotation, allowing them to be wound up by differential rotation, which generates strong toroidal magnetic fields.

For the PE wind in model \texttt{XUV}, the angular velocity $\Omega$ is slower than \texttt{strongB} (Figure~\ref{fig:lkwe}c), and the specific angular momentum $l=Rv_\phi$, is perfectly conserved all the way from the anchor point (Figure~\ref{fig:lkwe}d). The thermal pressure is the sole wind driving force (Figure~\ref{fig:lkwe}e) and is much stronger than the pressure gradient in model \texttt{strongB}. Compared to the MHD wind, the PE wind is more than ten times hotter (Figure~\ref{fig:lkwe}i),  three times much faster (Figure~\ref{fig:lkwe}f), and ten times lower density (Figure~\ref{fig:lkwe}g), even though the density at the anchor point is very similar (model \texttt{strongB} has a slightly denser inner disk because of accretion). In Figure~\ref{fig:lkwe}h, we find that the MHD wind is super-Keplerian and the PE wind is significantly sub-Keplerian. Angular momentum conservation of PE wind yields $v_\phi\sim1/R$ which decreases faster than $v_K$ when moving outward. For MHD wind, conservation of $l$ in Eq.~\ref{eq:l} means $B_\phi$ converts to $v_\phi$ in the wind to make it super-Keplerian. 

\begin{figure}
    \centering
    \includegraphics[width=0.45\textwidth]{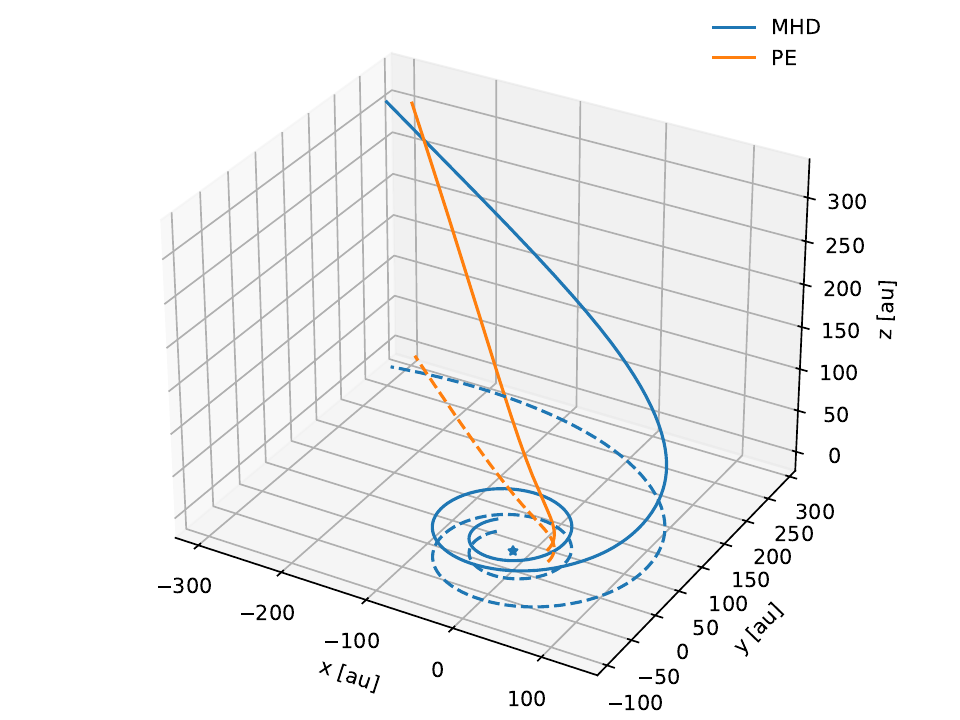}
    \caption{3D view of the wind streamlines, anchored at the same location at disk surface. The blue line is the MHD wind and the orange line is the photoevaporative wind. Note we rotated the MHD streamline azimuthally so the outer part of the streamline is closer to the photoevaporative wind for better comparison. The star marker at the origin represents the central star.}
    \label{fig:stream3d}
\end{figure}

Due to the 2.5D axisymmetric nature of our simulation domain, we can combine $\mathbf{v}_p$ and $\mathbf{v}_\phi$ to construct the 3D view of the two streamlines in Figure~\ref{fig:stream3d}. The PE wind follows an almost straight trajectory, while the streamline of the MHD wind winds up into approximately two full circles. From the projection at the midplane, after the initial one and three-quarters of a circle, the MHD wind streamline starts to ``straighten up'' roughly between 110 to 150 AU, which aligns well with the Alfv\'{e}n radius of 140 AU.

\section{LOS velocity map of Parametric Disks}
\label{sec:LOS}
\begin{figure}[htbp!]
    \centering
    \includegraphics[width=0.45\textwidth]{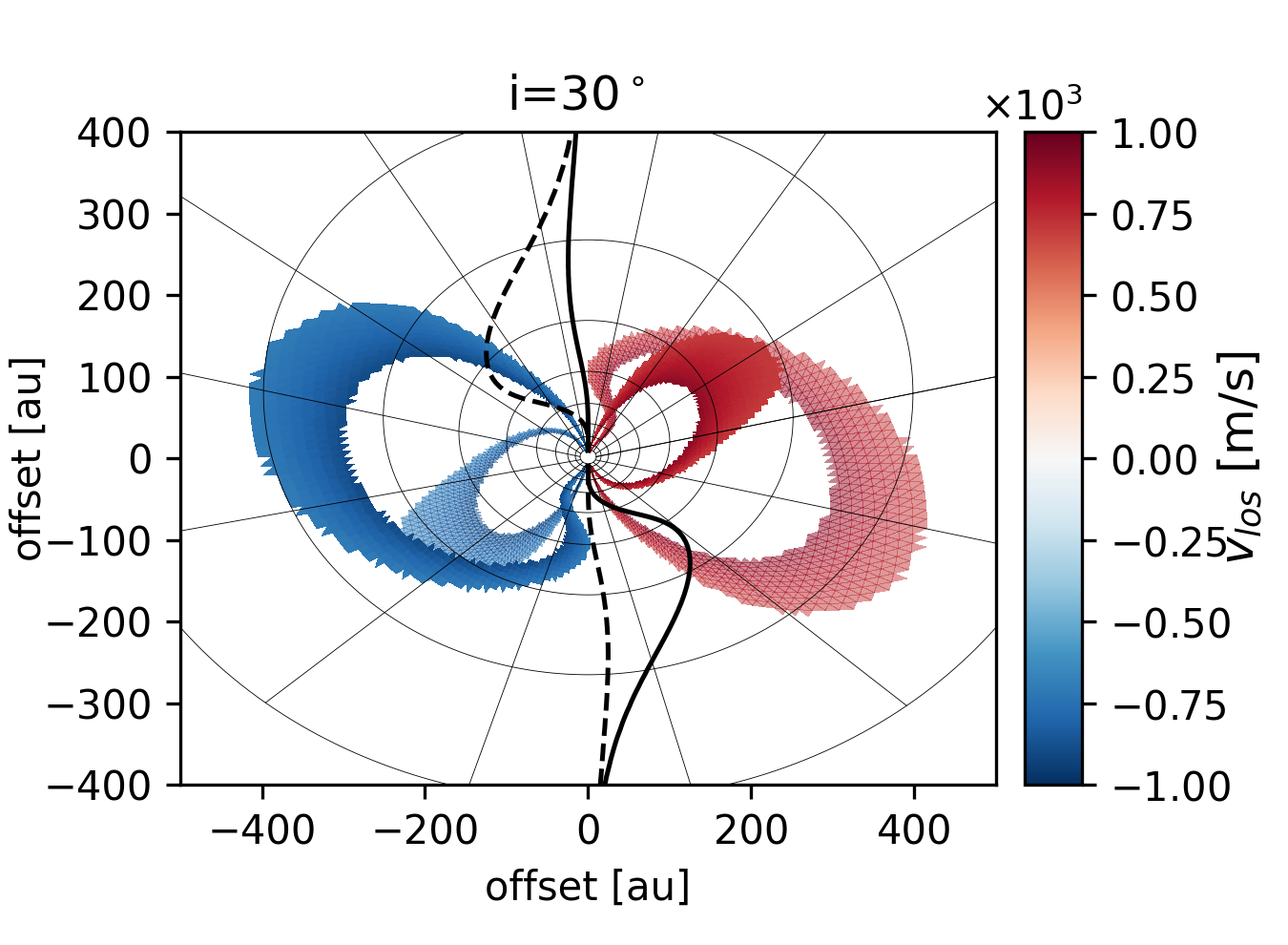} 
    \includegraphics[width=0.45\textwidth]{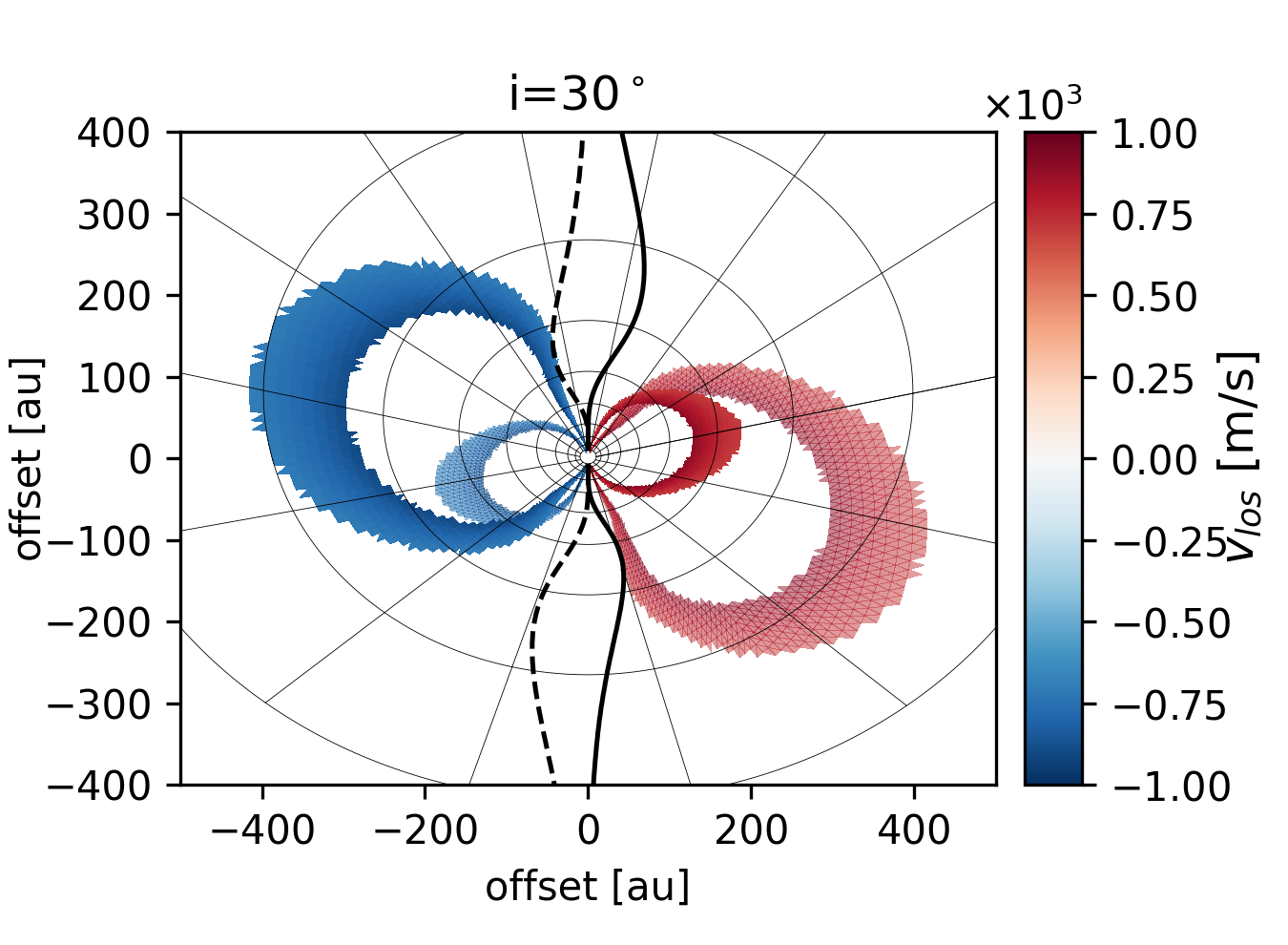} 
    \caption{Line-of-sight velocity map of a Keplerian disk with $v_r$ only (top) and $v_\theta$ only (bottom) on the z/R=$\pm0.4$ surfaces, with the top surface outlined by grids. The black and dashed lines through the center are the zero-velocity contours of the top and bottom disk surfaces, respectively. We masked the color map outside $700~m/s<\abs{v_{LOS}}<900~m/s$. The top surface is in solid color and the bottom is less saturated. Both panels share the same $v_\phi=v_K$, while the top panel only includes a non-zero $v_r$ profile, and the bottom panel only includes $v_\theta$.}
    \label{fig:model2face}
\end{figure}

Before making synthetic line emission observations, we would like to understand the line-of-sight (LOS) velocity distribution at a given emission surface. The inclination of the disk $i$ is defined as the angle between the disk angular momentum axis and the LOS. We set $i=30^\circ$, so the top surface is also the front surface closer to the observer, and the disk rotation is counter-clockwise. In this section, we focus on the MHD wind since the velocity structure is more complicated than the PE wind, and the latter is covered in the appendix. We also cover the effect of collapsing flow at the disk surface of the PE wind model. The probed emission surface could intersect the MHD wind having significant $v_r$ or $v_\theta$ velocity components simultaneously, making the butterfly pattern in channel maps deviate from the classical shape. As shown in Figure~\ref{fig:lkwe}f, the poloidal velocity plateaus quickly after the wind launching point. Because of the flared disk, a flat (cone in 3D, flat in 2D) surface with fixed $z/R=0.4$ would intercept the disk surface twice at the two ends, with the middle section in the wind region. The radial profile of the poloidal velocity in this flat surface should increase quickly as the gas is accelerated into the atmosphere, then decrease slowly as it approaches the outer disk surface. We use a simple lognormal function to mimic this behavior:
\begin{equation}
    f(x; \mu, \sigma) = v_0\frac{1}{x \sigma \sqrt{2 \pi}} \exp\left(-\frac{(\ln x - \mu)^2}{2 \sigma^2}\right), \quad x = r/r_0,
\end{equation}
where $\mu=0$, $\sigma=0.5$, and $r_0=200$~au. For $v_r$, $v_0=10^3$~m/s, and for $v_\theta$, $v_0=\pm400$~m/s for lower and upper surfaces. The rotation velocity is set to be $v_K=\sqrt{GM R^2/(R^2+z^2)^{3/2}}$, the Keplerian velocity with vertical pressure balance. The result is shown in Figure~\ref{fig:model2face}, symmetry breaking between both the redshifted and blueshifted side, and the front and backside disk surfaces. At the front surface, the blue-shifted pattern is enlarged, the redshifted pattern shrinks, and vice versa for the backside. With $i=30^\circ$ and $z/R=0.4$, the LOS projection of $v_\theta$ is always negative (blueshift) and $v_r$ is mostly negative except for the top central sector. The area with a certain redshifted velocity must move inward to achieve the same LOS velocity as a pure Keplerian disk, resulting in a smaller redshifted pattern. With positive $v_r$ perturbation, the entire pattern is twisted counter-clockwise. Take the zero velocity channel as an example: at the bottom sector, $v_r$ adds blueshift so the zero velocity area moves to the left, the redshifted half from Keplerian motion; at the top sector, $v_r$ contributes to redshift that moves the zero velocity area to the right. Similarly, the effect of $v_\theta$ perturbation bends the zero velocity pattern on the front surface to the right side. Because $v_\theta$ is better aligned with LOS at the top half sector, the stronger ``bending'' moves the blueshifted pattern much closer to the minor axis, which could be mistakenly interpreted as a higher $z/R$.  Both ``twisting'' and ``bending'' are not uniform as the projection effect depends on the position angle. In the real world, the shape of the emission pattern could be further complicated by the different velocities integrated over the whole emission region. 

\begin{figure}[t!]
    \centering
    \includegraphics[width=0.45\textwidth]{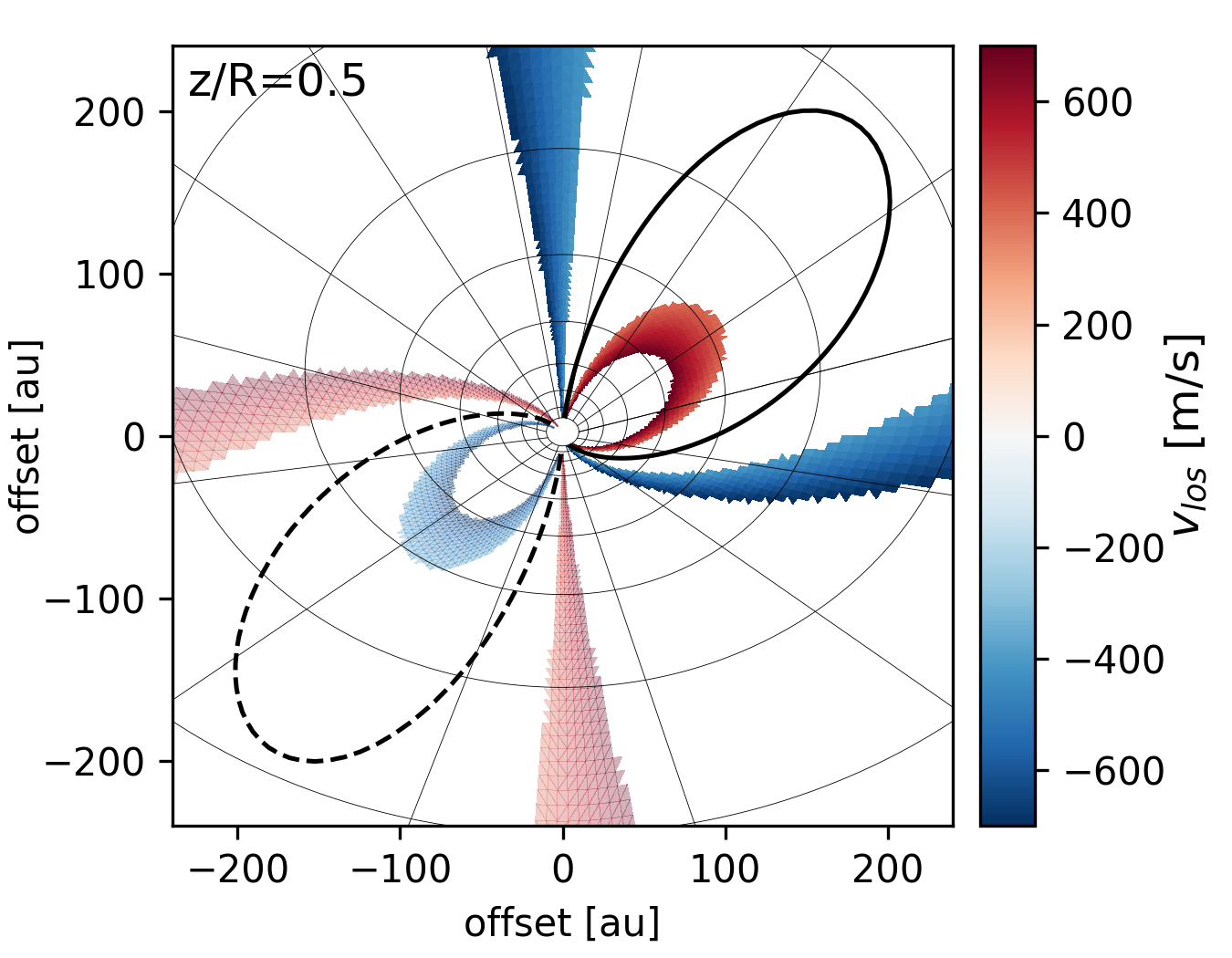}
    \includegraphics[width=0.45\textwidth]{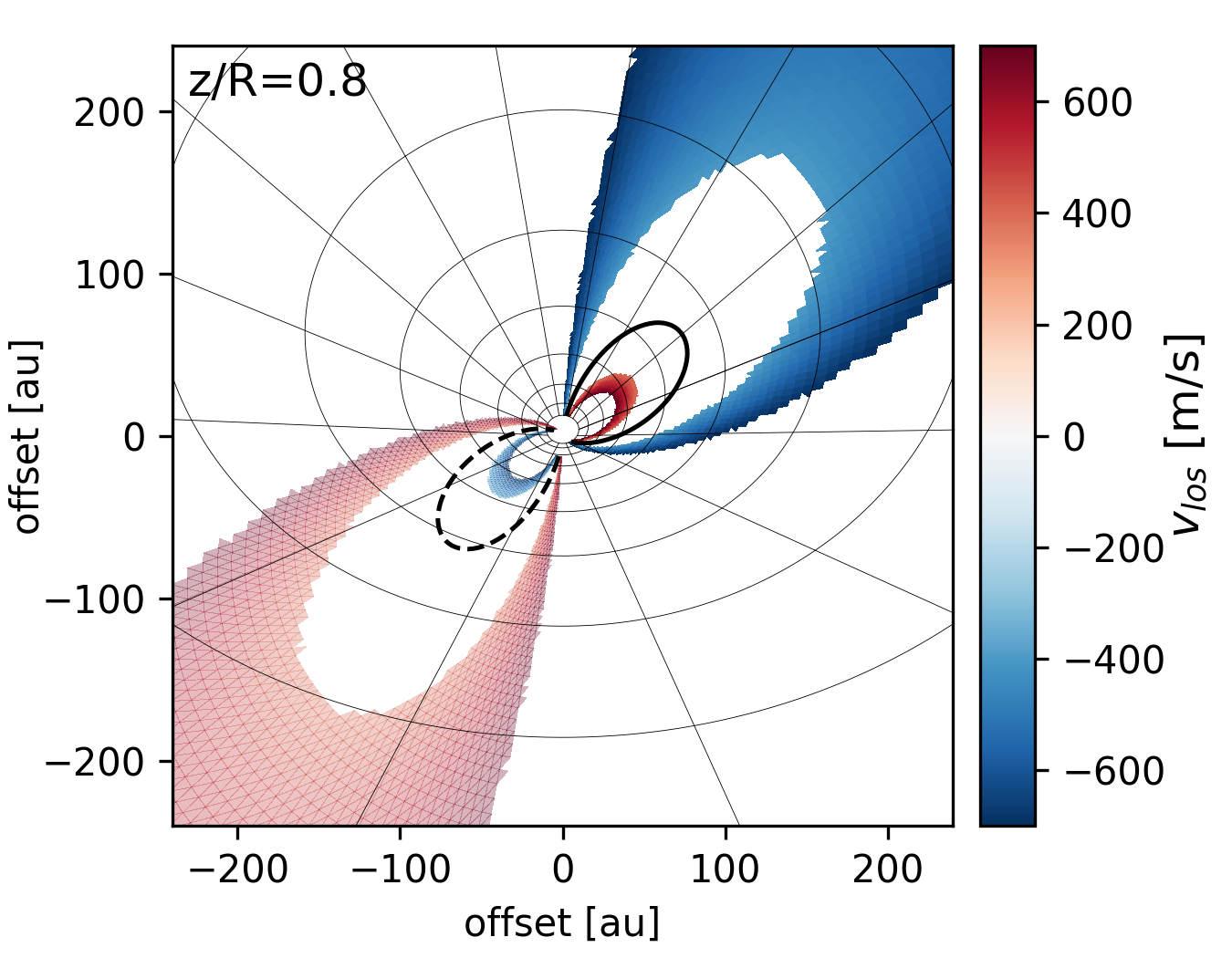} 

    \caption{Similar to Figure~\ref{fig:model2face}, now on two more elevated surfaces, mediated by global $v_r$ and $v_\theta$ at the same time. We masked the color map outside $400~m/s<\abs{v_{LOS}}<700~m/s$. }
    \label{fig:model_elevated}
\end{figure}

To better understand emission morphology from the denser MHD wind, we adjust $z/R$ to 0.5 and 0.8. For simplicity we set $v_r=1500~{\rm m~s}^{-1}$ and $v_\theta=-500~{\rm m~s}^{-1}$ to be constant as a rough approximation for the upper atmosphere (above wind launching point). We set $v_\phi=1.2v_K$ to account for the super-Keplerian rotation. For $z/R=0.5$, we see the top panel in Figure~\ref{fig:model_elevated} that the redshifted side on the front emission surface is squeezed into the first quadrant while the blueshifted side occupies the rest. With higher $z/R$, the far side of the elevated emission surface is almost perpendicular to or tilted towards the LOS. Combined with an overall higher wind velocity, the radial outflow is more effective in canceling the redshift from disk rotation. Secondly, negative $v_\theta$ would still contribute to blueshift at least in the top sector. These two factors combined to squeeze the zero velocity channel into a closed loop. The $z/R=0.8$ case is to account for emissions from atomic lines like [C I] that could trace higher layers. Only the very inner regions could rotate fast enough to counter the blue shift from $v_r$, so the zero velocity channel is about half the size of $z/R=0.5$. Now we could detect two loops located in opposite directions in one channel, e.g., for the $\sim-0.7~{\rm km~s}^{-1}$ channel, a large loop at the first quadrant from the front surface and a much smaller loop at the third quadrant is the emission from the backside.

\begin{figure}[hbtp!]
    \centering
    \includegraphics[width=0.45\textwidth]{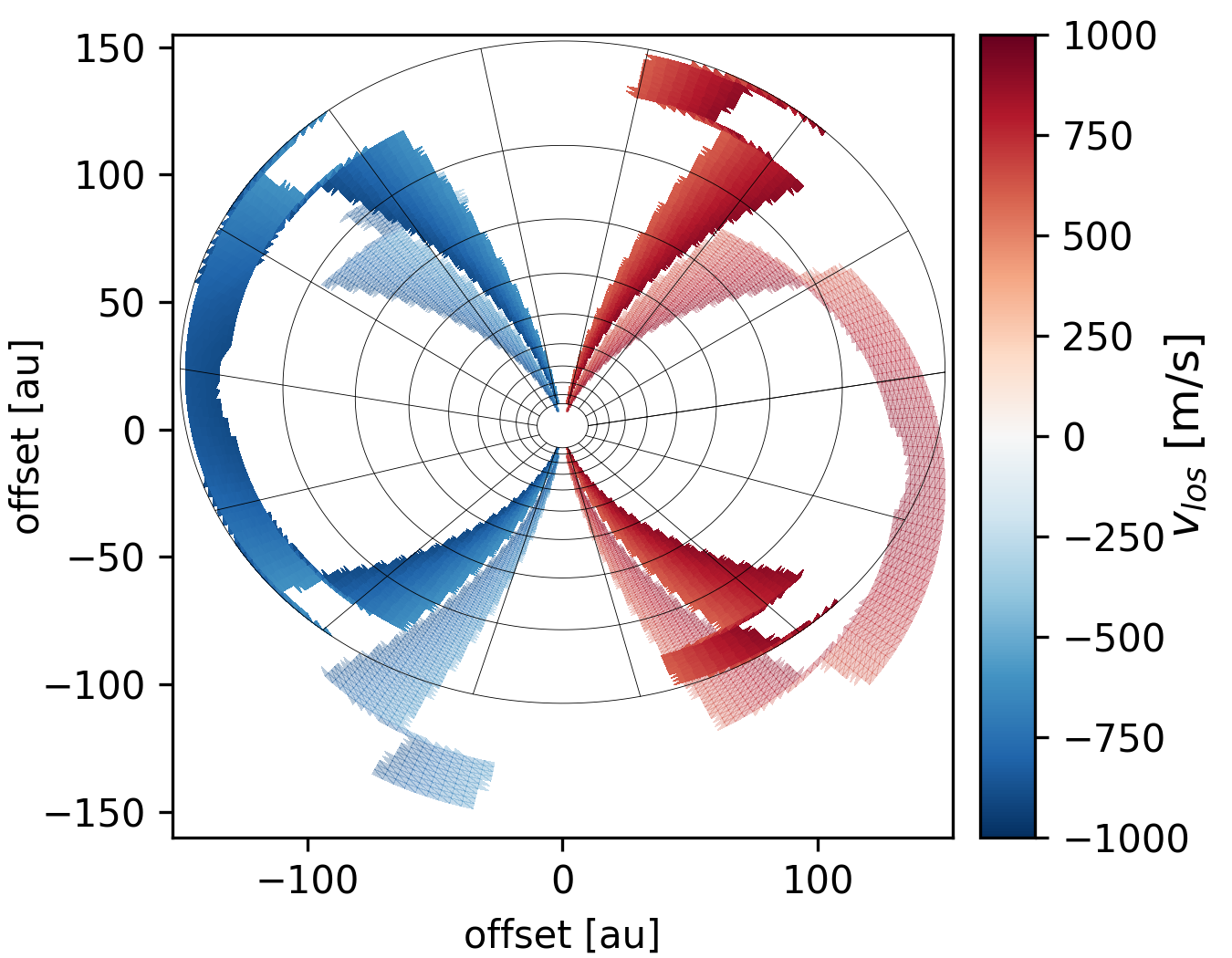}
    
    \caption{Line-of-sight velocity map mediated by a local $v_\theta$. To highlight the effect of perturbation in a narrow velocity range, we masked the disk outside $600~m/s<\abs{v_{LOS}}<900~m/s$.}
    \label{fig:model}
\end{figure}

Apart from the global velocity structures, a local $v_\theta$ perturbation can also alter the appearance of channel maps. To mimic the effect of the collapsing flow at the disk surface in model \texttt{XUV}, we add a localized $v_\theta$ bump of $0.15v_\phi$ to the Keplerian rotating surface at z/R=0.3, within 130 au $<$ R $<$ 150 au. In Figure~\ref{fig:model}, we show the disk's front surface with $600~\text{m/s} < \abs{v_{LOS}} < 900~\text{m/s}$ at an inclination of $30^{\circ}$. The most distinct feature is the ring that connects the two tips of the butterfly pattern in the blueshifted channel. Without the collapsing flow that mitigates the blueshift from $v_\phi$, the area of the ring should appear in a more blueshifted channel (as the tip of a smaller ring) because of rotation. The same ring would appear on the redshifted half if there is a flow of updraft. Since the collapsing flow always moves away from the observer, it makes the area that should be on a slower channel (closer to zero velocity) appear on this channel. This is why we see the two red tips closer to the center line than the main pattern. 

\section{Morphology of disk winds in ALMA line observations}
\label{sec:obs}

\subsection{CO emission}
\begin{figure*}[htbp!]
\centering
    \includegraphics[width=1\textwidth]{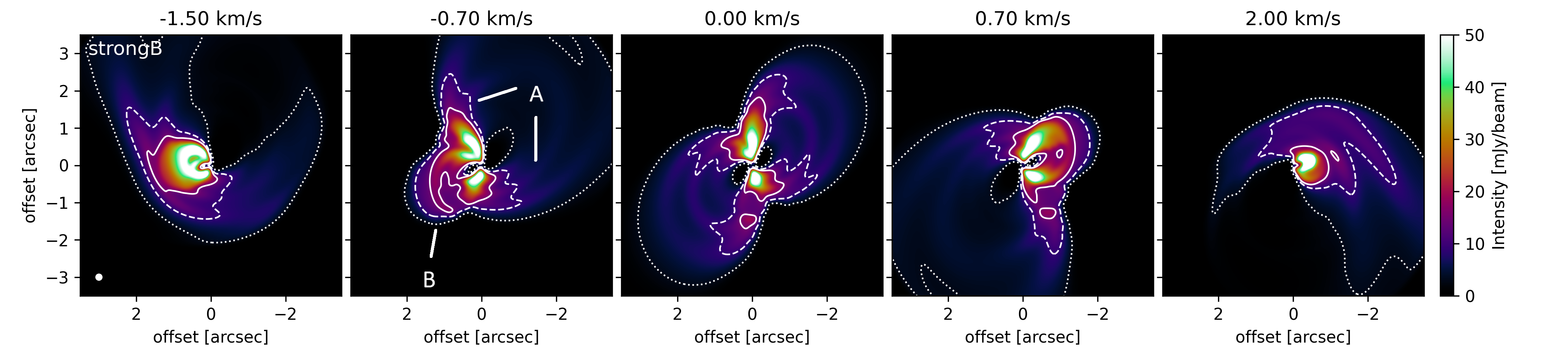}
    \includegraphics[width=1\textwidth]{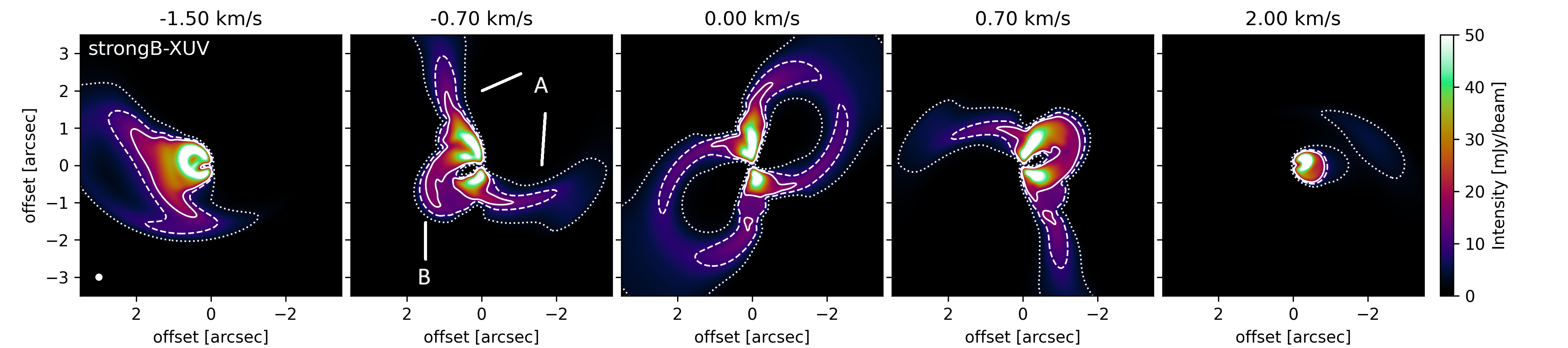}
    \includegraphics[width=1\textwidth]{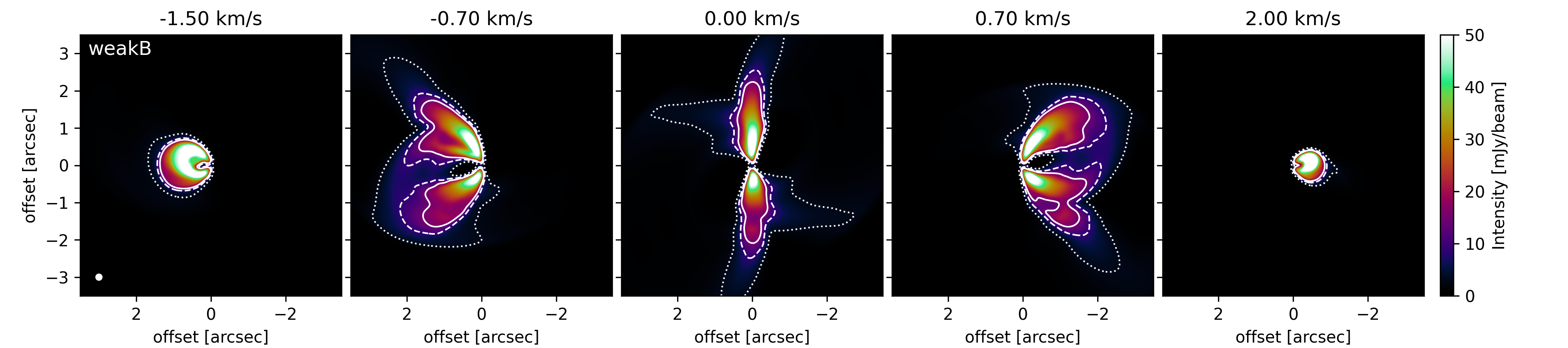}
    \includegraphics[width=1\textwidth]{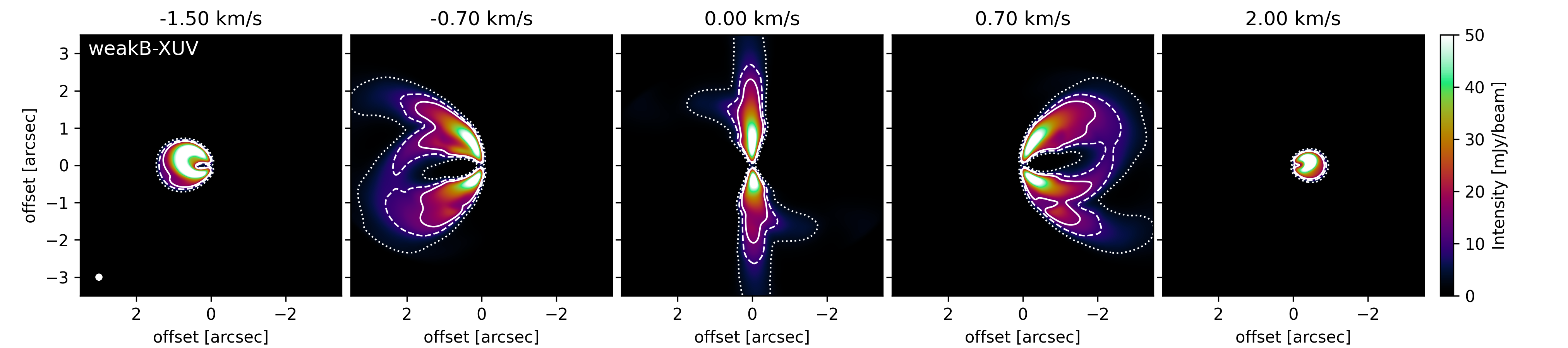}
    \includegraphics[width=1\textwidth]{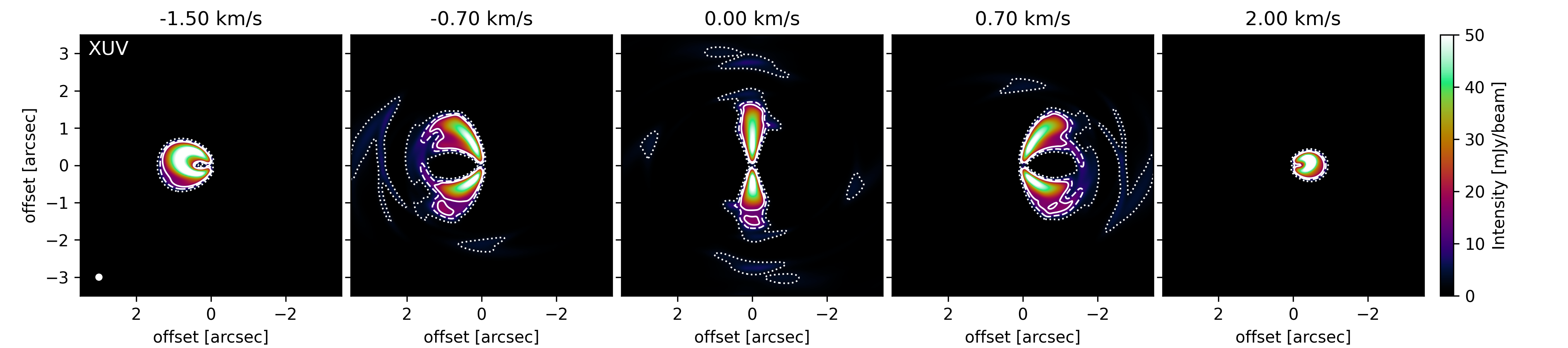}
    \caption{CO J=2-1 channel maps convolved with a $0\farcs15$ beam (white spot at the bottom left corner). We saturated part of the disk to highlight the emission from the wind. The white contours represent three detection limits: $5\sigma$ (solid), $3\sigma$ (dashed), and $1\sigma$ (dotted). Letter A labels the horn-like structures that originate from the front surface, and letter B labels the emission from the backside.} 
    \label{fig:peakchan}
\end{figure*}

\begin{figure}[htbp!]
    \centering
    \includegraphics[width=0.45\textwidth]{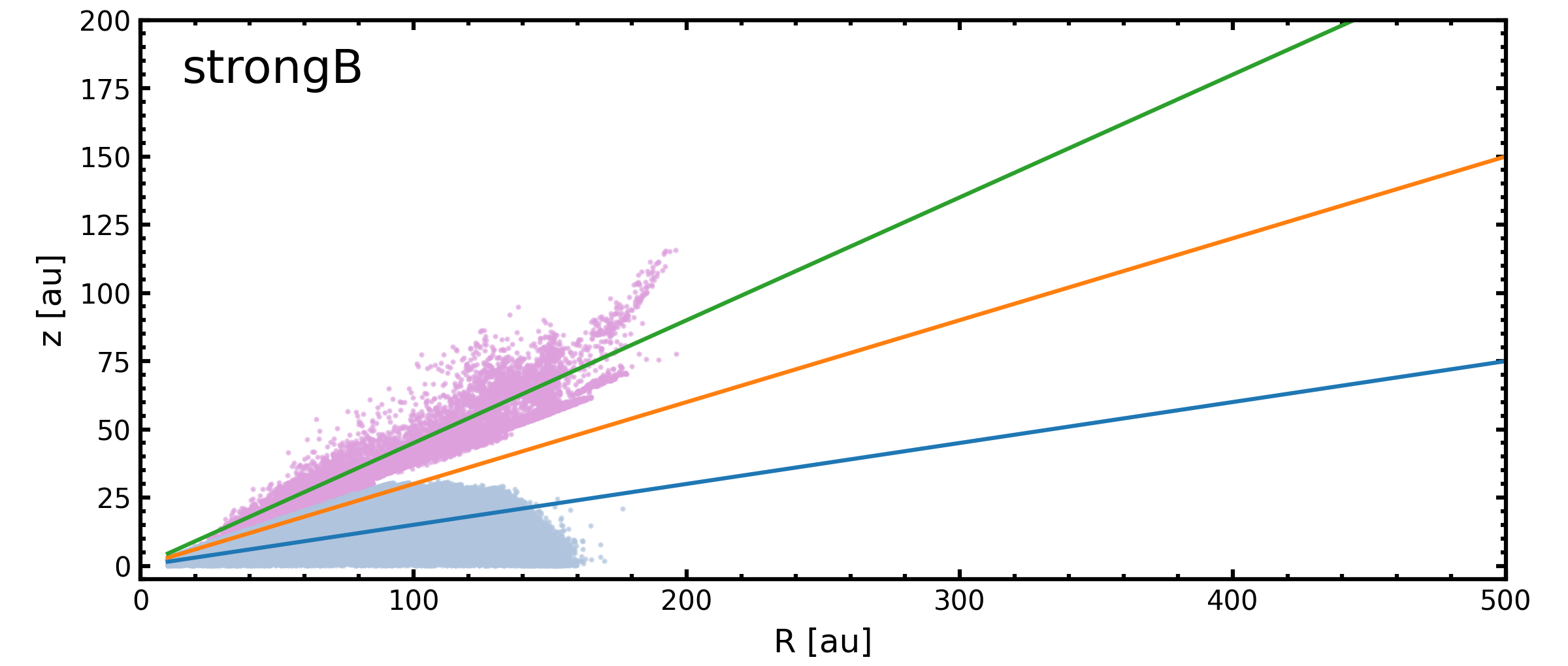}
    \includegraphics[width=0.45\textwidth]{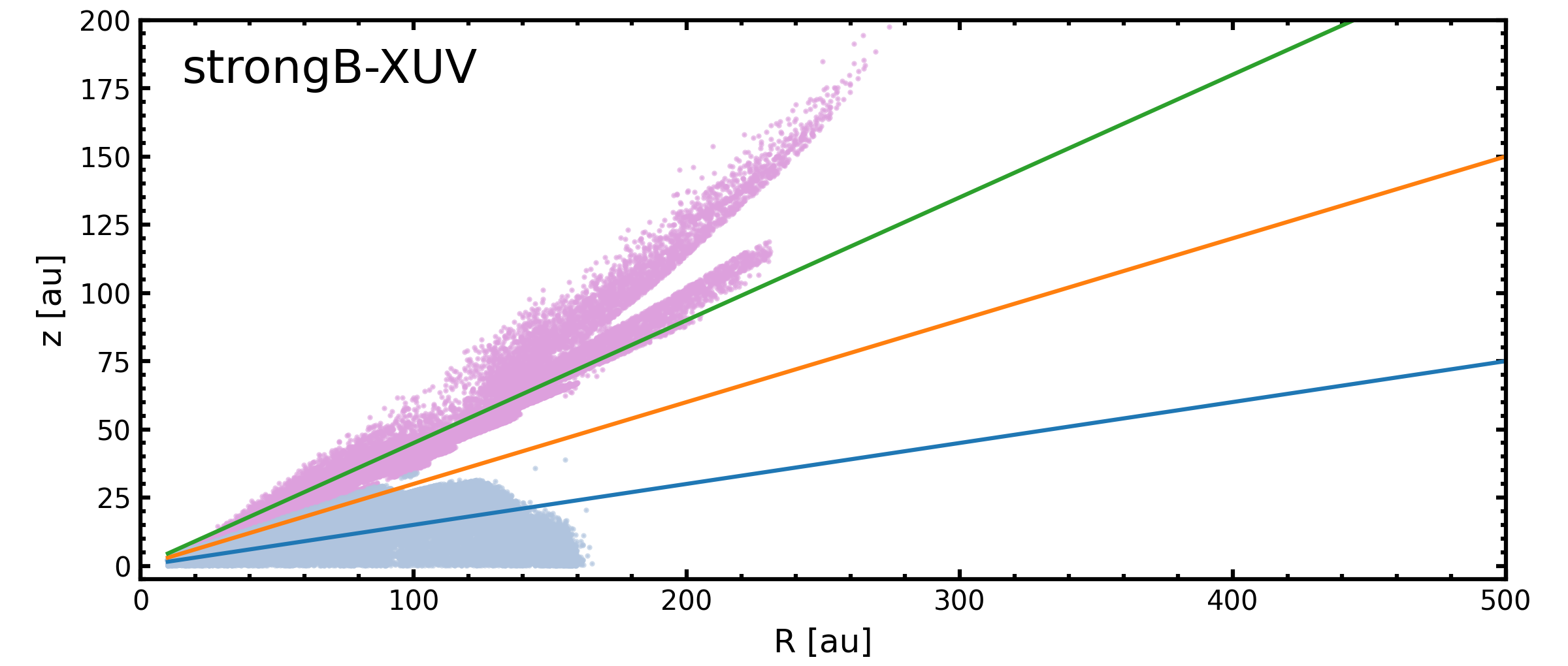}
    \includegraphics[width=0.45\textwidth]{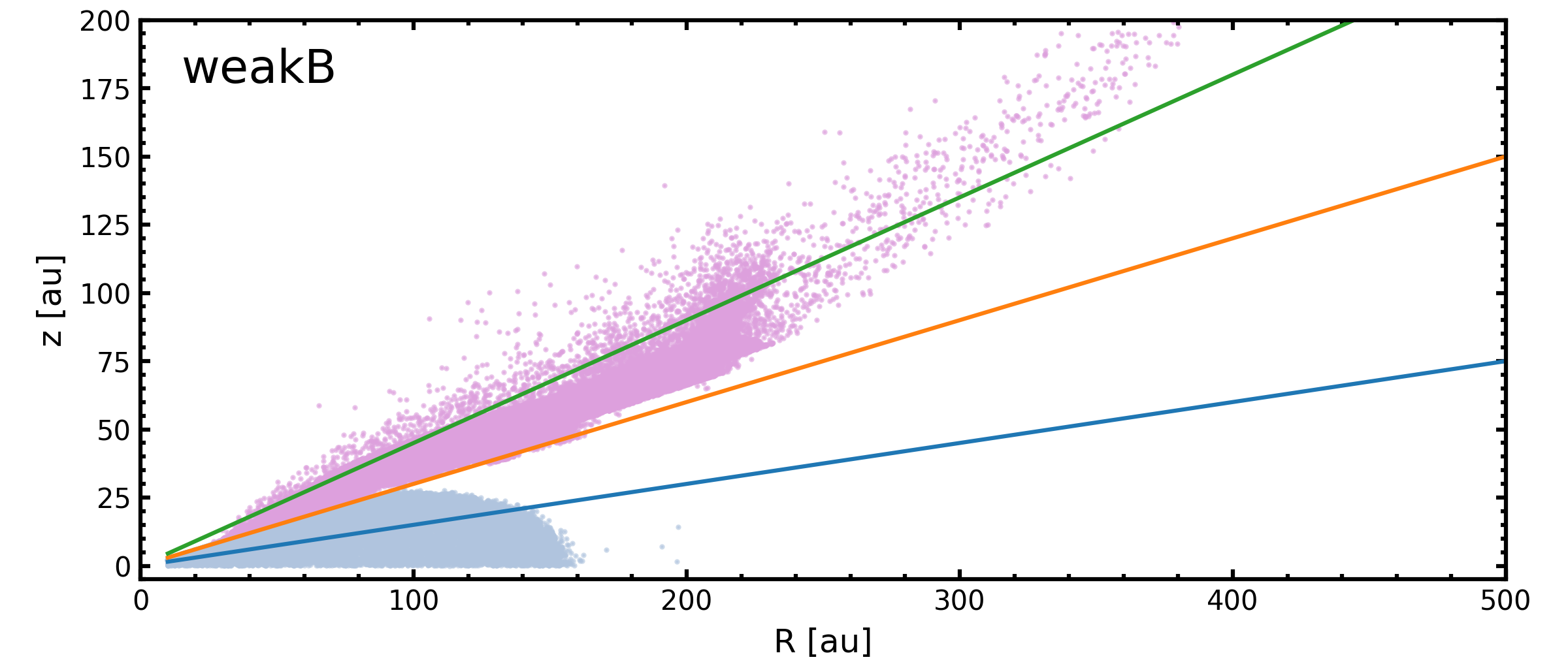}
    \includegraphics[width=0.45\textwidth]{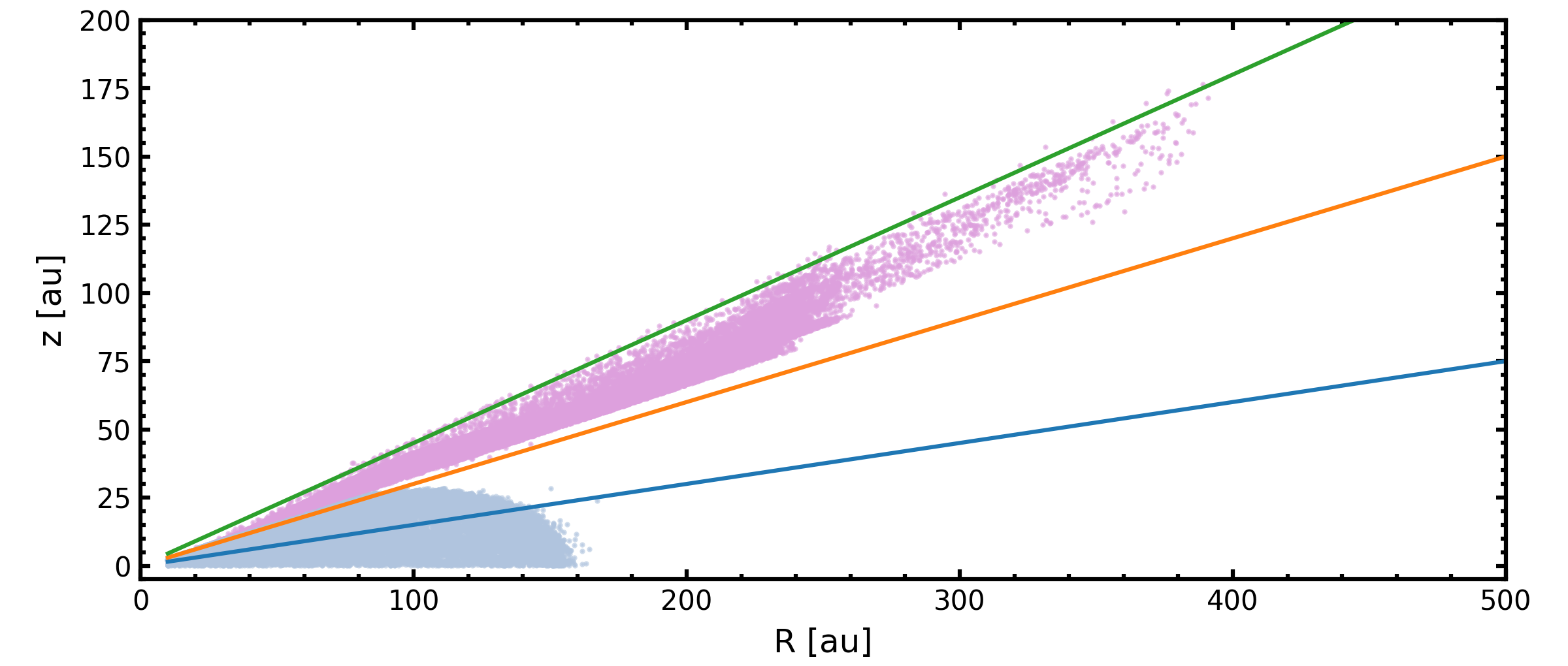}
    \includegraphics[width=0.45\textwidth]{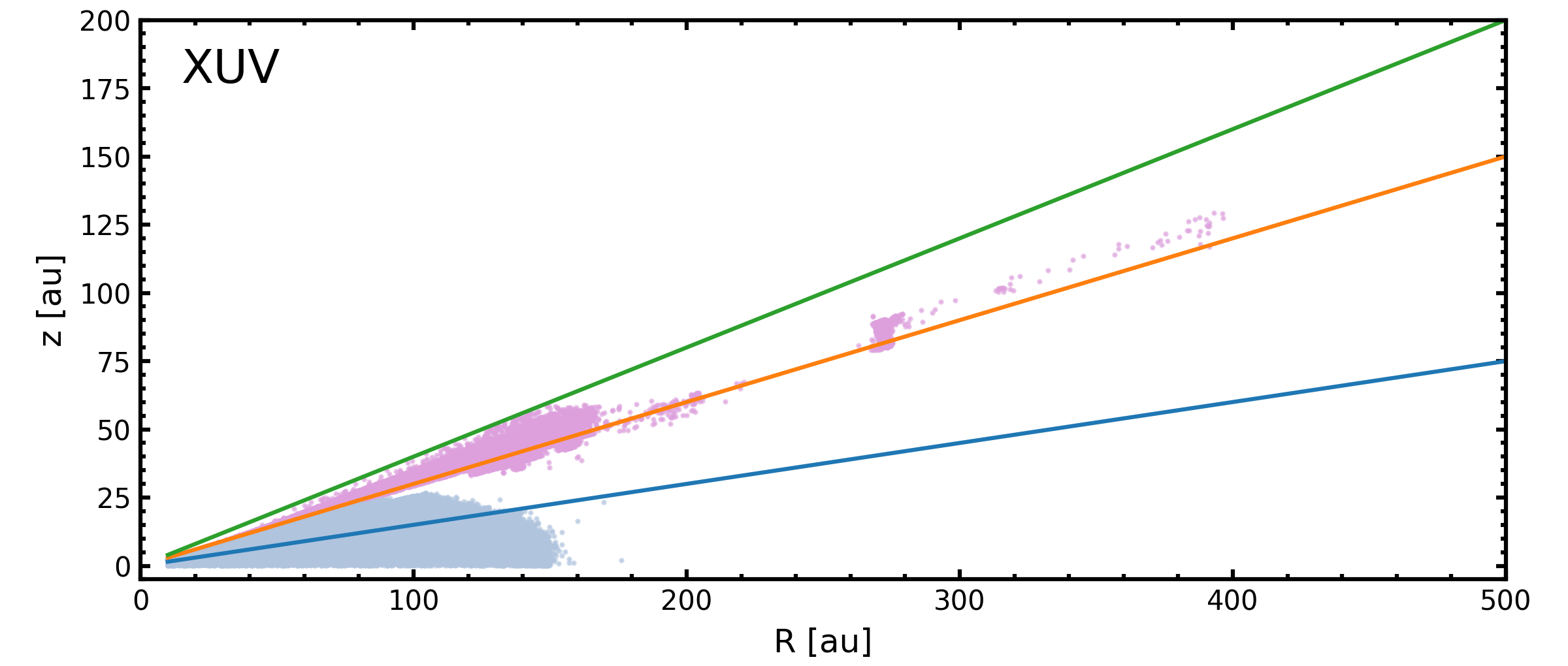}    
    \caption{Emission surface of optical depth~$\tau=1$. From top to bottom: model \texttt{strongB} ($\beta=10^4$), model \texttt{strongB-XUV} ($\beta=10^4$ with 300eV), model \texttt{weakB} ($\beta=10^5$), model \texttt{weakB-XUV} ($\beta=10^5$ with 300 eV), model \texttt{XUV} (no magnetic field, 300 eV). The pink dots indicate emissions from the atmosphere/wind region, and the light blue dots are from the disk. The three lines correspond to z/R=0.15, 0.3, and 0.45 in all panels except in panel \texttt{XUV}, where the top surface has z/R=0.4.}
    \label{fig:surf}
\end{figure}

In the previous section, we found that PE winds and MHD winds differ significantly in density, velocity, and directionality. The PE wind follows a straight line at higher velocities, while the MHD wind maintains rotational motion even at high altitudes. How do these features manifest in ALMA observations? To investigate the observability of disk winds in line observations, we generated $^{12}$CO $J=2-1$ emission maps using \texttt{RADMC-3D} \citep{2012ascl.soft02015D}, assuming a disk inclination of $30^\circ$. The temperature used for these calculations was directly taken from our simulations, as the dust temperature estimated by the \texttt{mctherm} module in \texttt{RADMC-3D} is not necessarily the same as the gas temperature in the low-density regions of the disk's atmosphere. Using temperature from the simulations also guarantees that the chemistry and line emission are treated self-consistently with the same temperature profile. Due to the absence of thermal Monte Carlo calculations, we only have two active parameters in \texttt{radmc3d.inp} control file. \verb|scattering_mode_max = 0|: dust scattering is not treated. \verb|tgas_eq_tdust = 1|: radmc3d reads the \verb|dust_temperature.inp| file and then equate the gas temperature to the dust temperature. Here the \verb|dust_temperature.inp| is written with gas temperature from (M)HD simulations. The central star is set to have solar properties. We created 121 velocity channels spanning from $-6~{\rm km~s}^{-1}$ to $6~{\rm km~s}^{-1}$, with a step size of $0.1~{\rm km~s}^{-1}$. All disks were placed at a distance of 100 pc, meaning 1 arcsecond in the channel map corresponds to 100 au in physical units. Each channel was convolved with a $0\farcs15\times0\farcs15$ beam using \texttt{syndisk}, a beam size representative of high-resolution ALMA line observations \citep[MAPS;][]{2021ApJS..257....1O}. The original pre-convolved channel maps are presented in the appendices. A two-hour integration would result in a noise level of $\sigma~\simeq 3~{\rm mJy~beam}^{-1}$\footnote{ALMA Sensitivity Calculator: asa.alma.cl/SensitivityCalculator}.  The choice of two-hour is also to match the typical integration time in the MAPS campaign~\citep{2021ApJS..257....1O}. The selected channel maps are displayed in Figure~\ref{fig:peakchan}, with the corresponding emission surfaces ($\tau = 1$) for the whole surface shown in Figure~\ref{fig:surf}. Each point is a $\tau=1$ point calculated by \texttt{RADMC-3D} and projected on the R-z plane. We used the ``tausurf'' command to generate the 3D coordinate of each pixel's $\tau=1$ location along the line-of-sight, and then combine all dots into a 3D unstructured array. Due to disk inclination, for dots along the same radius, the axisymmetry of the top emission surface is not perfect, as the density and velocity profile along each line-of-sight are different. To better illustrate the emission surfaces, we flatten the 3D array via $\phi$ direction and plot the $\phi$-averaged emission surfaces in Figure~\ref{fig:surf}. Note the non-axisymmetry is more prominent in the wind region but much less so at the disk surface. To distinguish the contributions from the disk and the wind, we color-coded the $\tau=1$ surfaces light blue for the disk and pink for the wind. The separation criterion between the two components is straightforward: the disk's top boundary is defined as a smooth, continuous surface, while the wind is represented by regions distinctly above this boundary. Although wind emission is always above the disk, it does not necessarily obscure the disk. Instead, the velocity difference between the wind and disk components allows us to disentangle their origins and identify each contribution in the channel maps.

All disks exhibit the characteristic butterfly pattern associated with Keplerian rotation \citep{2013ApJ...774...16R, 2013A&A...557A.133D}. The differences mainly lie in non-Keplerian features and the vertical location of emission surfaces. Only the strong magnetic field cases ($\beta = 10^{4}$) display distinct disk wind signatures, characterized by excess emissions from higher altitudes above the disk surface, as shown in the top two rows of Figure~\ref{fig:peakchan}. In the zero velocity channel, two closed loops, anchored at the disk's top and bottom surfaces, form a tilted figure-eight structure, as illustrated in Figure~\ref{fig:model_elevated}. In the \texttt{strongB-XUV} model, the loop has a similar outer size but appears thinner due to an empty interior. This is the result of XUV radiation photodissociating CO in the polar regions. At lower altitudes, the MHD wind is dense enough to block XUV. The distance from the central star also helps reduce photodissociation, causing the upper emission surface in \texttt{strongB-XUV} to flare at a larger radius, as seen in Figure~\ref{fig:surf}. Emissions from the extended regions originate from elevated surfaces, creating a ``halo''-like conic structure visible in high-velocity channels, especially in \texttt{strongB}. However, these spatially extended emissions are challenging to detect as they are just above the $1\sigma$ level, while wind emissions closer to the inner disk exceed the $3\sigma$ limit. For instance, a significant portion of the figure-eight structure in \texttt{strongB-XUV} is well above this threshold. In the $-0.7~{\rm km~s}^{-1}$ channel, the elevated emission surface from the backside (labeled \texttt{B}) reaches a $5\sigma$ significance, explained by a high radial velocity ($>1~{\rm km~s}^{-1}$), as shown in Figure~\ref{fig:model2face}. These horn-like structures, extending away from the Keplerian disk (labeled \texttt{A}), originate from the same conic surface as the figure-eight structure, as depicted by the blue-dashed contour in Figure~\ref{fig:model_elevated}. Notably, both strong magnetic field models exhibit wind loss rates of $4\times10^{-8}~M_\odot~\text{yr}^{-1}$, which is exceptionally high compared to the typical range for Class II objects, spanning $10^{-12}$ to $10^{-8}~M_\odot~\text{yr}^{-1}$ \citep{2023ASPC..534..567P}.

A moderate level of wind signatures is present in models \texttt{weakB} and \texttt{weakB-XUV}. In \texttt{weakB-XUV}, the photodissociation-dominated polar region extends further toward the disk compared to \texttt{weakB}. The emission surface primarily resides below $z/R = 0.45$ (Figure~\ref{fig:surf}) due to the sparser wind. Being closer to the disk, the radially extended emission gives the Keplerian disk the appearance of being larger than in the more magnetized models. However, the actual CO disk, beneath the bright surface, maintains the same radial extent in all setups, constrained by CO freeze-out beyond $\sim$150 au at the midplane (Figure~\ref{fig:surf}). The zero velocity channel is not perfectly "straight" but shows slightly twisted outer tips caused by poloidal motion, consistent with predictions in Figure~\ref{fig:model2face}. These radial and azimuthal extensions at the tips correspond to the central portion of the figure-eight structure, forming part of the conic emission surface similar to more magnetized models. However, the signal strength at $1\sigma$ makes direct detection challenging. The $\pm0.7~{\rm km~s}^{-1}$ channels show more radially extended emission, though the back surface appears less elevated than in the stronger field cases, as confirmed by the $\tau=1$ points in Figure~\ref{fig:surf}. The detection of wind is further hampered by low wind loss rates of $4.5\times10^{-9}~M_\odot~\text{yr}^{-1}$ for the MHD wind, or $1.8\times10^{-8}~M_\odot~\text{yr}^{-1}$ when combined with photoevaporation. The upper atmosphere lacks sufficient CO to trace the wind directly. Nevertheless, features like the twisted zero velocity map could serve as indirect evidence of a global radial outflow, offering a potential observational signature of the wind in these weaker field models. 

Model \texttt{XUV} does not exhibit direct wind signatures, as there is no detectable emission from regions significantly above $z/R = 0.4$, with the emission surface dropping to $z/R \sim 0.3$ beyond 180 au. The disk surface is less extended compared to the weak field models, only slightly protruding above the cold disk. As a result, this model produces the most Keplerian disk pattern among all five cases. The only detectable ($\geq 3\sigma$) non-Keplerian features are the partial ring-like structures at the outer edges of the disk surfaces in the $\pm$0.7 km/s channels. These features connect the two tips of the emission and correspond to the $v_\theta$/$v_z$ local perturbations seen in Figure~\ref{fig:model}, where the expanding wind interacts with the disk surface just beyond the puffed-up layer. Despite the PE wind having a loss rate close to $1\times10^{-8}~M_\odot~\text{yr}^{-1}$, directly detecting the wind remains challenging in this model.

\begin{figure}[ht]
    \centering
    \includegraphics[width=0.45\textwidth]{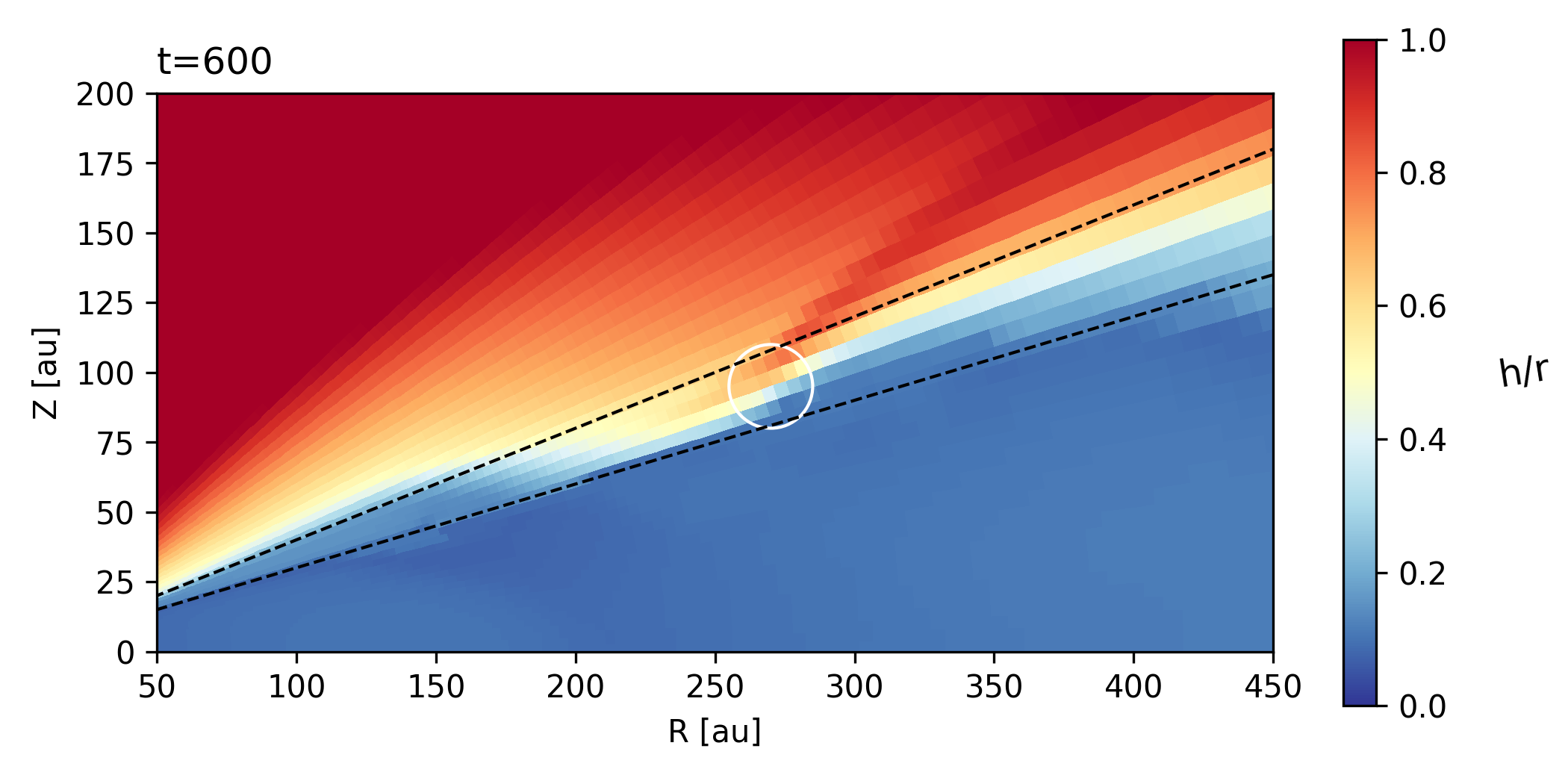}
    \caption{2D $h/r$ map of model \texttt{XUV}, the two dashed lines are $z/R=0.3$ and $z/R=0.4$. The white circle marks the origin of the outer ring in the channel map.}
    \label{fig:hr}
\end{figure}

\begin{figure}[h!]
    \centering
    \includegraphics[width=0.45\textwidth]{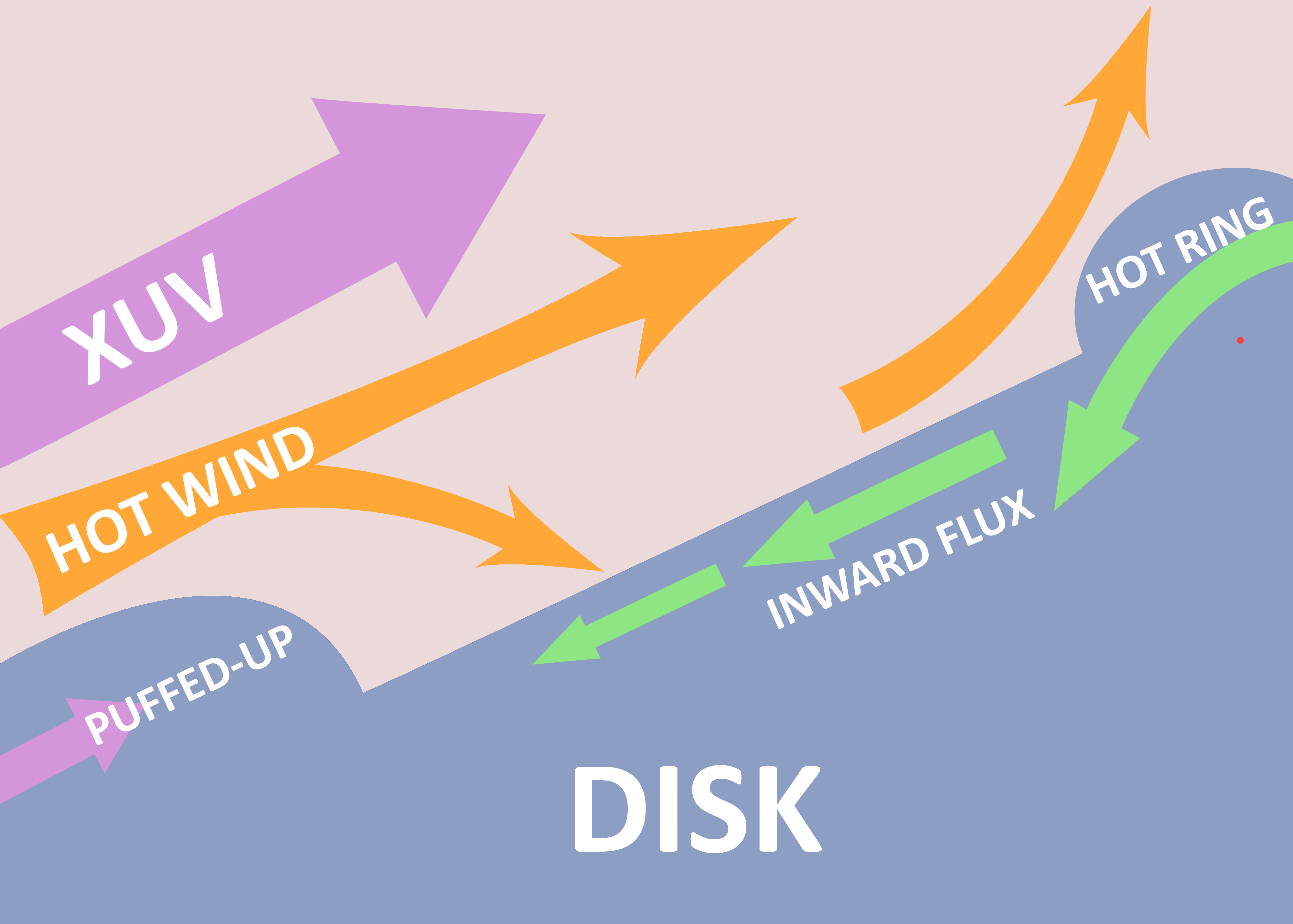} 
    \caption{Diagram showing the major physical processes involved in forming ring-like structures in model \texttt{XUV}.}
    \label{fig:diagram}
\end{figure}

In Model \texttt{XUV}, a distinct thin ring appears, separated from the main emission surface at $R \sim$ 270 au, with a signal strength of approximately $1\sigma$. This structure consists of two rings, each originating from the front and back surfaces of the disk, and spans a wide range of velocities. Additionally, it exhibits a significant azimuthal width, particularly in the zero velocity channel. The observed feature cannot be explained by $v_z$ alone. At $z/R = 0.3$, between 200 and 270 au, the expanding wind reduces the CO abundance but simultaneously increases the local temperature, as illustrated in Figure~\ref{fig:Trhovzco}. At 270 au, the wind collides with inward-moving gas, causing a rebound back into the atmosphere. The inward-moving gas from the outer disk is shielded from XUV radiation, leading to a higher CO abundance in this region. This interaction creates a localized temperature bump accompanied by CO enhancement. The elevated temperature in this region contributes to significant thermal broadening, allowing the emission to span a wide range of velocities. The aspect ratio, $h/r = c_s/v_k$, can be used to quantify this thermal broadening relative to the local Keplerian velocity. As shown in Figure~\ref{fig:hr}, $h/r$ at the temperature bump approaches unity, significantly exceeding the values in the rest of the disk below $z/R = 0.4$.

We summarize the processes of partial ring formation in model \texttt{XUV} in Figure~\ref{fig:diagram}. A slight collapsing flow between 150 and 200 au, and a notable up draft beyond 220 au, reaching almost 0.4 ${\rm km~s^{-1}}$. This is the signature of the ``puffed-up wind base'' discussed in Section~\ref{sec:2dquant}. Just beyond the wind base between 200 au and 270 au, the super-heated disk surface together with the ram pressure of radial drive photoevaporative wind, provides extra pressure support that results in sub-keplerian rotation. The slightly collapsed disk wind then deflected back up into the wind region, causing the updraft.

\begin{figure}[t!]
    \centering
    \includegraphics[width=0.45\textwidth]{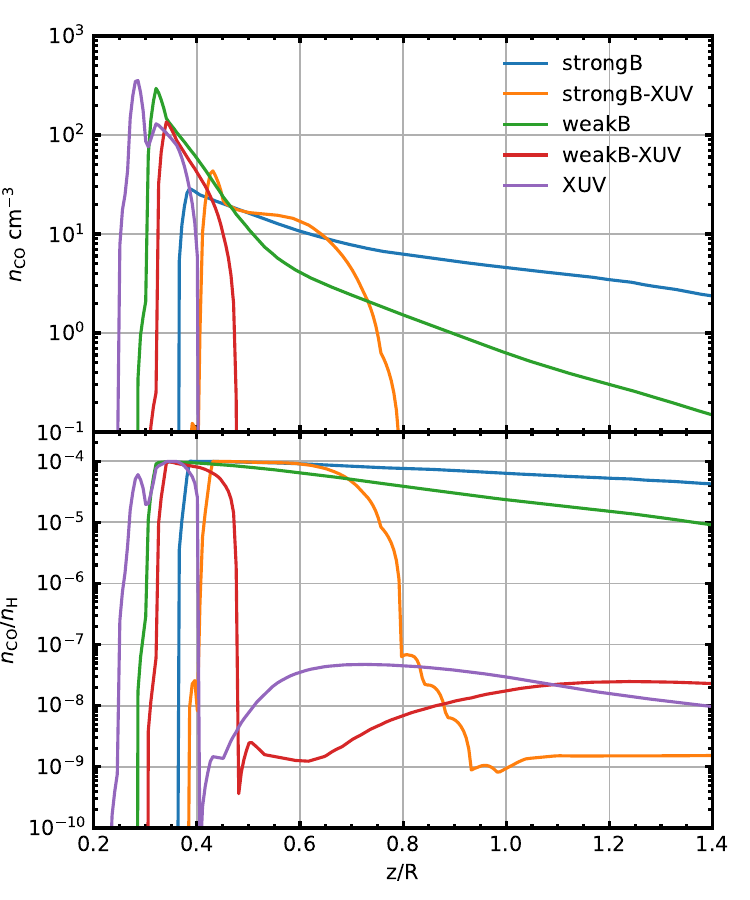}
    \caption{CO number density {\bf (top) and fraction (bottom)} along a vertical slice at R=140 au. }
    \label{fig:vertCO}
\end{figure}

\begin{figure*}[htbp!]
\centering
    \includegraphics[width=1\textwidth]{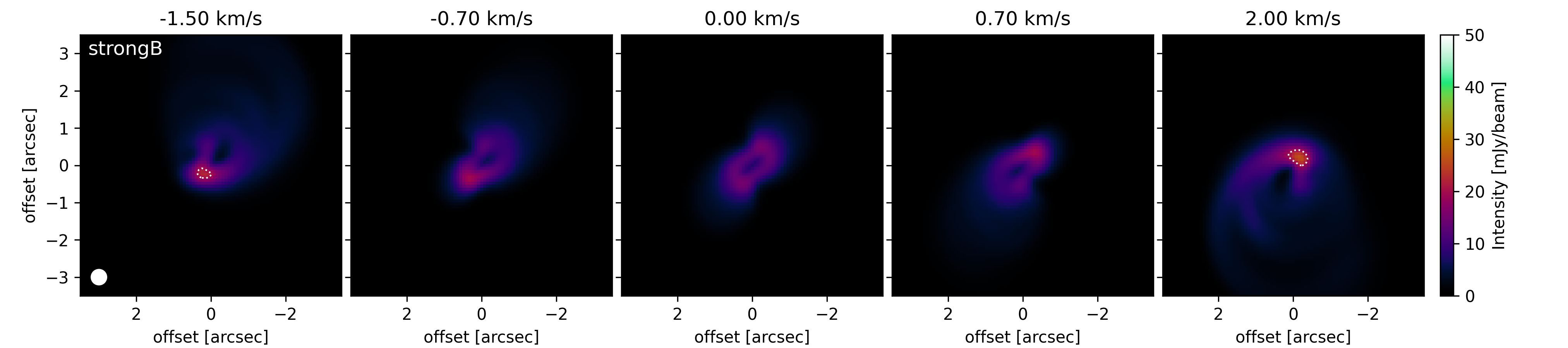}
    \includegraphics[width=1\textwidth]{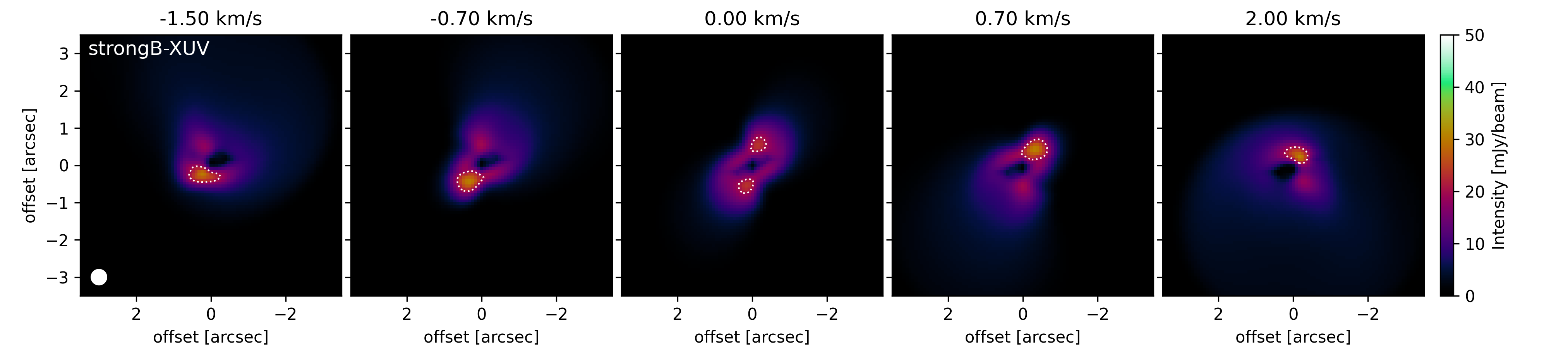}
    \includegraphics[width=1\textwidth]{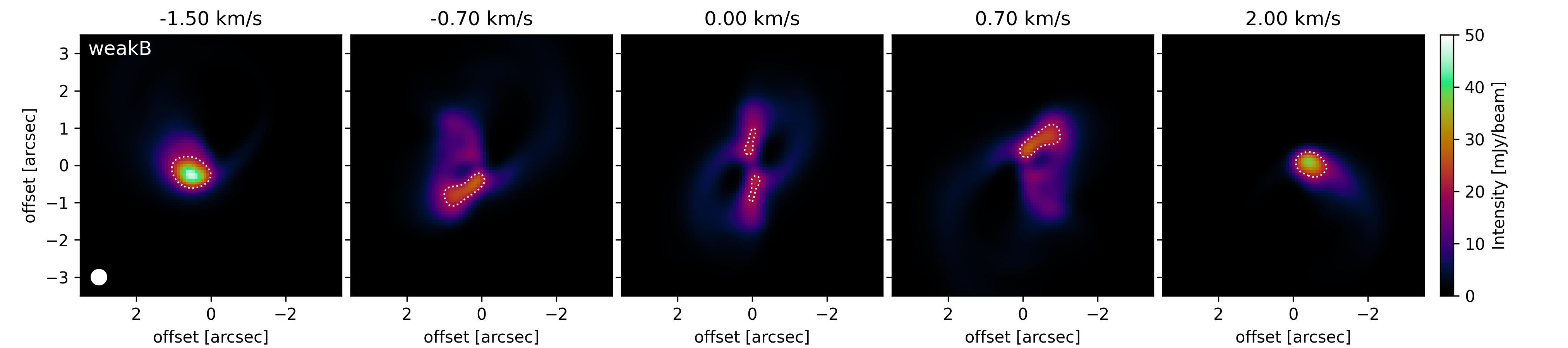}
    \includegraphics[width=1\textwidth]{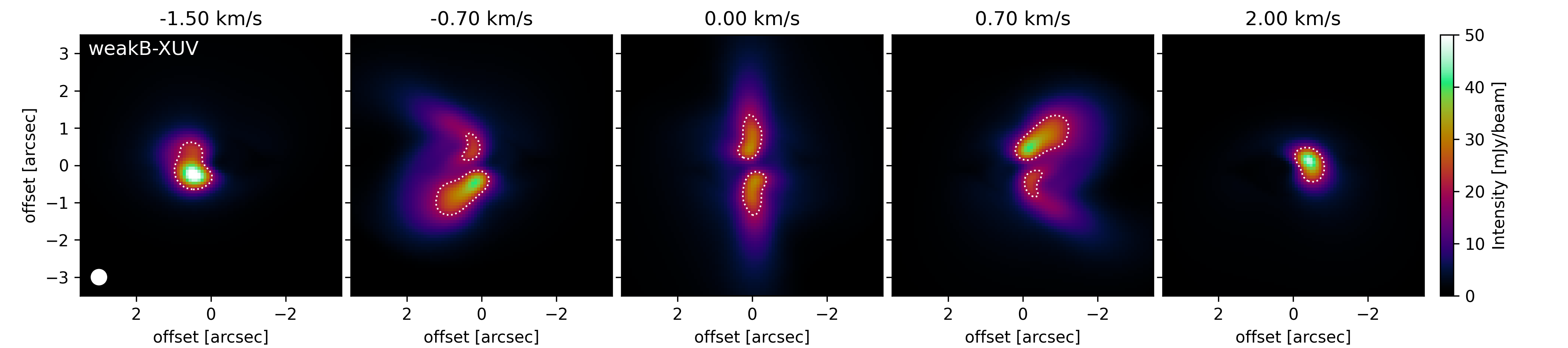}
    \includegraphics[width=1\textwidth]{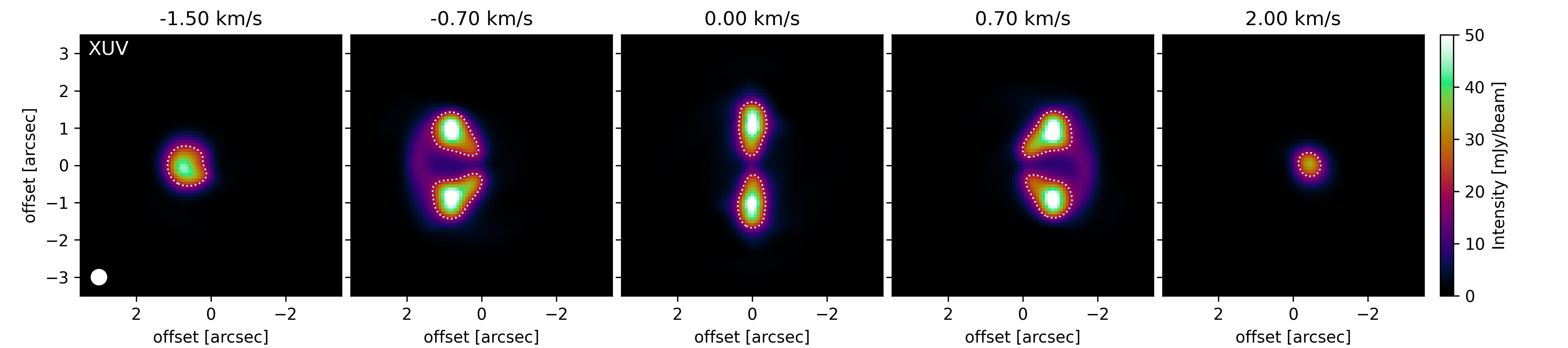}
    \caption{[C I] $^3P_1-^3P_0$ channel maps convolved with a $0\farcs4$ beam. The dotted contours mark the signal strength as equal to 1 $\sigma$.}
    \label{fig:peakchanC10}
\end{figure*}

The varying levels of wind signatures can primarily be attributed to differences in CO number density. Figure~\ref{fig:vertCO} shows vertical slices of CO number density and CO fraction $n_{\rm CO}/n_{\rm H}$ at R = 140 au, with $n_{\rm H}$ is the density expressed in the number of hydrogen nuclei. Only the two non-XUV models maintain the $10^{-4}$ CO abundance in the wind region within one order of magnitude, and Model \texttt{strongB-XUV} has a similar abundance within z/R=0.7, because of the heavy shielding of dense MHD wind. Model \texttt{XUV} exhibits the highest peak CO density because the disk surface, being exposed to unobstructed radiation, absorbs more energy and subsequently releases CO from grain surfaces at a lower layer (hence higher density) with $z/R=0.3$. The CO layer in this model ends at $z/R = 0.4$, as photodissociation dominates everywhere above. Thanks to the shielding from a weak MHD wind, model \texttt{weakB-XUV} extends this limit to 0.45 and the stronger MHD wind in \texttt{strongB-XUV} pushes to 0.8. The CO layers in both \texttt{strongB} and \texttt{weakB} extend beyond $z/R=1$, but only \texttt{strongB} showed an elevated wind signature. Compared to the $\tau = 1$ surfaces in Figure~\ref{fig:surf}, it reveals that the minimum CO number density required for detectable emission is approximately $10~{\rm cm}^{-3}$ (see the horizontal grid line in Figure~\ref{fig:vertCO}, also the violet contours in $n_{\rm CO}$ panels of Figure~\ref{fig:Trhovzco}). This explains the wide loops visible in the zero velocity channel and the ``halo'' conic surface seen in higher velocity ($2~{\rm km~s}^{-1}$) channels in \texttt{strongB} and \texttt{strongB-XUV}. Lastly, among the magnetized cases, regardless of the magnetic field strength, the lower boundary of the CO layer is slightly higher in models with XUV radiation. This is due to the additional absorption caused by the puffed-up layer at smaller radii, which not only absorbs the 300 eV XUV photons but also blocks soft FUV (7 eV) and LW (12 eV) photons.

\subsection{{\rm [C I]} emission}

Neutral atomic carbon (C I) is expected to trace the layer above the CO photodissociation regime \citep{2006FaDi..133..231V} and thus a likely more suitable disk wind tracer than CO.  The C I gas is observable in the fine-structure transitions [C I] ($2p^2:~^3P_1-^3P_0$), hereafter [C I] (1-0), and [C I] ($2p^2:~^3P_2-^3P_1$) in the submillimeter regime \citep{1980ApJ...238L.103P}. Similar to CO, we produced [C I] (1-0) channel maps with \texttt{RADMC-3D} and \texttt{syndisk}. The assumed beam size is also $0\farcs4\times0\farcs4$, similar to $0\farcs41\times0\farcs35$ in an ALMA archival observation of IM Lup \citep{2023ApJ...959L..27L}. The original pre-convolved channel maps are listed in the appendices. The 9-minute integration produced a noise level of $\sigma=65~{\rm mJy~beam}^{-1}$, thus a two-hour integration would yield $\sigma\simeq20~{\rm mJy~beam}^{-1}$.

\begin{figure*}[ht!]
    \centering
    \includegraphics[width=1\textwidth]{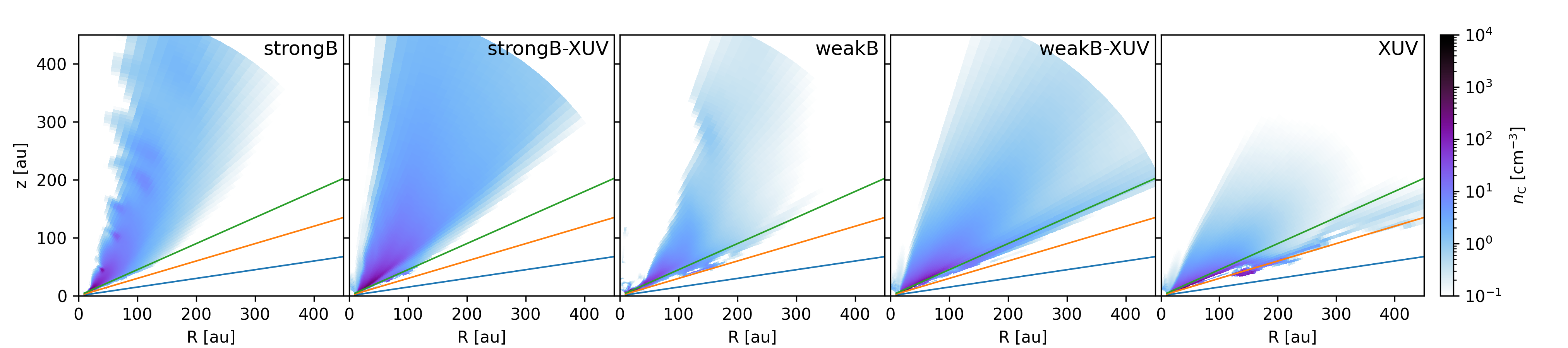}
    \caption{Neutral atomic carbon number density maps. The three diagonal lines correspond to $z/R=0.15$, 0.3, and 0.45 in all panels.}
    \label{fig:2dcompC}
\end{figure*}

In general, the [C I] emission is significantly weaker than CO, and only a tiny portion of the emission is above $1\sigma$. [C I] is mostly optically thin and originates from a higher layer, though the extent of this difference varies across our setups. As a product of photodissociation, the radial distribution of [C I] has a steeper gradient in the wind than CO. This is especially notable in the magnetized models without XUV, and the more centrally concentrated [C I] appears smaller in emission maps. In the more magnetized models, i.e., \texttt{strongB} and \texttt{strongB-XUV}, the signature butterfly pattern of Keplerian rotation is almost gone. Instead, we see two loops on the opposite side of the star from $-0.7$ to $+0.7~{\rm km~s}^{-1}$ channels. This behavior has been shown in the high $z/R$ panel of Figure~\ref{fig:model_elevated}. At the zero velocity channel, it's easily recognizable as a smaller version of the figure-eight pattern from CO emission, i.e., the loop in the first quadrant is from the front surface, and the one in the third quadrant is from the backside. This is also true for $\pm0.7~{\rm km~s}^{-1}$ channels. For $-0.7~{\rm km~s}^{-1}$ channel, the small loop at the third quadrant is the back surface, similar to pattern \texttt{B} in Figure~\ref{fig:peakchan}. The large loop at the first quadrant can be seen as a more extreme case of pattern \texttt{A} in Figure~\ref{fig:peakchan}. At a more elevated surface, the two wings are bent more towards the top right corner and eventually become a closed loop. This ``two-sided-loop'' is unique to MHD wind, because PE wind does not have a $v_\theta$ component that is comparable to $v_r$. All patterns have a smaller radial extension than the CO wind, not just because of the geometric effect of the higher [C I] layer that's closer to the central star, but also due to the projection of velocity vectors. The bright spots are denser [C I] regions closer to the disk surfaces, and they are more prominent in \texttt{strongB-XUV} because the extra XUV radiation penetrates deeper and photodissociates CO. The 2D [C I] maps in Figure~\ref{fig:2dcompC} reflect this trend. Another feature due to different [C I] distribution is shown in the $2~{\rm km~s}^{-1}$ channel. The [C I] cone in \texttt{strongB} is narrower than that in \texttt{strongB-XUV}, and it only occupies the lower half of the image with a 30$^\circ$ inclination. 

In the less magnetized setups, more disk patterns are visible as [C I] emission originates from an even lower altitude. The weaker wind absorbs less UV radiation, thereby shifting the lower boundary of [C I] closer to the disk surface, as shown in Figure~\ref{fig:2dcompC}. The [C I] layer in \texttt{weakB} largely overlaps with the CO emission layer in \texttt{strongB} and \texttt{strongB-XUV}, resulting in similar patterns in channel maps, especially from in $\pm0.7~{\rm km~s}^{-1}$  and zero velocity channels. Model \texttt{weakB-XUV} exhibits a more diffuse and blurry wind profile because the XUV radiation expands the photodissociating region and [C I] occupies the largest space among the five setups. The bright disk surface in the middle three channels shows unique acute angles at the inner edge. It's the innermost portion of the figure-eight structure of the disk wind, and the outer portion is strongly blurred. The moderate MHD wind lifts just enough gas to absorb ionizing radiation, maintaining a largely neutral atmosphere, where XUV photons continue to photodissociate most of the CO. This creates a disk atmosphere rich in neutral atomic carbon but depleted of CO. As a result, the optically thin [C I] emission arises from all vertical locations, producing blurry wind patterns.

Model \texttt{XUV} has the brightest [C I] emission. The disk patterns are above $1\sigma$ and resemble that of CO, featuring a ring connecting ``wing-tips'' in the $\pm0.7~{\rm km~s}^{-1}$ channels. Figure~\ref{fig:2dcompC} shows that atomic carbon's lower boundary lies just above and overlaps with the CO layer. The thin upper atmosphere contributes minimally to emission, explaining the butterfly pattern similarity. Expanding winds mix atomic carbon with the CO surface just beyond the ``puffed-up'' layer, while some carbon concentrates below this outer edge, first brought down by winds and then transported inward near the disk surface. This may account for observational interpretations of [C I] surfaces appearing below $^{12}$CO \citep[e.g., HD163296 in][]{2024A&A...686A.120U}. Overall, under the same radiation field, [C I] brightness anti-correlates with wind loss rate. Lower wind density and/or stronger XUV radiation dissociate CO at lower altitudes, forming denser atomic carbon layers.

\section{Kinematics of disk winds in ALMA line observations}
\label{sec:velocity}

\subsection{Radial profiles of the gas velocity}

\begin{figure*}[htbp!]
\centering
    \includegraphics[width=1\textwidth]{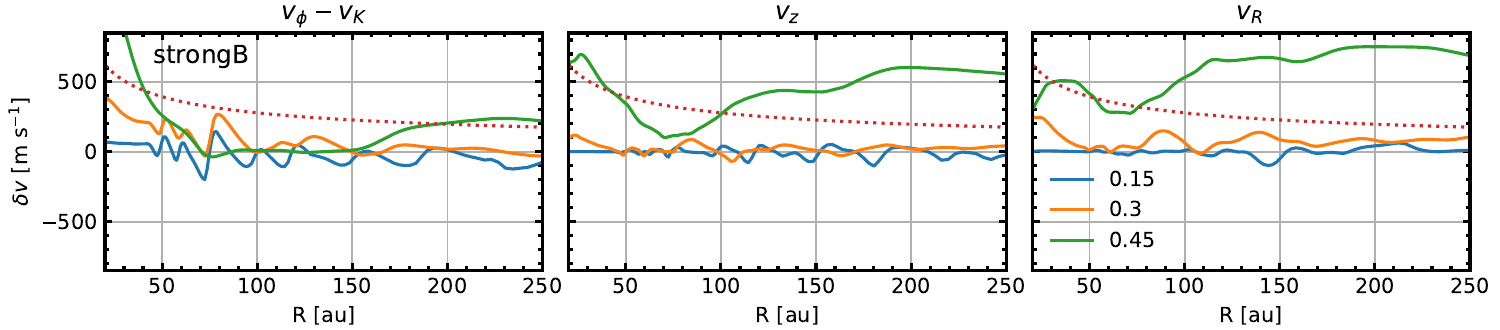}
    \includegraphics[width=1\textwidth]{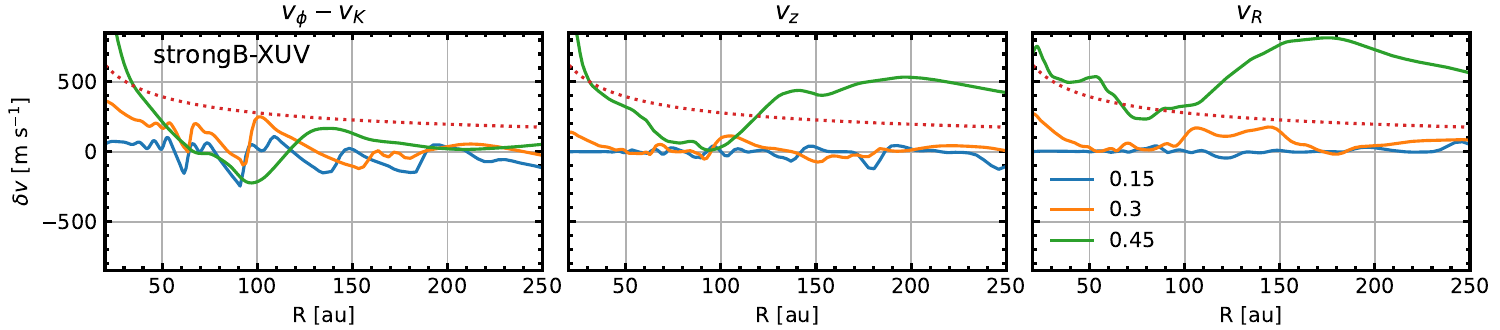}
    \includegraphics[width=1\textwidth]{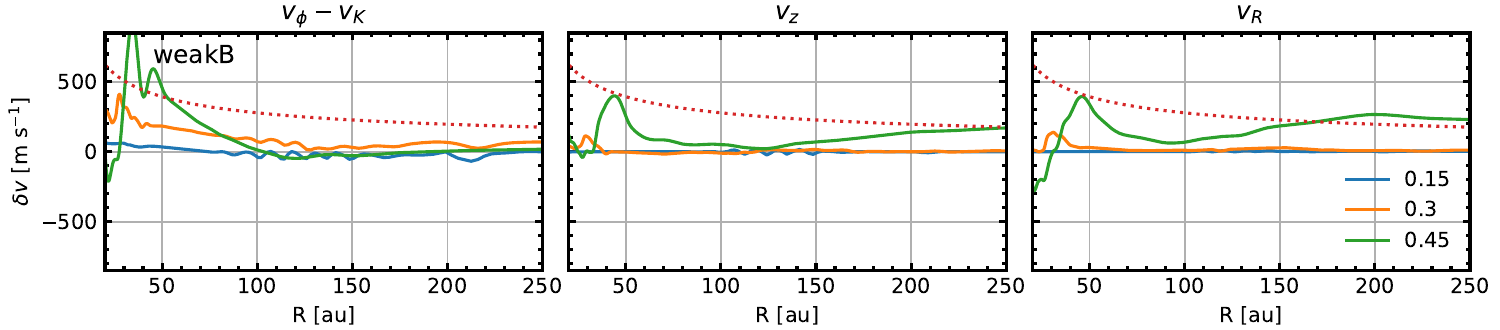}
    \includegraphics[width=1\textwidth]{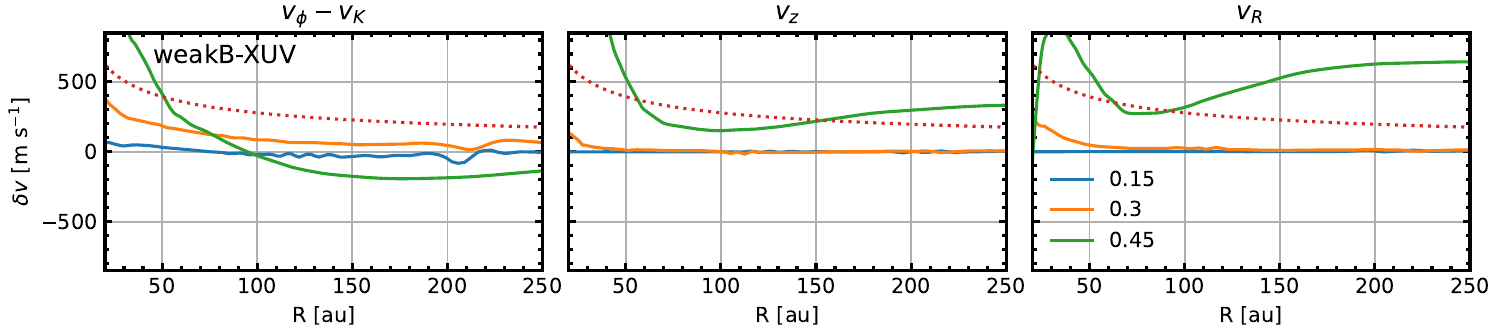}
    \includegraphics[width=1\textwidth]{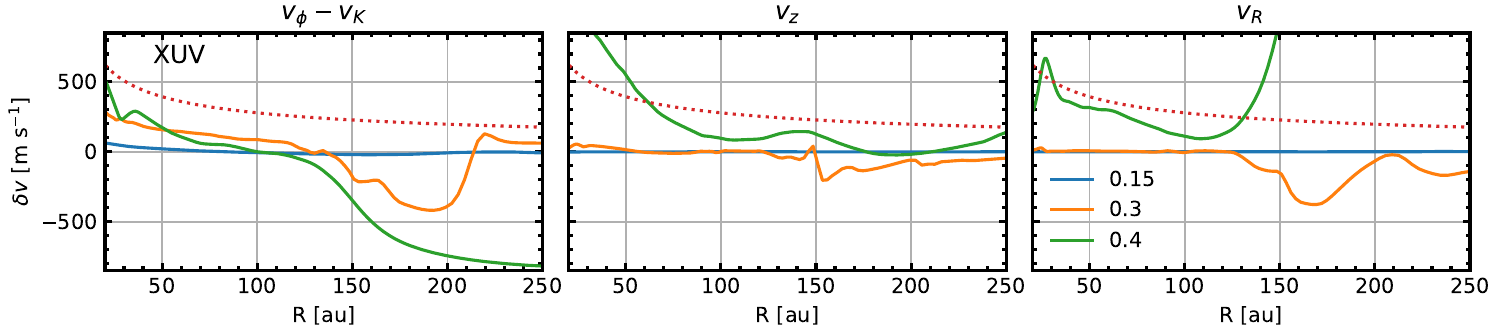}
    \caption{Velocity profiles measured at three different surfaces (three different colors). From top to bottom: $\beta=10^4$, $\beta=10^4$ with 300 eV, $\beta=10^5$, $\beta=10^5$ with 300 eV, and PE model with 300 eV. The red dotted lines are 10\% of the Keplerian velocity at z/R=0.3 surface.}
    \label{fig:3v3h}
\end{figure*}

As seen in Figure~\ref{fig:surf}, the $\tau=1$ surface varies from different models. To facilitate a quantitave comparison, we extract 1D velocity profiles from three different layers, with $z/R=0.15$, 0.3, 0.45 (except for \texttt{XUV}, where 0.4 is used as the top layer), to mimic emission surfaces of molecules with different abundances or exciting temperatures: within the disk (e.g., $^{13}$CO 2-1), near the disk surface (e.g., $^{12}$CO 2-1), and above the surface (e.g., $^{12}$CO 7-6). The three velocity components are plotted in three columns of Figure~\ref{fig:3v3h}, with three colors representing different surfaces.

Starting with the azimuthal velocity (left column in Figure~\ref{fig:3v3h}), we observe a general trend where the inner disk consistently rotates at super-Keplerian velocities, with deviations increasing with height. In a state of vertical hydrostatic equilibrium, such super-Keplerian motion arises from a positive $\d p/\d R$ due to the flared disk surface. This trend is also reflected in the $\delta v_p$ panel of Figure~4 from \citet{2024ApJ...970..153A}. At $z/R = 0.15$, all magnetized models exhibit some degree of perturbation, as a result of spontaneous substructures formation via the redistribution of magnetic flux, as noted in previous works \citep[e.g.,][]{2018MNRAS.477.1239S, 2021MNRAS.507.1106C, 2022MNRAS.516.2006H} involving ambipolar diffusion. In models with stronger magnetic fields ($\beta = 10^4$), such as \texttt{strongB} and \texttt{strongB-XUV}, $\delta v_\phi$ reaches up to 10\% of the local Keplerian velocity, and the radial width of each ``velocity bump'' spans from 10 to 40 au. At $z/R=0.3$, the amplitude of velocity variation is similar but some small-scale ``wiggles'' are ``smoothed''. The overall profile has become more super-Keplerian. The magnetic-driven substructures have little effect in the wind region since the magnetic flux is evenly distributed at this level (see also Figure~\ref{fig:am}). Due to intense XUV radiation, model \texttt{strongB-XUV} also exhibits a puffed-up layer, similar to model \texttt{XUV} but at a smaller radius. The increased radial pressure gradient supports sub-Keplerian motion between 80 and 120 au at even $z/R=0.45$. Beyond 130 au, the high-density MHD atmosphere is able to shield the XUV radiation, and the wind resumes super-Keplerian motion beyond as magnetic fields regain dominance. With weaker magnetic fields ($\beta = 10^5$), the disk's rotation curve is notably smoother, as $v_\phi$ deviations remain under 1\% of $v_K$ at $z/R=0.15$. XUV radiation penetrates deeper into the less massive atmosphere in \texttt{weakB-XUV}, so the thermal pressure gradient suppresses the magnetic pressure variations. The result is a smoother rotation curve at $z/R=0.3$, and the sub-Keplerian rotation at $z/R = 0.45$ begins at 100 au, extending beyond 300 au.

Model \texttt{XUV} has a smooth rotation profile at $z/R=0.15$, but for $z/R=0.4$, it drops abruptly at $\sim$120 au. This is expected at the outer edge of the puffed-up layer, as the hot wind above expands downwards to the disk surface. The super-sonic wind breaks the force balance and accelerates gas radially. Being pushed outward, angular momentum conservation reduces the rotation to sub-Keplerian rotation. The downward expansion also reduces $v_z$, though $v_z$ is not completely negative due to the projection angle (from $\hat{\theta}$ to $\hat{z}$) and a much faster radial (spherical $r$) dominant motion: $v_R$ quickly jumps to $1~{\rm km~s}^{-1}$ level beyond 150 au in this layer. 

The condition at $z/R = 0.3$ for model \texttt{XUV} is more complicated, and we have already seen the effects on channel maps (e.g., the bottom row of Figure~\ref{fig:peakchan}). All three velocity components remain relatively ``normal'' until approximately 130 au, the outer edge of the puffed-up layer. Several competing factors exist at this transition layer: the expanding sub-Keplerian wind, the denser disk below, and an inward flux ($v_R < 0$) just beneath the surface. This inward flux results from a flatter pressure profile due to insufficient dust heating. Within the CO freeze-out radius, the midplane usually possesses enough dust grains to maintain a power-law temperature profile, by efficient heat exchange between gas and dust. When density gets lower, the number density of dust is not enough to maintain the heat exchange rate to compensate for line cooling. Additionally, the density just below the disk surface can still block most of the radiation heating from the central star. In the absence of efficient angular momentum transport (without magnetic fields or high viscosity), gas from the outer disk rotates slightly faster than $v_K$ due to positive $\d p/\d R$ from the flared disk surface, as seen for $R > 215$ au. Between 140 and 215 au, $v_\phi$ is more influenced by the expanding wind, resulting in sub-Keplerian motion. Note that $z/R = 0.3$ lies just below the wind, not within it, so wind effects arise through advection terms along the $z$ direction, i.e., $v_z\frac{\partial v_R}{\partial z}$ for $v_R$ and $v_z\frac{\partial v_\phi}{\partial z}$ for $v_\phi$. The $v_\phi$ term prevails over the slight super-Keplerian inward flux, whereas the $v_R$ term does not. In the following section, we'll have a more detailed analysis of the advection terms. The extent of the wind-dominated region is also evident from the $v_z$ profile at $z/R = 0.4$: $v_z$ decreases near R = 150 au but rises again around R = 210 au, where the expanding wind starts to be deflected by the denser disk (also check the $v_z$ panel in Figure~\ref{fig:Trhovzco}.

The $v_z$ and $v_R$ profiles in the magnetized models are easier to understand: they are both around a couple of hundred ${\rm m~s}^{-1}$ at $z/R=0.45$ and are typical for regions near the wind base. There are no significant differences between model \texttt{strongB} and \texttt{strongB-XUV}, as strong magnetic fields dominate the dynamics of the wind. For weaker magnetic fields, XUV radiation could boost the wind speed by at least twice. Due to more absorbed radiation, the radial velocity in \texttt{weakB-XUV} is still notably slower than the pure PE wind in model \texttt{XUV}. 

\subsection{Force balance and disk mass}

\begin{figure}[htbp!]
    \centering    
    \includegraphics[width=0.45\textwidth]{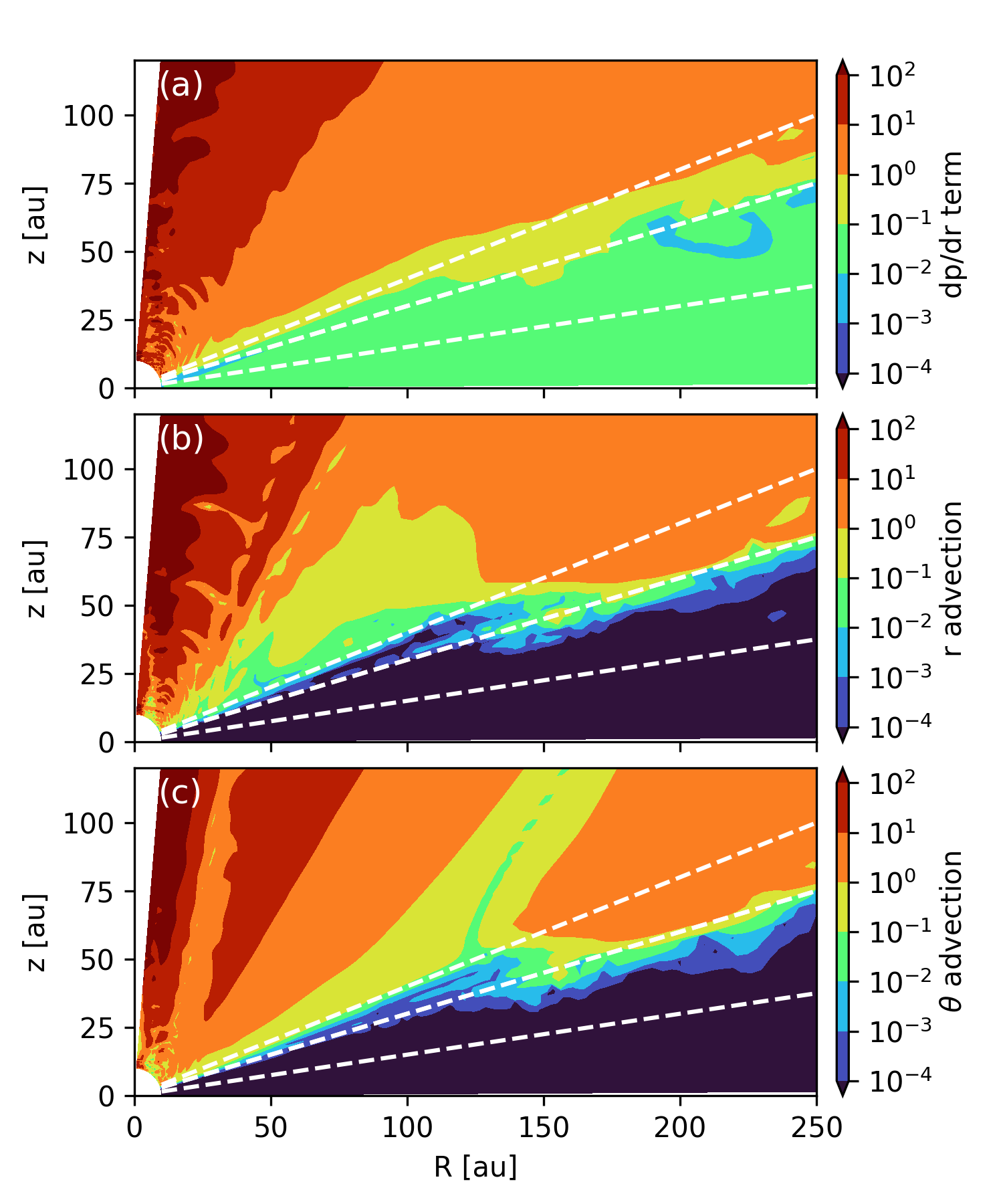}
    \caption{Relative contributions (normalized by $v_K^2$) from different correction terms when calculating $v_\phi$ in model \texttt{XUV}. Panel (a) is the thermal pressure gradient term $\d p/\d r$, (b) is radial advection $v_r \frac{\partial v_r}{\partial r}$, and (c) is polar advection  $\frac{v_\theta}{r} \frac{\partial v_r}{\partial \theta}$. The three dashed lines in each panel are z/R=0.15, 0.3, and 0.4.}
    \label{fig:vphiterm300ev}
\end{figure}

\begin{figure*}[htbp!]
    \centering    
    \includegraphics[width=0.45\textwidth]{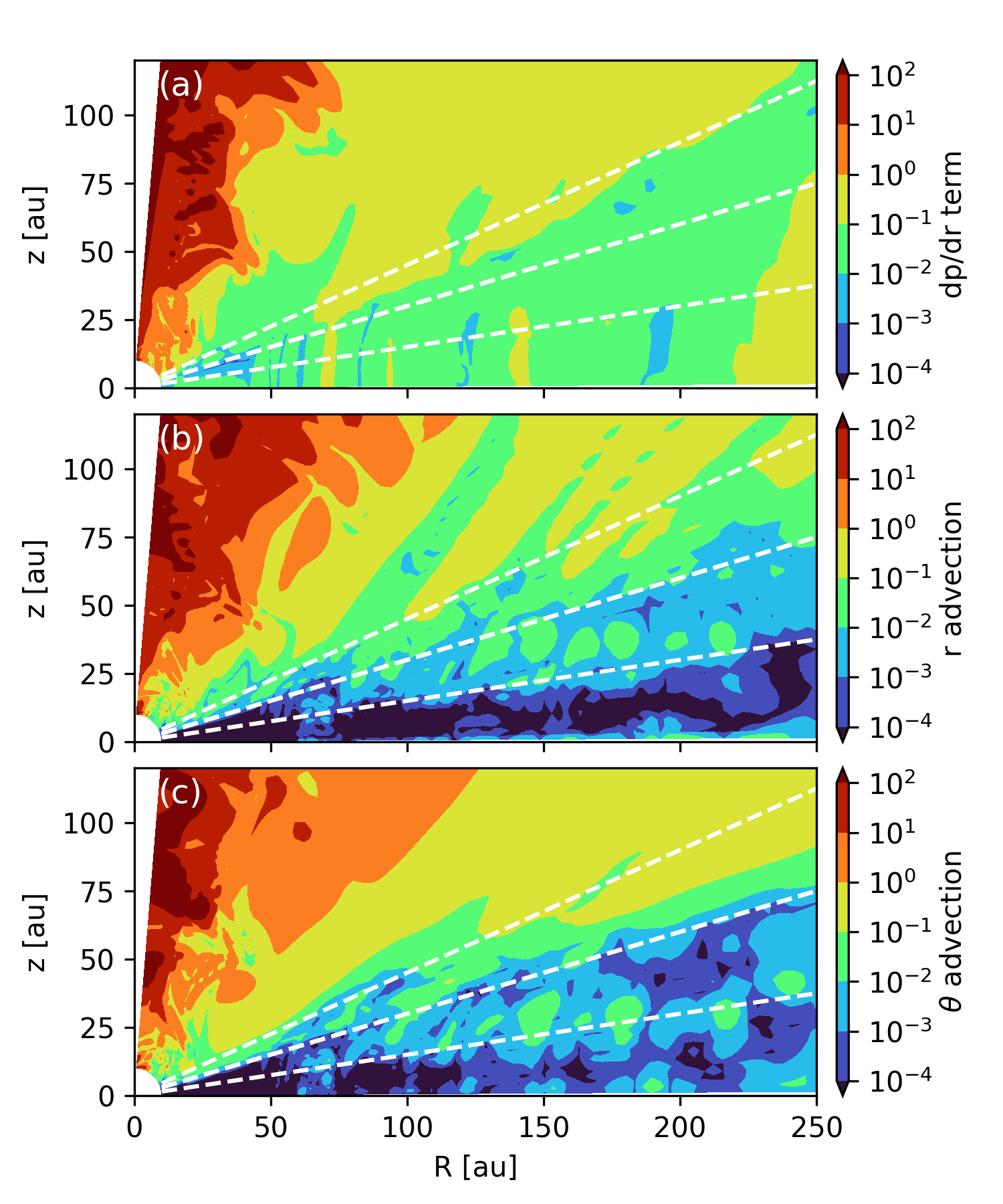}
    \includegraphics[width=0.45\textwidth]{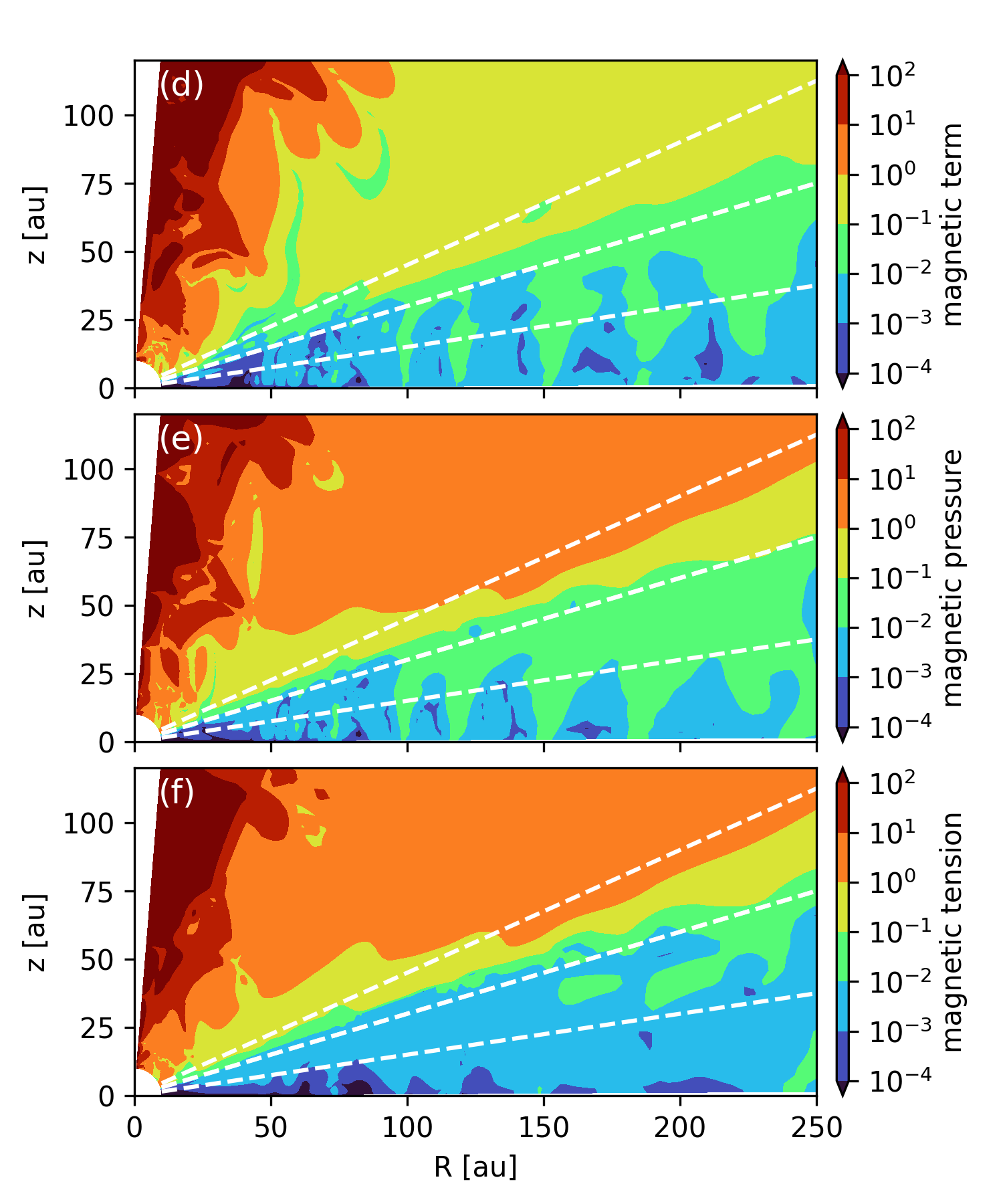}
    \caption{Relative contributions (normalized by $v_K^2$) from different correction terms when calculating $v_\phi$ in model \texttt{strongB}. Panel (a),(b), and (c) are the same as Figure~\ref{fig:vphiterm300ev}, and panel (d) is the combined magnetic terms, (e) only includes magnetic pressure, and (f) only includes magnetic tension. The three dashed lines in each panel are z/R=0.15, 0.3, and 0.45.}
    \label{fig:vphitermb4}
\end{figure*}

\begin{figure*}
    \centering    
    \includegraphics[width=1\textwidth]{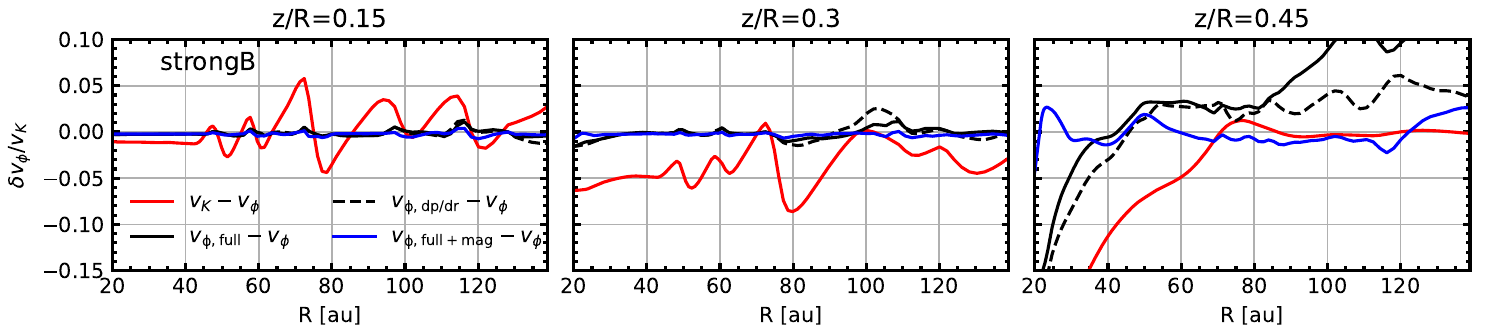}
    \includegraphics[width=1\textwidth]{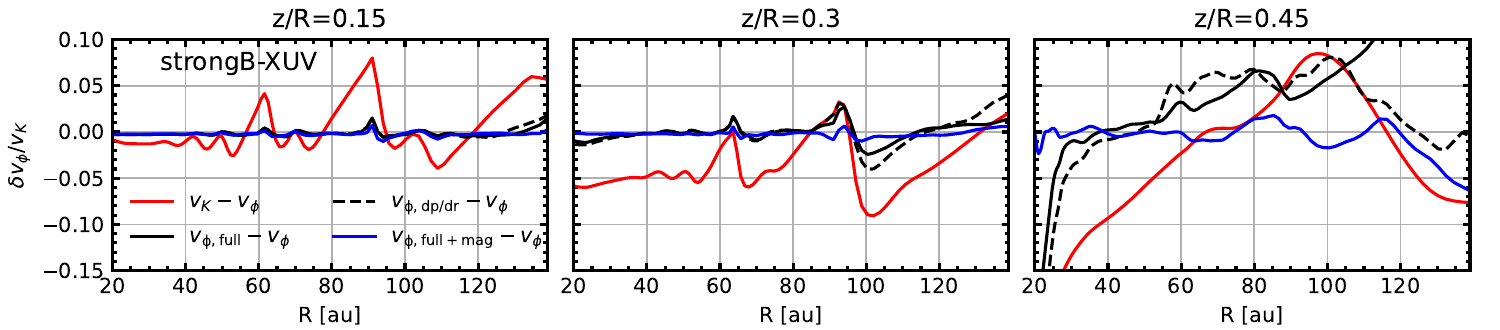}
    \includegraphics[width=1\textwidth]{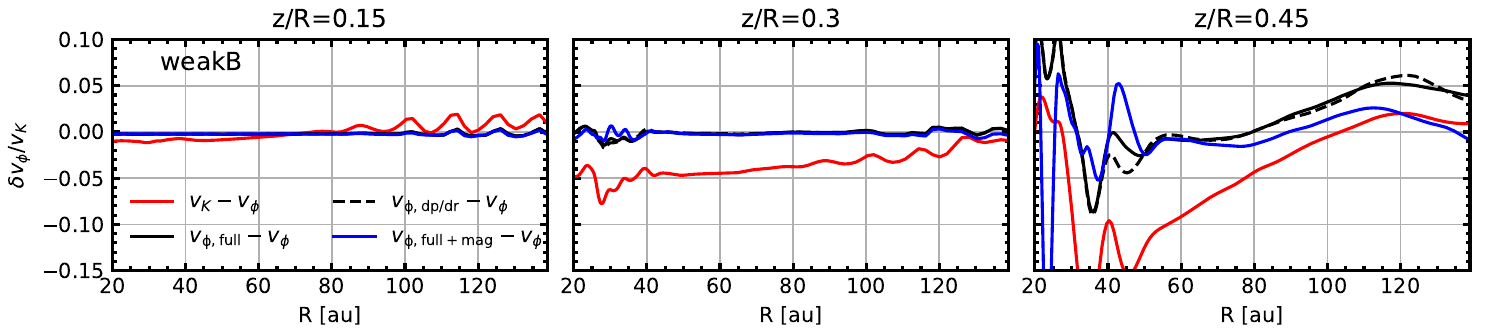}
    \includegraphics[width=1\textwidth]{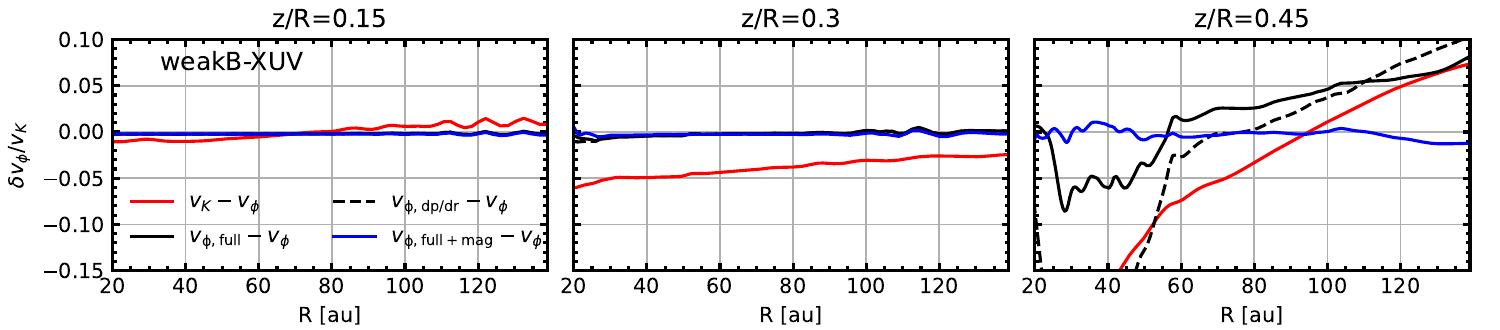}
    \includegraphics[width=1\textwidth]{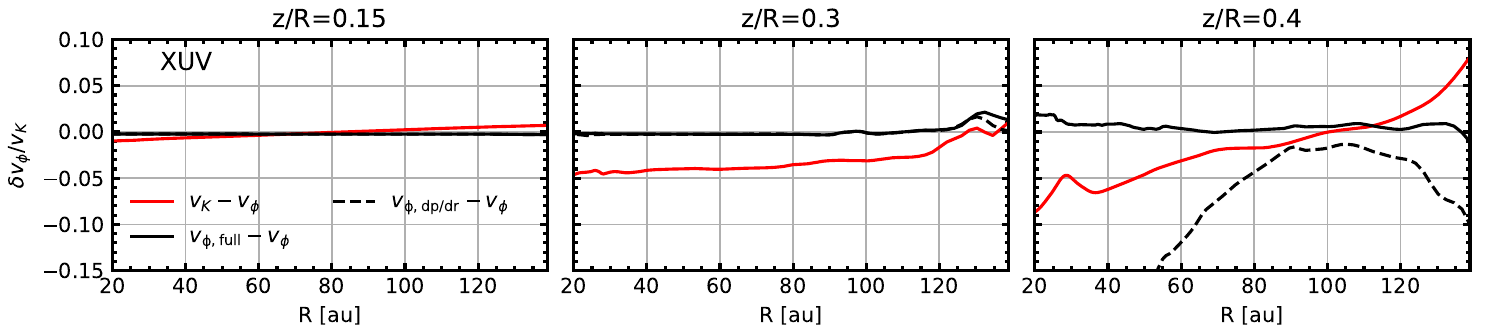}
    \caption{Deviations from the measured $v_\phi$ with different force balance models, scaled by $v_K$ at three different heights. The red solid line is the Keplerian velocity vertical pressure support. The black dashed line includes the correction by the radial pressure gradient. The black solid line }
    \label{fig:vphicalc}
\end{figure*}
What can we learn about the disk from the measured three-dimensional velocities and the corresponding vertical layers? The deviation between the measured velocity and pressure-gradient corrected Keplerian velocity could measure the effect of the disk's self-gravity, thus the disk mass \citep[e.g.,][]{2021ApJ...914L..27V,2023MNRAS.518.4481L,2024ApJ...970..153A}. Here we follow the same idea with the addition of magnetic terms. The momentum equation in MHD can be written as:
\begin{equation}
\rho \left(\frac{\partial}{\partial t} + \mathbf{v}\cdot \nabla \right)\mathbf{v} = \left(\mathbf{B}\cdot \nabla \right)\mathbf{B} - \nabla \left(\frac{B^2}{2} + p\right)-\nabla \Phi
\end{equation}
where $(\mathbf{v}\cdot \nabla )\mathbf{v}$ is the advection term, $(\mathbf{B}\cdot \nabla)\mathbf{B}$ is the magnetic tension, the second term on the right-hand side is the pressure gradient, and the third is gravity which should include both the central star and the disk's self-gravity. Since we neglect self-gravity in our simulations, we will only include the star's gravity in the equations. We can evaluate the significance of each term and compare them to the dynamics masses from self-gravity in previous works. The magnetic tension term is equivalent to the magnetic stress term $\nabla \cdot (\mathbf{B}\mathbf{B})$ since $\nabla \cdot \mathbf{B}=0$.
In a steady axisymmetric system $\partial/\partial t=0$, and $\partial/\partial \phi=0$. The momentum equation along the $R$ direction of the cylindrical coordinate system is:
\begin{eqnarray}
\rho \left(v_R \frac{\partial v_R}{\partial R}  + v_z \frac{\partial v_R}{\partial z} - \frac{v_\phi^2}{R}\right) &=& F_R -\frac{\partial p}{\partial R} \nonumber \\
-\frac{\partial (B^2/2)}{\partial R} +  B_R\frac{\partial B_R}{\partial R} &+& {B_z}\frac{\partial B_R}{\partial z}-\frac{B_\phi^2}{R}
\end{eqnarray}
Here $F_R$ is the gravity projected along the $R$ axis. Since our simulation used a spherical polar grid, to reduce numerical errors, we'll perform the analysis in spherical polar coordinates:
\begin{eqnarray}
\rho \left( v_r \frac{\partial v_r}{\partial r} + \frac{v_\theta}{r} \frac{\partial v_r}{\partial \theta} - \frac{v_\theta^2 + v_\phi^2}{r} \right) &=& F_r -\frac{\partial p}{\partial r} \nonumber \\ 
-\frac{\partial (B^2/2)}{\partial r} +  B_r\frac{\partial B_r}{\partial r} + \frac{B_\theta}{r}\frac{\partial B_r}{\partial \theta}&-&\frac{B_\theta^2+B_\phi^2}{r}
\end{eqnarray}
The advection terms have the same form as the magnetic tension terms but with different signs. Fluid field and magnetic field tend to constrain each other (the reason for different signs) to maintain the current form. Of the three parts of the advection terms, the last one $-(v^2_\theta+v^2_\phi)/r$ is the centrifugal force, which is the main part of radial force balance. For model \texttt{XUV}, $v_\theta$ is always very small compared to $v_\phi$ even at $z/R=0.4$, since the PE wind is mostly radial. Thus, $v_\theta$ contributes more to the centrifugal force in the magnetized models. The other two advection terms describe the radial convective acceleration, i.e., the variation of radial momentum carried by radial ($v_r \cdot \partial v_r/\partial r$) and polar ($v_\theta/r \cdot \partial v_r/\partial \theta$) advection.

Reorganizing the terms:
\begin{eqnarray}
\frac{1}{v^2_{K,r}}(&v^2_\phi&+v^2_\theta-v^2_{K,r})=\nonumber\\
\frac{1}{v^2_{K,r}}(&\delta v_{th}& + \delta v_{bp} +\delta v_{bt}+\delta v_{ad,r}+\delta v_{ad,\theta})
\label{eq:vphirterm}
\end{eqnarray}
where
\begin{eqnarray}
\delta v_{th}&=&\frac{r}{\rho}\frac{\partial p}{\partial r}\nonumber\\
\delta v_{bp}&=&\frac{r}{\rho}\frac{\partial(B^2/2)}{\partial r}\nonumber\\
\delta v_{bt}&=&\frac{r}{\rho}\bigg(- B_r\frac{\partial B_r}{\partial r} -\frac{B_\theta}{r}\frac{\partial B_r}{\partial \theta} + \frac{B_\theta^2+B_\phi^2}{r}\bigg)\nonumber\\
\delta v_{ad,r}&=&rv_r \frac{\partial v_r}{\partial r}\nonumber\\
\delta v_{ad,\theta}&=&v_\theta \frac{\partial v_r}{\partial \theta}
\label{eq:vphirterm2}
\end{eqnarray}
and $v^2_{K,r}=-F_r r=GM_*/r$. Now we can compare the contribution of these terms quantitatively as we scale each of them to $v_K^2$. 

The 2D distributions of the contribution from each term are shown in Figures~\ref{fig:vphiterm300ev} and \ref{fig:vphitermb4}. Starting with the non-magnetized model \texttt{XUV}, inside the disk (below z/R=0.3), the radial pressure gradient is the dominant effect, with a correction level of several \%. Both radial advection and polar advection come into play above z/R=0.4. Within 150 au, the polar term $\frac{v_\theta}{r} \frac{\partial v_r}{\partial \theta}$ is also important in the puffed-up layer, and because of the expanding wind colliding with the disk surface, both advection terms become more significant even below $z/R=0.3$ beyond 150 au. For \texttt{strongB} in Figure~\ref{fig:vphitermb4}, the thermal pressure term's contribution is similar in the disk and the atmosphere, while the two advection terms are more significant in the disk than \texttt{XUV}, thanks to a more active midplane, especially the meridional flows \citep[see also~][]{2022MNRAS.516.2006H}. Also in the disk, there is a coarse anti-correlation between the thermal pressure term and the magnetic term, and this aligns with the formation mechanism of magnetic-induced substructures: the highly concentrated magnetic flux drives faster accretion that depletes local gas, which also reduces the contribution from thermal pressure. For comparison, in a 0.1$M_\odot$ disk, the $v_\phi$ deviation from the Keplerian rotation caused by the disk's self-gravity is about $100~{\rm m~s}^{-1}$ \citep{2024ApJ...970..153A} at the CO emission surface ($z/R=0.3$), which translates to 2.5\% to 5\% of $v_K$, and 5\% to 10\% when scaled by $(v^2_\phi-v_K^2)/v_K^2$ between 50 and 200 au. At this layer, both the advection terms and the magnetic terms in \texttt{strongB} have similar level of effect on $v_\phi$ as the massive disk's gravity. In model \texttt{XUV}, the advection terms are similarly important beyond the puffed-up layer but much less so within it.

Now we can try to recover the $v_\phi$ with different terms and compare them with the $v_\phi$ from the simulations. In Figure~\ref{fig:vphicalc} we used four models: Keplerian velocity assuming vertical pressure balance $v_K$; radial gradient of thermal pressure correction $v_{\rm \phi, dp/dr}$; pressure correction plus advection terms, i.e., all hydrodynamic terms $v_{\rm \phi, full}$; all hydrodynamic terms plus magnetic terms (magnetic pressure and magnetic tension). The plotted quantities are the deviations from these models to the measured $v_\phi$ from the simulation, then divided by $v_K$. The second model, 
$v_{\rm \phi, dp/dr}$, is not exactly the same force balance analysis performed on observational data, where the bulk motions of the disk models are assumed to be solely in the azimuthal direction~\citep[e.g.,][]{2024ApJ...970..153A}, and the pressure gradient is along the cylindrical radial direction. Because interpolating our spherical polar coordinate into the cylindrical coordinate introduces extra errors, we use $v_{\rm \phi, dp/dr}$ as a close approximation. Additionally, $v_\theta^2/r$ is also included when calculating $v_{\rm \phi, dp/dr}$, and we find this term has minimum effect at $z/R=0.15$ and 0.3.

In model \texttt{XUV}, the azimuthal velocity inside the disk (z/R=0.15) and the surface layer (z/R=0.3) can be corrected by the pressure gradient term alone. But $dp/dr$ term increases the error at $z/R=0.4$ to over 10\%, adding the advection reduces the error to below 1\% in the majority of the disk surface. The correction from the advection term at z/R=0.4 is dominated by $\theta$ component, i.e., $v_\theta/r \cdot \partial v_r/\partial \theta$. That implies the super Keplerian $v_\phi$ at this height is mainly supported by the combined effects of (1) the large velocity shear as gas is quickly accelerated by radiative heating at the top of the puffed-up layer, and (2) negative $v_\theta$ as gas leaving the disk and entering the wind region. The dominance of the $\theta$ advection over $r$ advection is evident when comparing panels (b) and (c) of Figure~\ref{fig:vphiterm300ev}, in the $R< 150$ au region where the $z/R=0.4$ line crosses. This momentum transport primarily involves polar movement, which brings slower gas into the fast wind region. In other words, a gas parcel from below is rapidly accelerated upon reaching the disk surface, rather than gradually speeding up within the surface layer.

In magnetized cases, deviations from Keplerian velocity at the midplane arise due to magnetically induced rings and gaps, which are well corrected by thermal pressure gradients at $z/R = 0.15$. The thermal pressure terms also do well at the disk surface for the weakly magnetized models, regardless of XUV strength. However, for more magnetized disks, thermal pressure corrections alone result in a 3-4\% error, which is reduced by half when advection terms are included. Incorporating magnetic field effects reduces this error to below 1\%. At $z/R = 0.45$, magnetic terms become critical for force balance, as hydrodynamic forces alone can produce errors exceeding 5\% or even 10\%, while magnetic contributions suppress these errors to below 3\% in most regions and often to as low as 1\%.

The velocities measured from the typical CO emission surface ($z/R=0.3$) are mainly affected by the pressure gradient except for the strongly magnetized (plasma $\beta=10^4$) cases. In model \texttt{strongB} and \texttt{strongB-XUV}, the ``extra'' contributions from the advection and magnetic terms at this layer are usually around 1\% to 4\% of $v_K$. This is to the similar level of a massive disk's self-gravity (2.5\% to 5\%). There is no clear trend that neglecting these terms would underestimate or overestimate the dynamical disk masses since correction from them can be both ways, but the errors are expected to be 40\% to 80\% of the real mass. If emission comes from $z/R=0.4$ and above, the advection terms can cause $>$5\% deviation alone, and so are the magnetic terms. In summary, the current method used in dynamical mass estimation is unlikely to be affected by gas advection or magnetic force at the typical CO emission surface. If significant $v_R$ or $v_z$ is detected, then the force balance needs to include the two extra terms. On the other hand, if we can extract kinematics from the wind region, we can put dynamic constraints on magnetic field strengths. 

\section{Discussion}
\label{sec:discuss}

\subsection{Low Mass Disks}
\label{sec:lowmass}
\begin{figure*}[!htbp]
\centering
\includegraphics[width=1\textwidth]{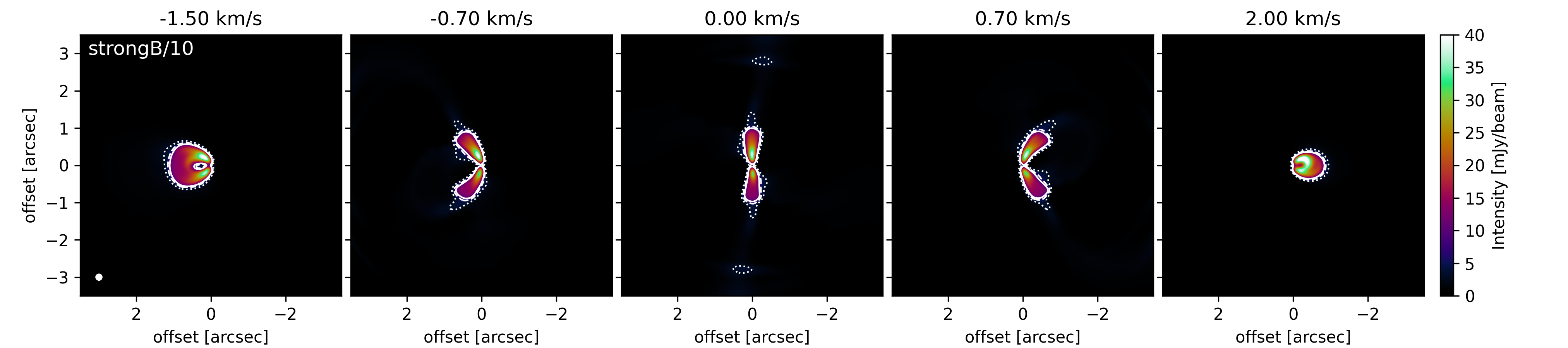}
\includegraphics[width=1\textwidth]{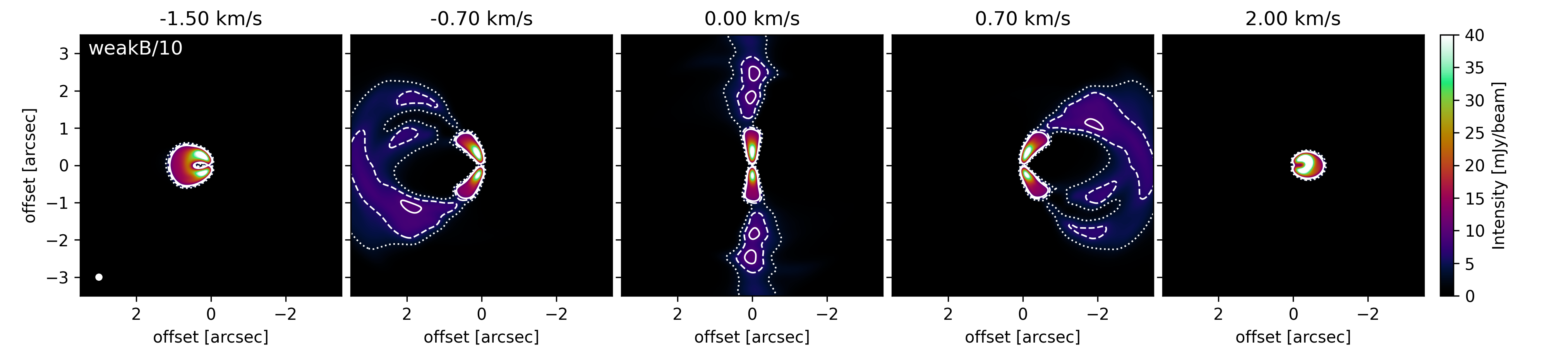}
\includegraphics[width=1\textwidth]{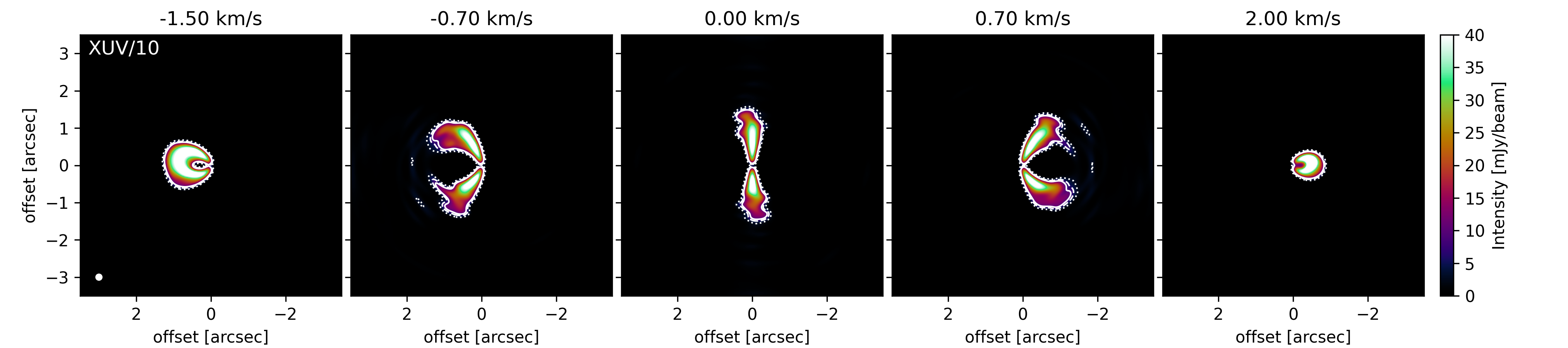}
    \caption{Channel map of a 0.002~$M_\odot$ mass disk, 1/10 of the fiducial disk mass. Because of the weaker signal, we reduced the color map range to 40 mJy/beam, and the white contours represent three different detection limits: $3\sigma$ (solid, instead of 5), $2\sigma$ (dashed, instead of 3), and $1\sigma$ (dotted).}
    \label{fig:lowmass}
\end{figure*}
We have performed similar simulations for a disk that is 1/10 the mass of the fiducial value, as many Class II disks are less massive than our fiducial model. The disk mass is now 0.002~$M_\odot$. We ran three models: two MHD models without XUV:$\beta=10^4$ as \texttt{strongB/10}, $\beta=10^5$ as \texttt{weakB/10}, and a PE wind model without magnetic field \texttt{XUV/10}. The wind loss rates of the three models are $3.5\times10^{-9}M_\odot~yr^{-1}$,$1.0\times10^{-9}M_\odot~yr^{-1}$, and $8\times10^{-9}M_\odot~yr^{-1}$. Note that due to the lower density, the temperate at the majority of the disk (except for the inner dense region) also deceased due to reduced heat exchange between gas and dust grains.

Again, we generated CO J=2-1 line emission channel maps, convolved with a $0\farcs15\times0\farcs15$ beam, and saturated the disk portion in Figure~\ref{fig:lowmass}. The original pre-convolved channel maps are listed in the appendices. A reduced CO abundance, e.g., a CO depletion factor of 10, would have a similar emission signature. Because of the weaker signal, we reduced the colormap upper limit and the three contours represent $3\sigma$, $2\sigma$, and $1\sigma$ thresholds instead. 

Model \texttt{weakB/10} is the only one that exhibits detectable signatures apart from the main disk. It features a bright disk surface that extends beyond the CO freeze-out radius. Similar to the fiducial disk, the weaker wind absorbs less UV photos in the upper atmosphere, leading to a more vertically confined CO layer. Notably, the outer disk, detectable at $2\sigma$ limit, surface in the zero velocity channel is twisted clockwise. This is a partial figure-eight structure when only the central ``vertical root'' is visible: the pattern in the top half sector is from the front surface and the one in the bottom sector comes from the backside surface. This is similar to the [C I] map of \texttt{weakB} in Figure~\ref{fig:peakchanC10}, where the ``horizontal root'' is dimmer in the central part of the figure-eight pattern. The small ``gap'' between the inner disk and the outer CO surface indicates a drop in CO density, which is likely due to (1) the sudden drop of CO density in the disk at the CO freeze-out radius, and (2) a slight shadowing effect from the inner disk wind. 

For the \texttt{XUV/10} model, no significant wind structures are observed. However, a partial ring feature is still evident, though only a small tip is detectable at $3\sigma$ level in the -0.7 km/s channel. Similar to \texttt{weakB/10}, the two tips at the top and bottom half in the zero velocity channel are from the front and backside surfaces, respectively. Because of the collapsing flow outward of the puffed-up region, they are twisted counter-clockwise instead. This is a robust outcome of photoevaporation. Its detectability indicates that even in the absence of large-scale winds, smaller substructures can provide valuable insights into disk dynamics and evolution. Notably, this indirect wind signature is less sensitive to disk mass and the density of tracer molecules in the wind.

\subsection{future observations}
\begin{figure}[hbtp!]
    \centering
    \includegraphics[width=0.45\textwidth]{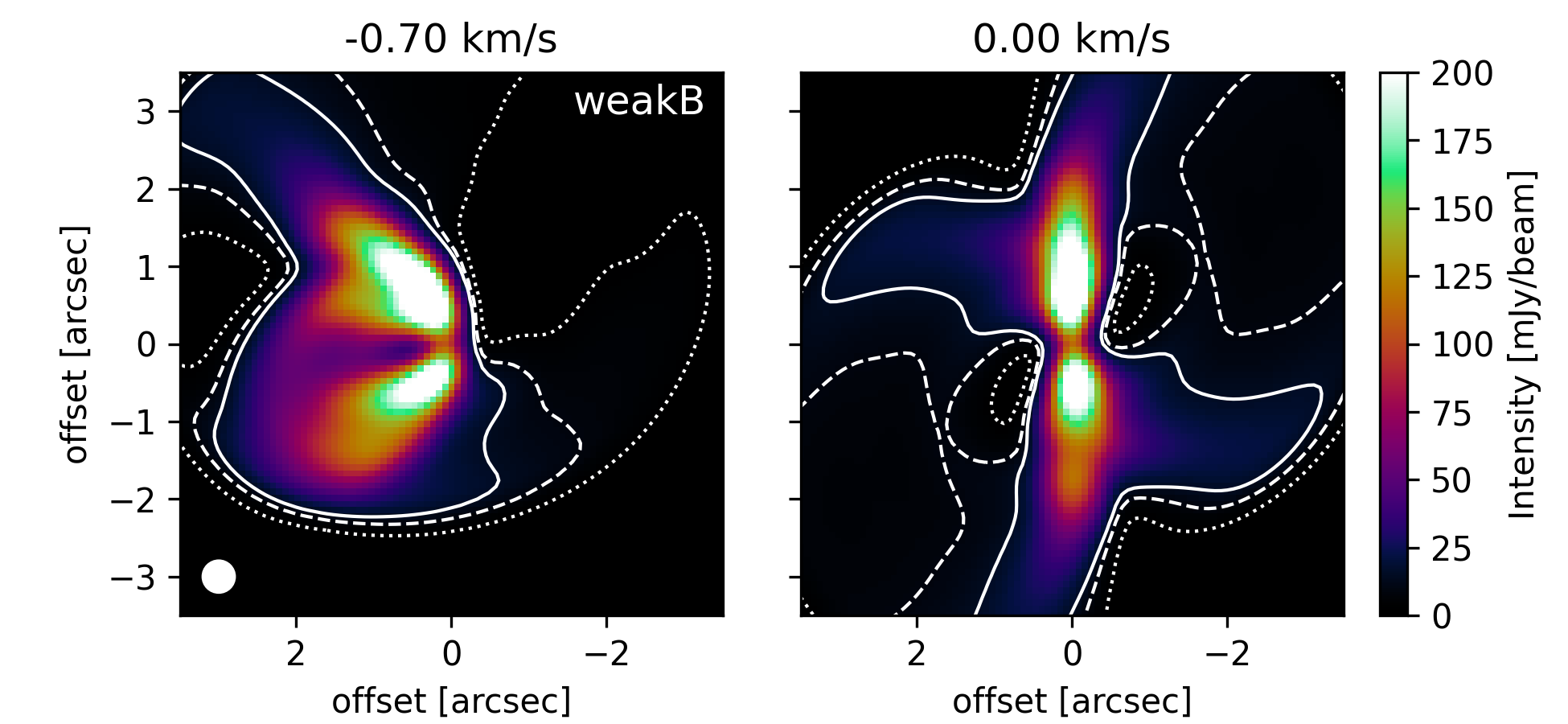} 
    \includegraphics[width=0.45\textwidth]{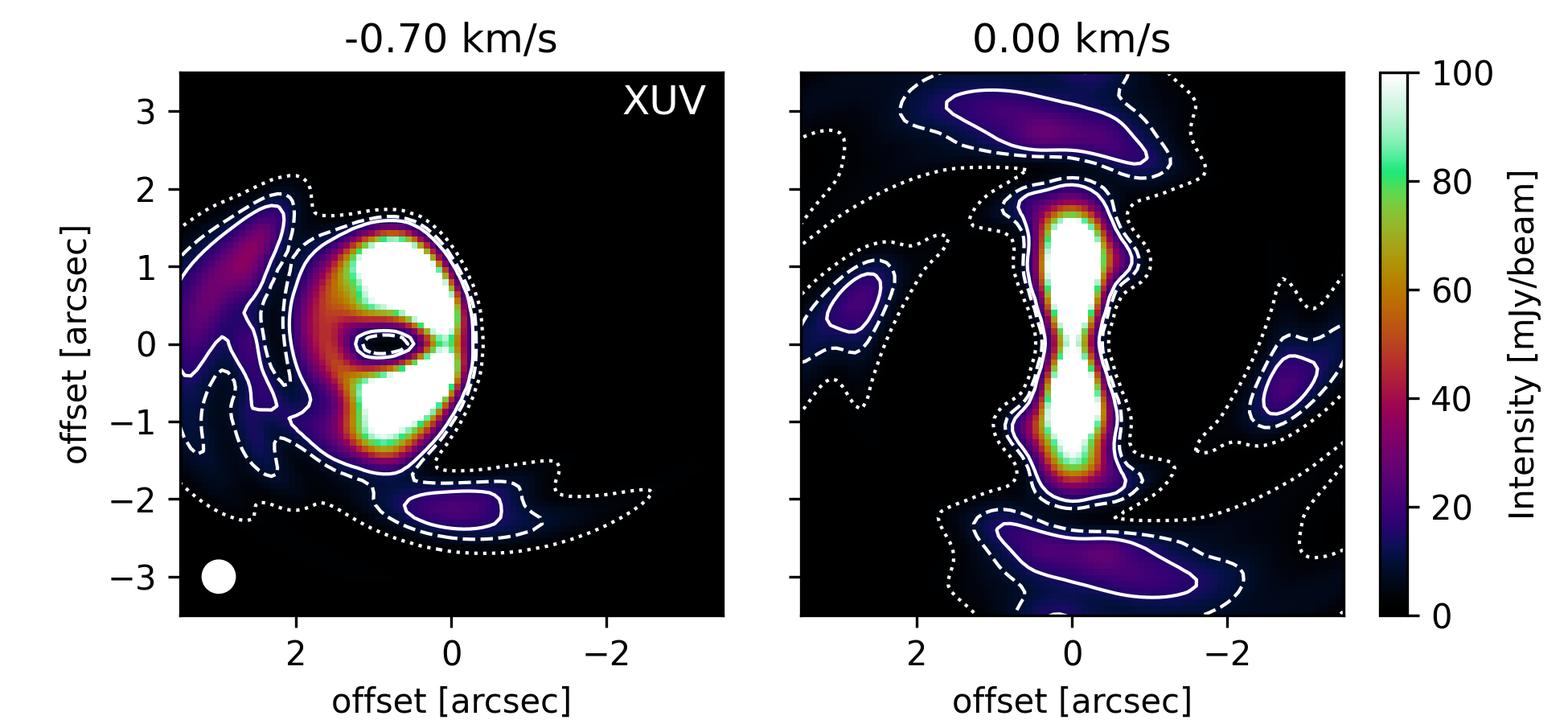}
    \includegraphics[width=0.45\textwidth]{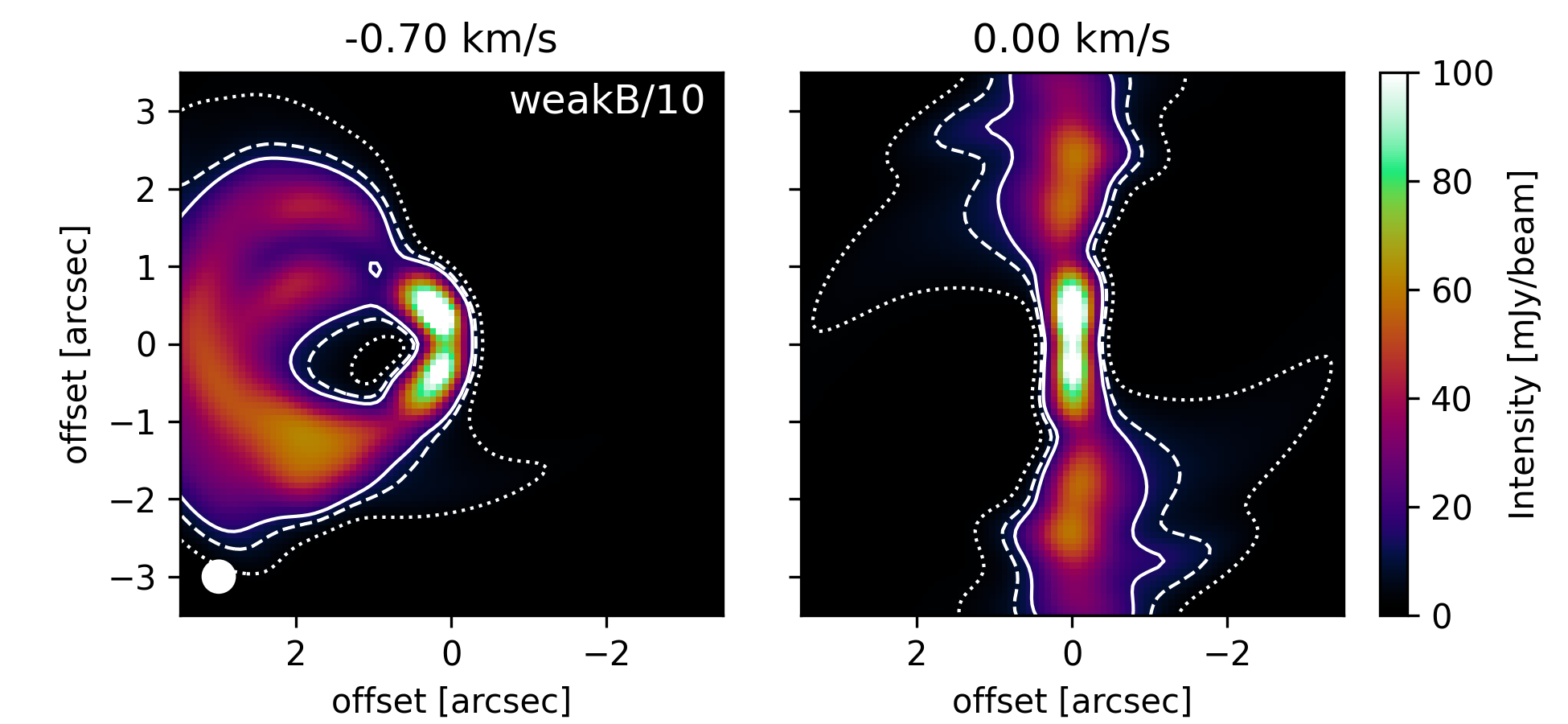}
    \includegraphics[width=0.45\textwidth]{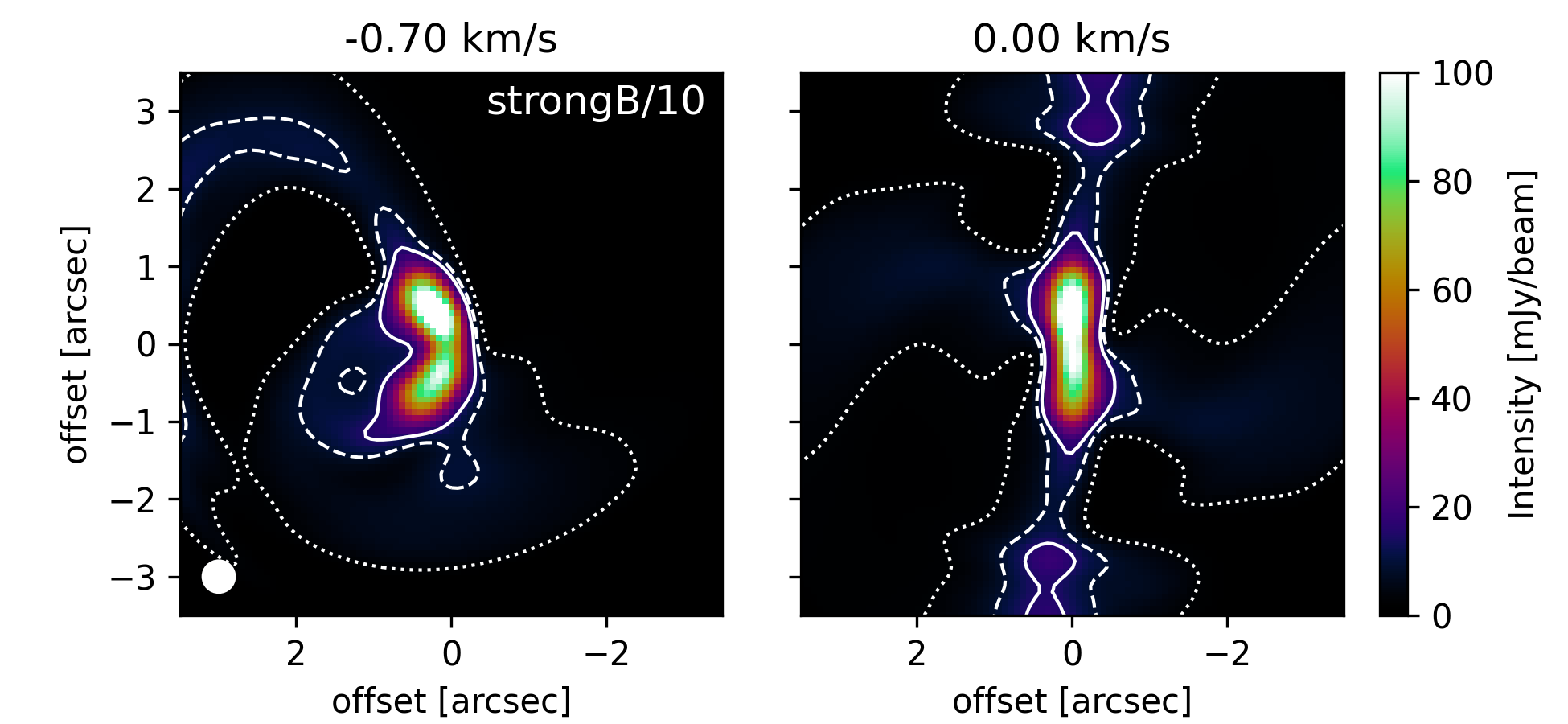}   
    \caption{CO J=2-1 channel maps convolved with a $0\farcs4\times0\farcs4$ beam (white spot at the bottom left corner). The white contours represent three detection limits: $5\sigma$ (solid), $3\sigma$ (dashed), and $1\sigma$ (dotted).}
    \label{fig:peakchan2p_0.4}
\end{figure}

Wind patterns in channel maps are typically more spatially extended than the Keplerian disk due to their high radial and polar velocities. While observations of CO J=2-1 emission often employ small beam sizes to resolve substructures, larger beams are recommended to better capture the dimmer, yet more extended wind signatures. To explore this, we convolved the CO emission maps for models \texttt{weakB}, \texttt{XUV}, \texttt{weakB/10}, and \texttt{strongB/10} with a $0\farcs4\times0\farcs4$ beam and presented the $-0.7~{\rm km~s}^{-1}$ and zero velocity channels in Figure~\ref{fig:peakchan2p_0.4}. A full list of $0\farcs4\times0\farcs4$ beam convolved channel maps of the eight setups (5 fiducial disk masses, 3 lower disk masses) is in the appendices. The larger beam significantly enhances the signal-to-noise ratio (SNR) of wind features. For the more massive disk (0.02~$M_\odot$) in \texttt{weakB}, the larger beam allows robust detection of the root of the figure-eight structure at $5\sigma$ in the zero velocity channel, even with a relatively low wind loss rate of $4.5 \times 10^{-9}~M_\odot~\text{yr}^{-1}$. In the $-0.7~{\rm km~s}^{-1}$ channel, the central portion (``root'') of the larger front surface and the smaller back emission surface, as illustrated in Figure~\ref{fig:model2face}, become visible. Both the partial ring from downward expanding wind and the outer ring from thermal broadening in \texttt{XUV} are above $5\sigma$ too. For less massive disks (0.002~$M_\odot$), the outer CO surface beyond the freeze-out radius becomes more prominent. While the CO surface is not a direct detection of disk wind, the twisted shape in the zero velocity channel serves as indirect evidence of strong vertical motions at the wind base. In model \texttt{strongB/10}, the outer CO surface is less bright, but most of the twisted zero velocity pattern remains detectable above the $3\sigma$ threshold. In the $-0.7~{\rm km~s}^{-1}$ channel, there is a clear contrast between the larger front surface and the smaller back surface, with portions of the front-side loop reaching the $3\sigma$ level. These observations underscore the importance of larger beam sizes for identifying extended wind structures and their associated kinematic signatures.

Compared to CO, [C I] can trace higher atmospheric layers in most cases, though achieving sufficient SNR remains challenging, even with a 2-hour integration. For example, while the peak C I column density in IM Lup exceeds \(10^{17}~\mathrm{cm}^{-2}\), our models yield no more than \(2 \times 10^{16}~\mathrm{cm}^{-2}\). A potential limitation of our setups is the 10 au inner boundary, which excludes the denser wind from the inner disk and could significantly increase atomic carbon density. Given that [C I] emission is largely optically thin, an order-of-magnitude increase in density would directly enhance brightness. Thus, [C I] is better suited for tracing winds in real observations, and longer integration times are strongly encouraged for studying disk winds.

\section{Summary}
\label{sec:summary}

We conducted two-dimensional axisymmetric simulations of MHD and photoevaporative (PE) winds in protoplanetary disks, with consistent thermochemistry. The main simulations included five different setups that combined varying magnetic field strengths and radiation fields (with and without XUV energy bins). A simple CO freeze-out mechanism is implemented to recover a realistic CO distribution in the disk. MHD winds, driven by a combination of centrifugal force and pressure gradient, exhibited a dense, slow outflow extending further into the polar regions, maintaining super Keplerian motion as it extracts angular from the disk. Conversely, photoevaporative winds, heated by high-energy XUV radiation, generated faster, less dense flows with primarily radial motion. The main results are as follows:

\begin{itemize}

\item The density and temperature of the two winds are different. In the disk atmosphere ($Z/R \gtrsim 0.45$), MHD winds are denser ($n_{\rm H} \simeq 10^6~{\rm cm}^{-3}$) and colder (few tens to hundreds Kelvin) than PE winds ($n_{\rm H} \simeq 10^5~{\rm cm}^{-3}$, $T\simeq$ few hundreds to thousands Kelvin; Figure~\ref{fig:Trhovzco}). 

\item The kinematics of the two winds are distinct. Once launched from the disk surface, magnetized winds quickly become super-Keplerian while photoevaporative winds become sub-Keplerian (Figure~\ref{fig:lkwe}). This is because the MHD wind is magnetically coupled to the rotating disk, allowing it to carry away the disk's ``extra'' angular momentum, whereas the PE wind only retains and conserves its own angular momentum. The poloidal velocity of magnetized winds remains low at $\simeq 1~{\rm km~s}^{-1}$, while that of photoevaporative winds reaches several ${\rm km~s}^{-1}$ as it feels the sustaining acceleration from the pressure gradient in the much hotter atmosphere. These differences lead to distinct streamlines (Figure~\ref{fig:stream3d}), suggesting that the ratio between azimuthal and poloidal velocities could be used to distinguish the origin of the two winds.

\item Assuming 2 hours of on-source integration with ALMA, direct detection of magnetized winds in CO channel maps is possible when wind loss rates are high ($\gtrsim 10^{-8}~M_\odot~{\rm yr}^{-1}$; Figure~\ref{fig:peakchan}). When wind loss rates are lower, weakly magnetized disk winds produce subtle morphological changes to the characteristic ``butterfly'' pattern in CO channel maps, which resemble the so-called velocity kinks induced by embedded protoplanets, e.g., slightly bent ``tips'' at the zero velocity channel (Figure~\ref{fig:peakchan}). Note that even for the less magnetized cases, the evolved midplane field strength reaches 1 mG at 100 au, and 0.2 mG at 200 au, just slightly weaker than the recently measured HD142527 disk \citep[0.3 mG at 200 au,][]{2025NatAs.tmp...37O}. In this context, it is likely that most disks with high-resolution ALMA data are only weakly magnetized. 

\item Direct detection of photoevaporative winds in CO channel maps appears to be generally challenging because strong XUV radiation dissociates CO; however, photoevaporative winds can create ring-like substructures which may be observable in deep CO observations (Figure~\ref{fig:peakchan}). The XUV photons heat up the disk surface within 150 au and create a puffed-up layer. The hot wind that expands towards the disk surface at the outer edge of this layer leaves $v_z$ perturbations.

\item Observed wind kinematics can be used to further distinguish MHD wind and PE wind. The ``two-sided-loop'' patterns (Figure~\ref{fig:model_elevated},\ref{fig:peakchanC10}) are unique to MHD winds when detected at z/R=0.5 or higher, as the PE wind does not have a $v_\theta$ component that is comparable to $v_r$. At the outer edge of the disk surface (usually $\sim$ CO emission surface), a PE wind dominated disk usually sees the $v_\phi$ drops below Keplerian, together with reduced or negative $v_z$ and increased $v_R$ (bottom panel of Figure~\ref{fig:3v3h}), though $v_z$ could increase again even further out. 

\item In both MHD and PE winds, [C I] emission is found to be optically thin and originates from higher layers than CO. [C I] in our models is significantly weaker than CO and with 2 hours of on-source integration, we predict that the [C I] emission is just above $1\sigma$ (Figure~\ref{fig:peakchanC10}). However, we note that the [C I] column density in our models may have been underestimated because we excluded the dense inner disk within 10~au.

\item {[C I] emission can appear brighter with less dense (weaker) wind and/or stronger XUV radiation. This is because atomic carbon is produced through CO photodissociation, and the photodissociation region is closer to the disk surface in a less dense atmosphere, then the [C I] emission would come from a denser region (rather than the thinner atmosphere).}

\item  Because both magnetized and photoevaporative winds are spatially extended, using a large beam (e.g., $\simeq0\farcs4$ for disks in nearby star-forming regions) will be helpful to observe disk winds, and direct wind detection is even possible for less massive disks (Figure~\ref{fig:peakchan2p_0.4}, see also Figure~\ref{fig:peakchanCO_0.4},~\ref{fig:peakchan_lowmass_0.4}).

\item Regardless of the origin, we found that disk winds significantly influence the rotational velocities of the gas in the atmosphere. In addition to thermal pressure gradient and disk's self-gravity (not included in our simulations), which previous studies showed to modulate the rotational velocity, magnetic pressure/tension and advection terms arising from spatial changes in radial velocities can significantly influence rotational velocities of the gas (Figures~\ref{fig:vphiterm300ev},\ref{fig:vphitermb4},\ref{fig:vphicalc}).

\end{itemize}

\begin{acknowledgements} 
We thank the referee for detailed and constructive comments, which
improved the presentation of the paper. We thank Andres Izquierdo for the helpful discussion, and Thomas Haworth for the suggestion of adding PV diagrams. Our simulations are made possible by an XSEDE allocation (AST200032). Z. Z. acknowledges support from NASA award
80NSSC22K1413 and NSF award 2408207. 
\end{acknowledgements}

\software{
Athena++ \citep{2020ApJS..249....4S}
RADMC-3D \citep{2012ascl.soft02015D}
Matplotlib \citep{Hunter2007}
NumPy \citep{2020Natur.585..357H}
CMasher \citep{2020JOSS....5.2004V}
PyVista \citep{2019JOSS....4.1450S}
}


\appendix

\restartappendixnumbering

\section{LOS velocity map of with constant radial motion}

\begin{figure*}[!h]
    \centering
    \includegraphics[width=0.3\linewidth]{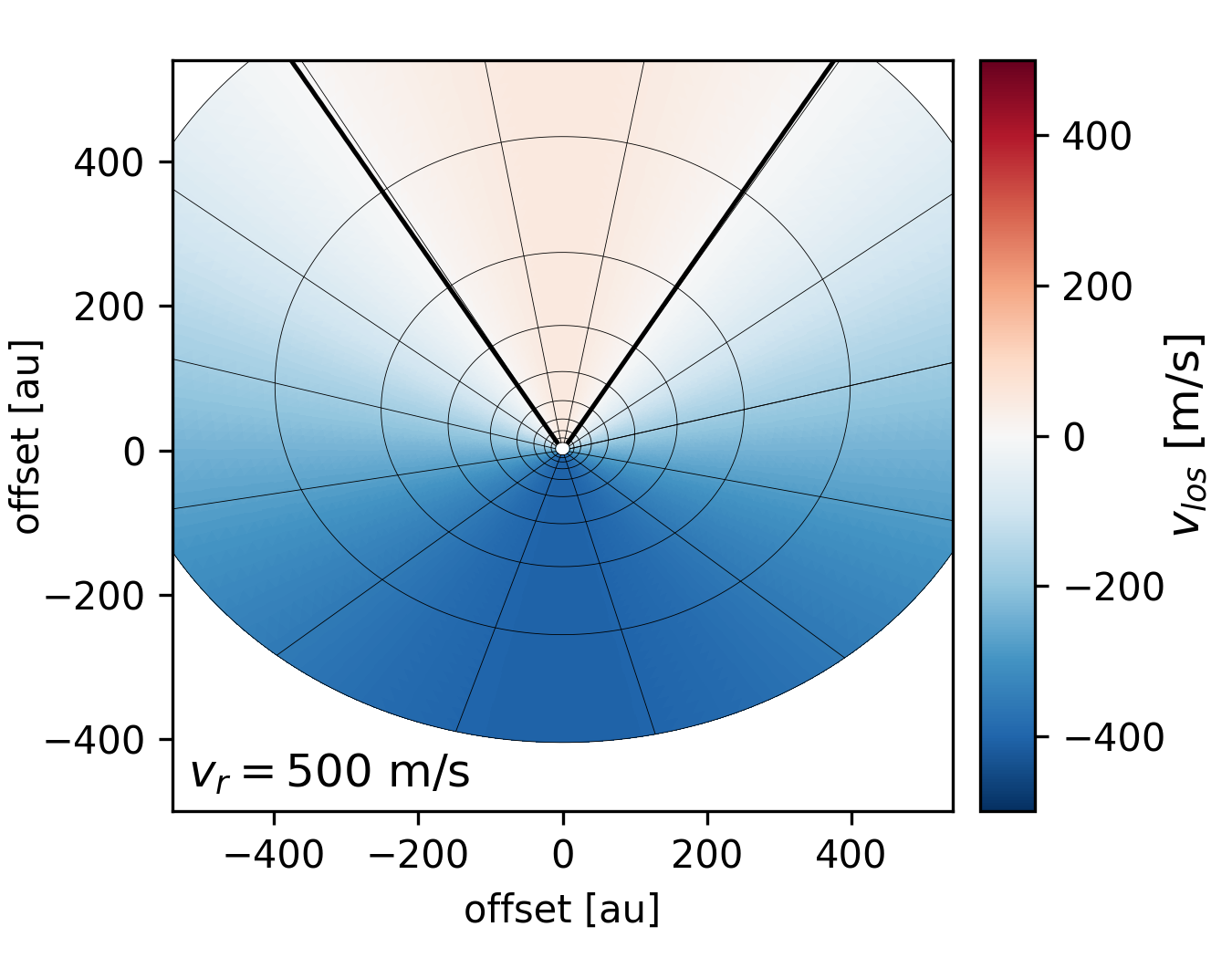} \includegraphics[width=0.3\linewidth]{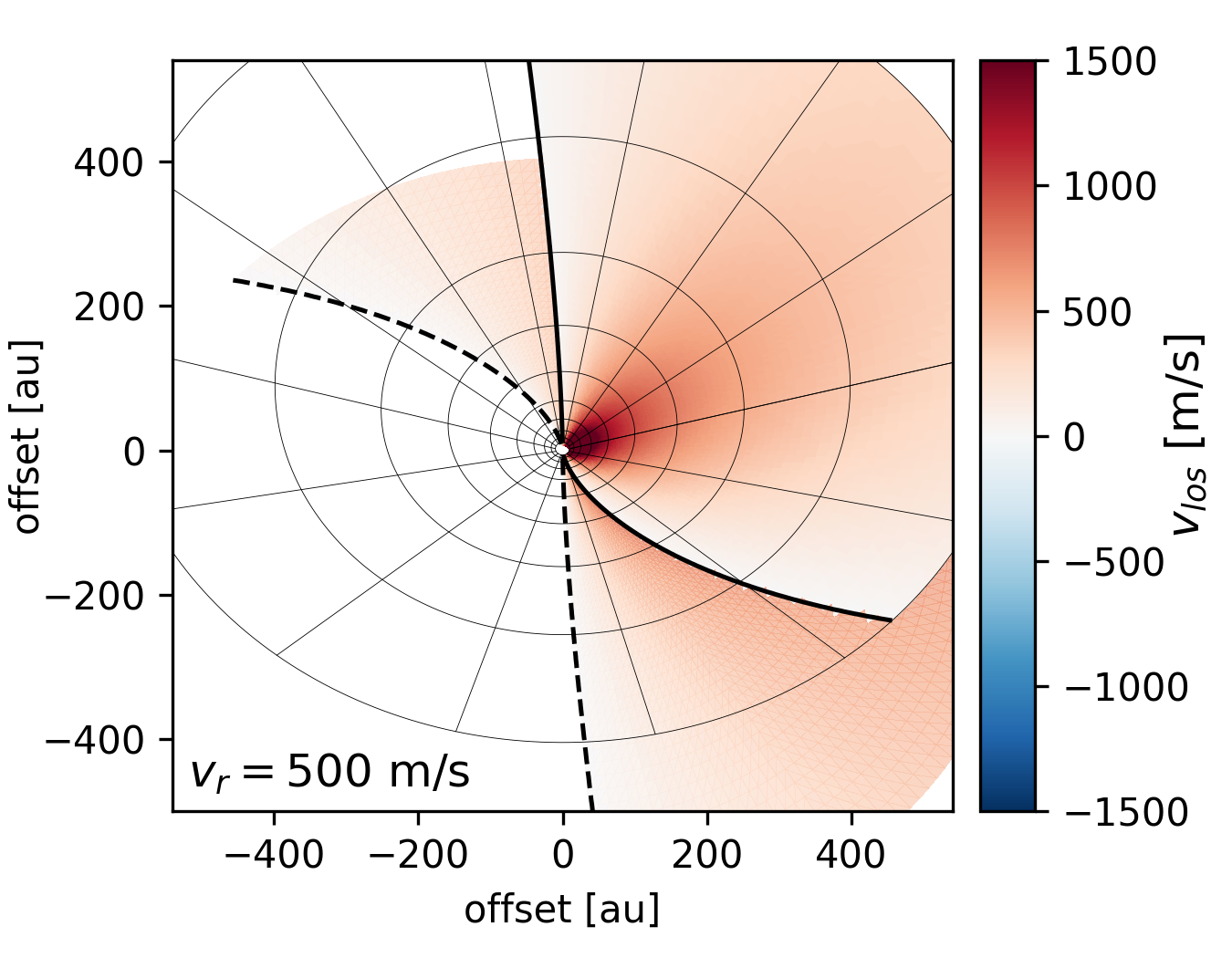}
    \includegraphics[width=0.3\linewidth]{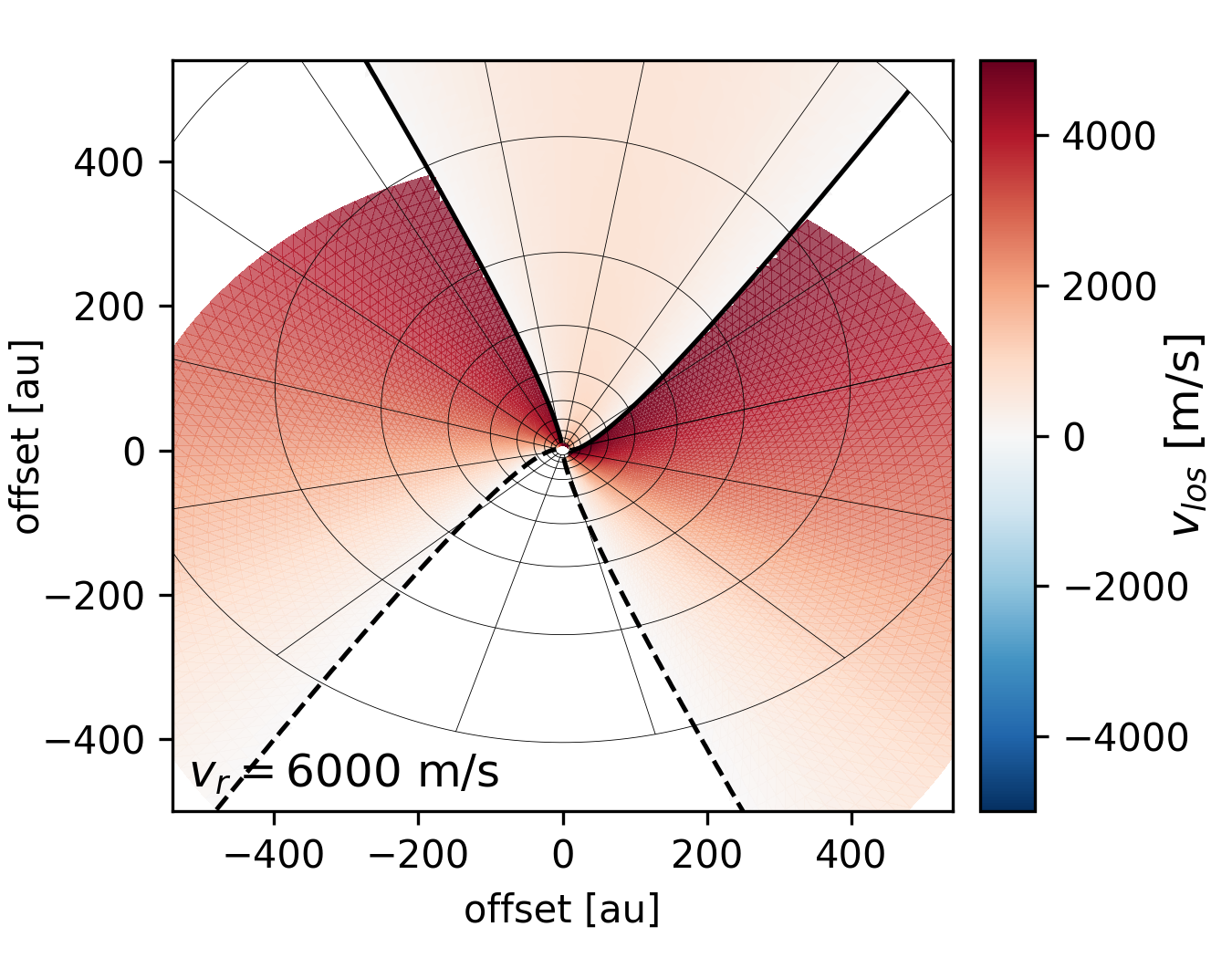}
    \caption{LOS velocity maps with different $v_r$ at z/R=0.45. The left panel is a disk without rotation ($v_\phi=0$) with a constant $v_r$=500 m/s, and only the front surface is shown. The middle and right panels have a sub-Keplerian (0.9$v_K$) surface with masked blueshift patterns. The middle panel also incorporates $v_r$=500 m/s while the right panel uses $v_r$=6000 m/s. The black solid lines are the front surface's zero velocity contours and the dashed ones are from the back surface. Note the color bar range is different in each panel.}
    \label{fig:vrmodel}
\end{figure*}

To illustrate how radial motion affects the LOS velocity, we set $v_\phi$ to zero in the left panel of Figure~\ref{fig:vrmodel}. The ``emission surface'' is set to $z/R=0.45$. As $\arctan{(0.45)}=24.2^\circ$ and the inclination is 30$^\circ$, the upper half along the major axis is still tilted ``backward''. Thus for a constant $v_r$=500~m/s, the upper middle quarter of the disk surface shows redshift. The zero velocity contour is where the emission surface intersects the plane perpendicular to the LOS, giving the V shape. The middle panel shows the same $v_r$ on a slightly sub-Keplerian ($v_\phi=0.9~v_K$) surface, where the zero velocity contour on the front surface is twisted counter-clockwise, making the redshifted portion smaller than the blueshifted half. Within the zero velocity contour, the top half is close to the minor axis as an unperturbed disk, as the projected LOS velocity is low. If we can probe the upper atmosphere where the PE wind is much faster, the LOS velocity map at that layer would be similar to the right panel. In reality, the emission is likely to be optically thin so the morphology in channel maps is the combination of the two extremes (middle and right panels). The resulting PE wind features will be more extended spatially.

\begin{figure*}[!h]
    \centering
    \includegraphics[width=1\linewidth]{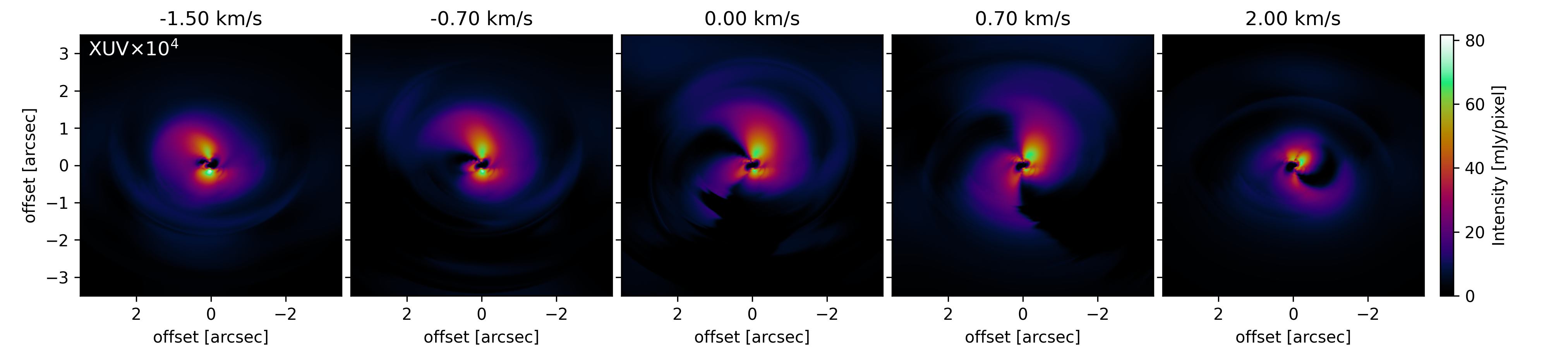}
    \caption{[C I] $^3P_1-^3P_0$ channel maps without beam convolution. The atomic carbon abundance is enhanced by $10^4$ times in model \texttt{XUV} in post-processing.}
    \label{fig:chan5pC10x1e4xuv}
\end{figure*}

Figure~\ref{fig:chan5pC10x1e4xuv}, demonstrates the morphology of the PE wind if it is directly detected. Since CO is effectively photodissociated above $z/R=0.4$, we enhance the atomic carbon abundance by a factor of $10^4$ to trace the wind region above. The radial motion-dominated PE wind looks quite different from the MHD wind. Due to the nature of pressure gradient acceleration, the PE wind has a larger velocity gradient along its trajectory compared to the MHD wind, as shown in Figure~\ref{fig:lkwe}f. The radial velocity near the wind base is an order of magnitude lower than in the atmosphere. For the front emission surface in Figure~\ref{fig:vrmodel}, the upper halves of the zero velocity channel are similar between different $v_r$s, while the lower halves are scattered around. The wide range of $v_r$ at different heights makes PE wind patterns particularly wide in channel maps.

\section{Position-velocity diagrams}

\begin{figure*}
    \centering
    \includegraphics[width=0.45\textwidth]{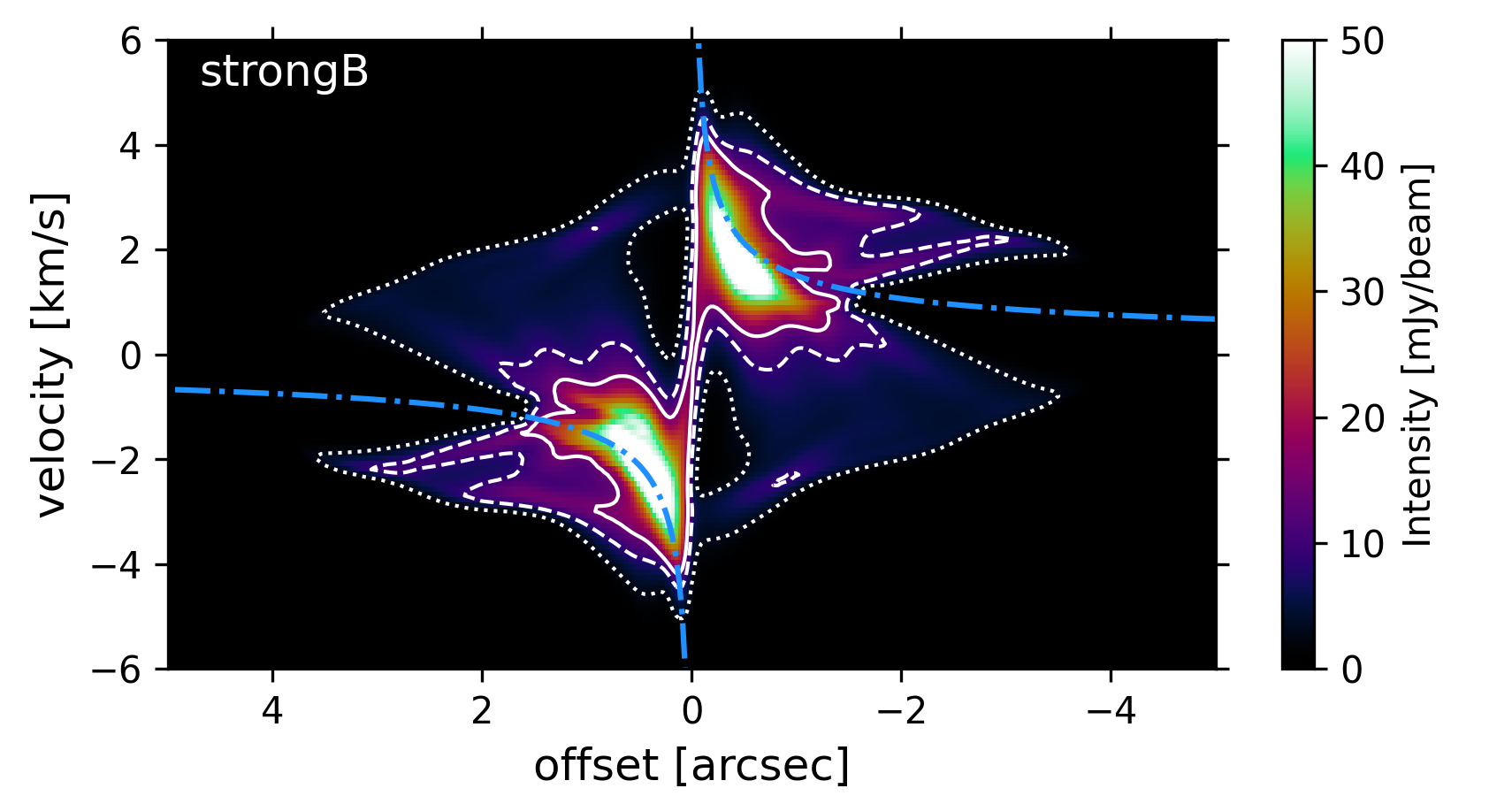}
    \includegraphics[width=0.45\textwidth]{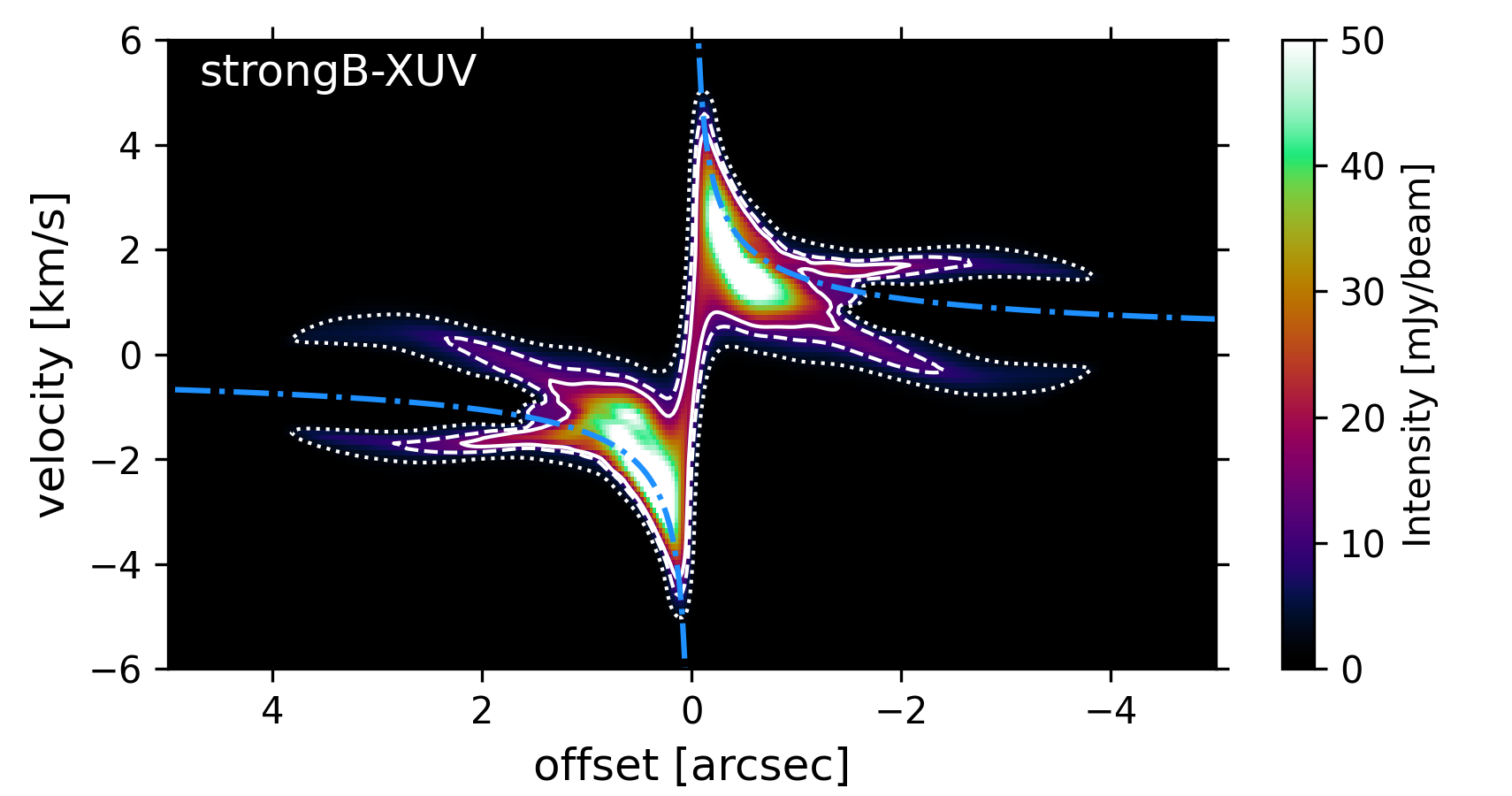}
    \includegraphics[width=0.45\textwidth]{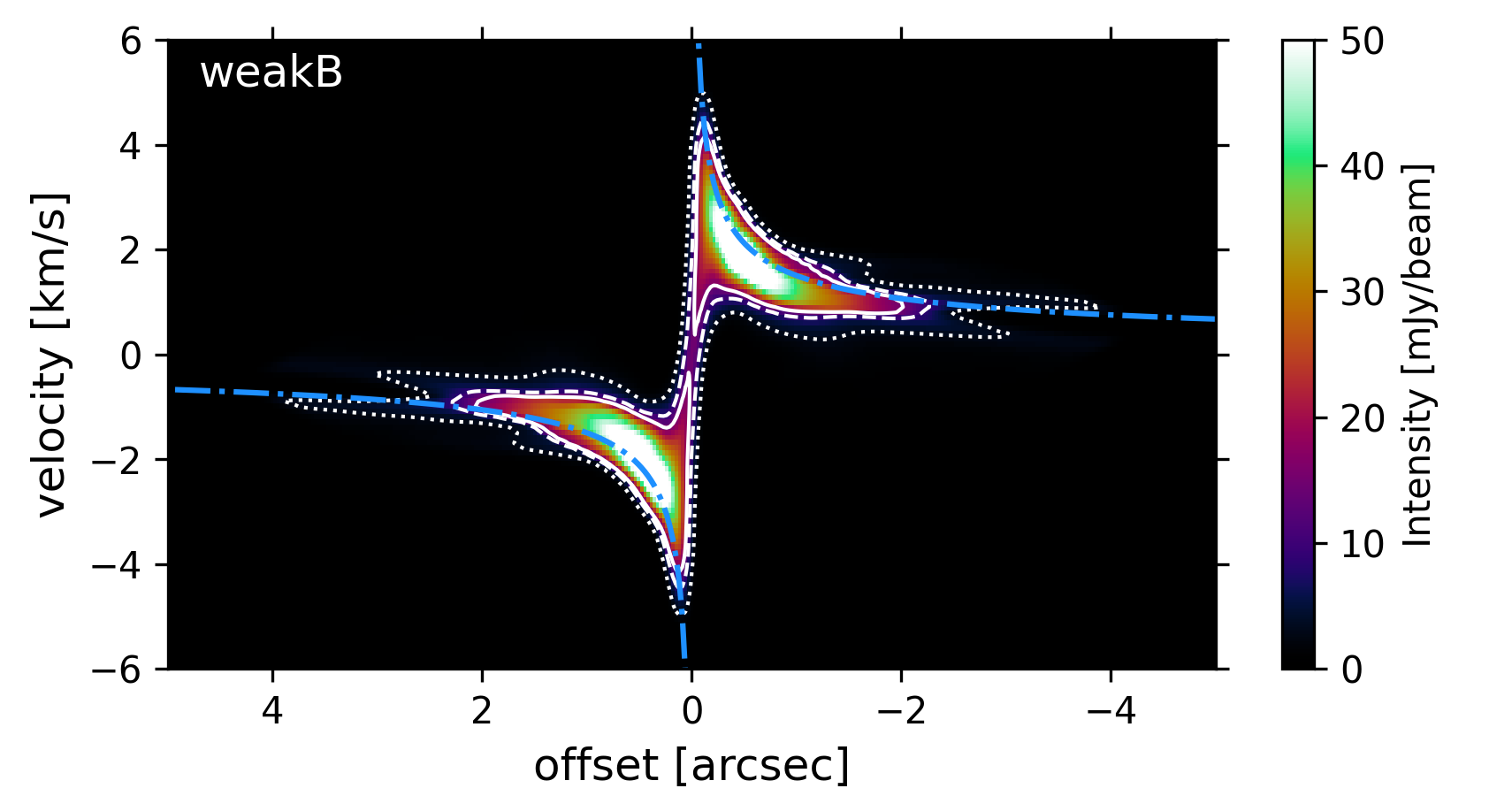}
    \includegraphics[width=0.45\textwidth]{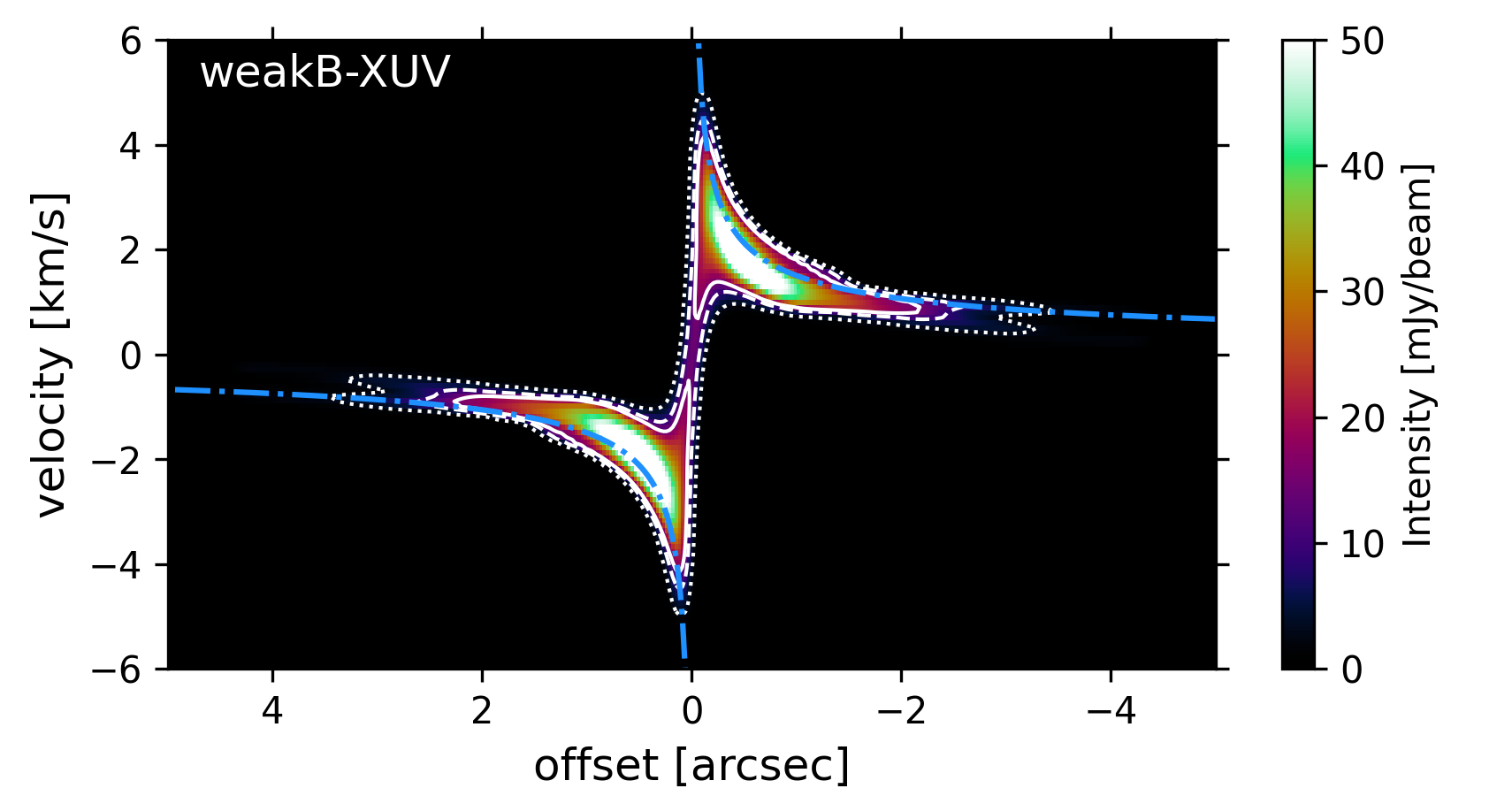}
    \includegraphics[width=0.45\textwidth]{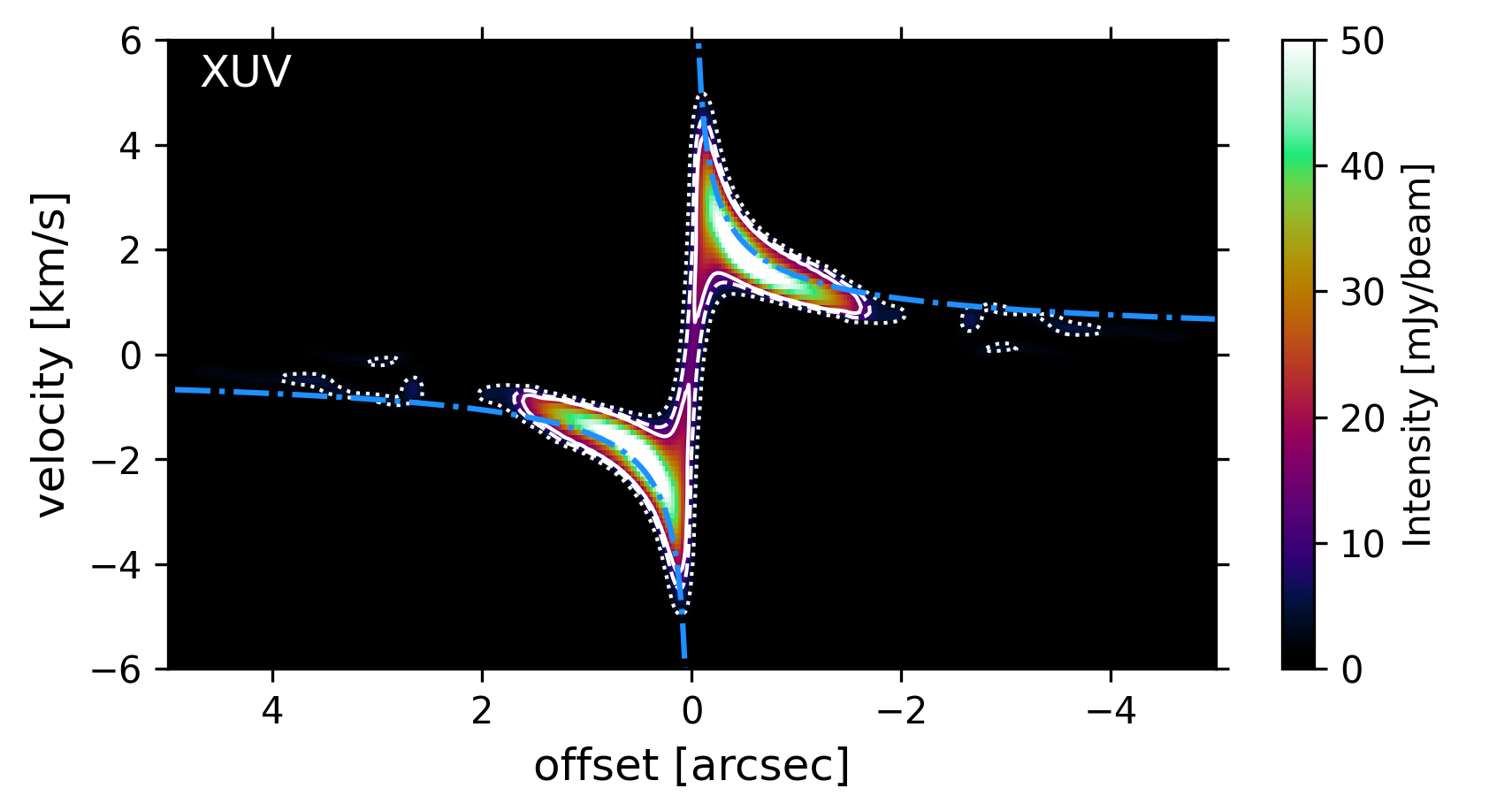}
    \caption{PV diagrams of CO J=2-1 channel maps convolved with a $0\farcs15$ beam. The blue dash-dotted lines outline the Keplerian rotation curve. We used the same colormap scale as Figure~\ref{fig:peakchan} for consistency, and the white contours still represent three detection limits: $5\sigma$ (solid), $3\sigma$ (dashed), and $1\sigma$ (dotted).}
    \label{fig:pv}   
\end{figure*}

Position-velocity (PV) diagrams are another useful tool for identifying non-Keplerian components in disks. Here we take a horizontal slice across the center of each image cubes used for the CO channel maps of Figure~\ref{fig:peakchan}. For every slice in each velocity channel, the 1D array is averaged within a height of 10 pixels. For reference, Keplerian rotation curves are over-plotted as blue dash-dotted lines in each panel. Here the Keplerian velocity is calculated at the midplane, i.e., the standard Keplerian values, so this is not a rigorous evaluation of the how much the deviation from being Keplerian but a quick scan of the overall rotation profiles, mainly because the wind emissions are way above midplane. Similar to Figure~\ref{fig:peakchan}, only model \texttt{strongB} and \texttt{strongB-XUV} show significant non-Keplerian components. Beyond $2\farcs$ offsets, the wind feature diverges into a double-winged structure. Take the left side of the PV diagram as an example, the super-Keplerian component comes from the MHD wind at the front side, and the other component that appears to be sub-Keplerian and even ``reverse-Keplerian'' comes from the wind at the backside. The reason for this ``reverse-Keplerian'' is not the rotation, but the radial (and vertical) motion in the wind. The backside wind moves away from the observer, making the redshifted wing stack above the blueshifted main component. This redshifted feature is also significantly weaker because streamlines in MHD wind only ``tilt-up'' at higher regions with lower densities (see Figure~\ref{fig:lkwe}(a)). Model \texttt{weakB} and \texttt{weakB-XUV} have some $1\sigma$ wind signals that lie slightly above and below the Keplerian curve. In model \texttt{XUV}, all signals beyond the disk are slightly sub-Keplerian.

\section{Additional channel maps}

\begin{figure*}[htbp!]
\centering
    \includegraphics[width=1\textwidth]{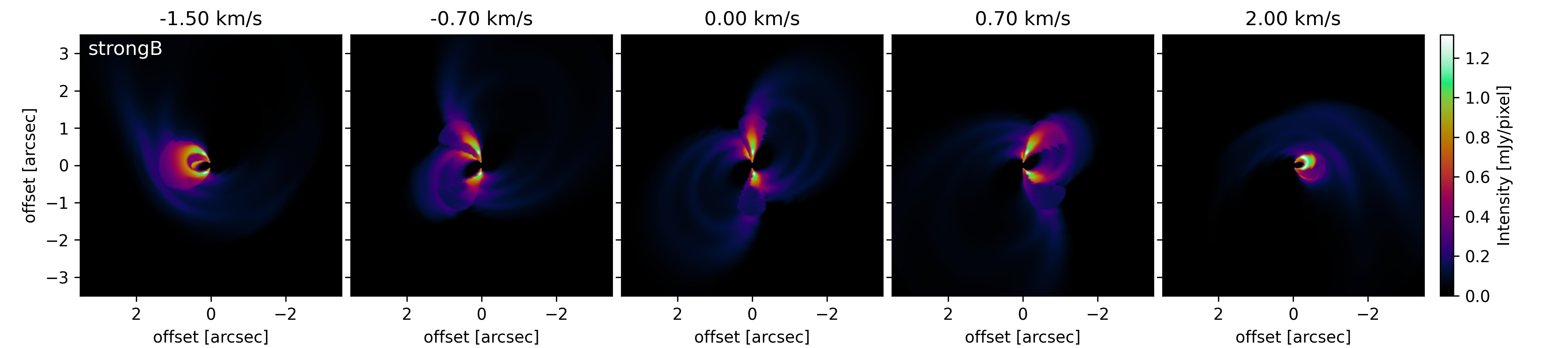}
    \includegraphics[width=1\textwidth]{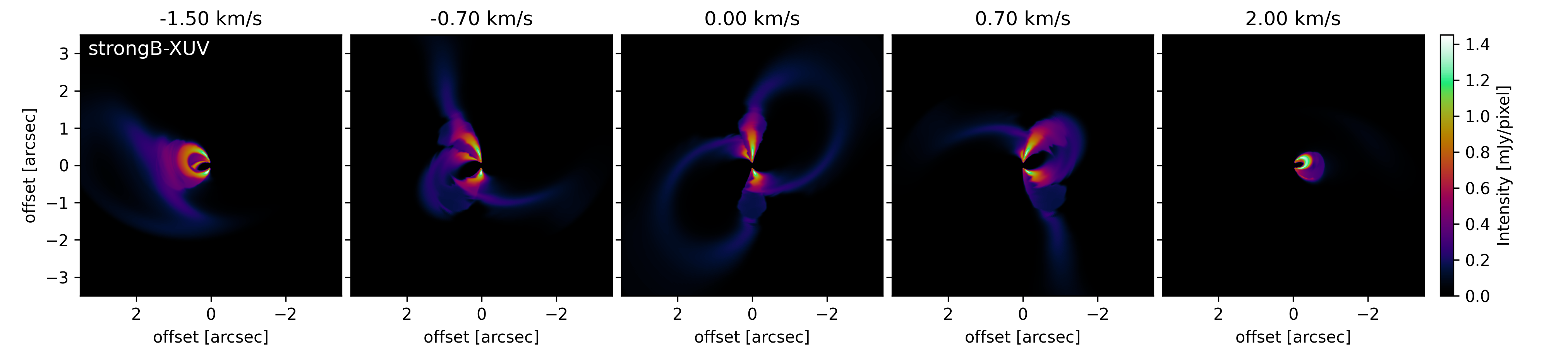}
    \includegraphics[width=1\textwidth]{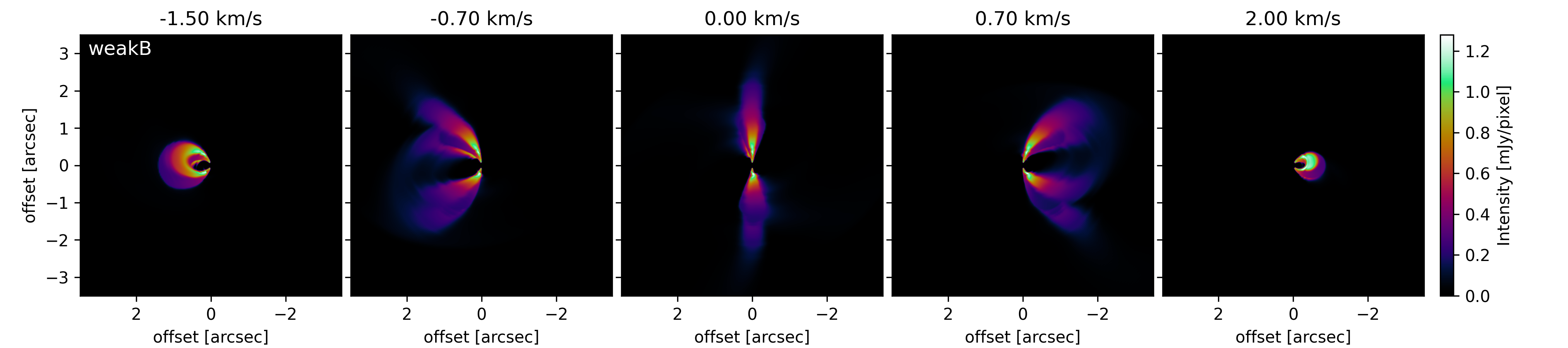}
    \includegraphics[width=1\textwidth]{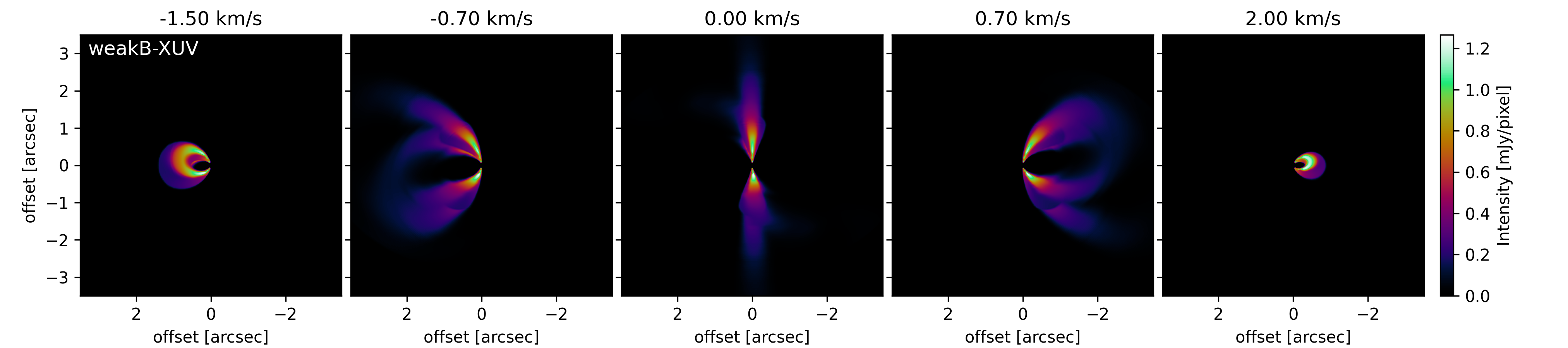}
    \includegraphics[width=1\textwidth]{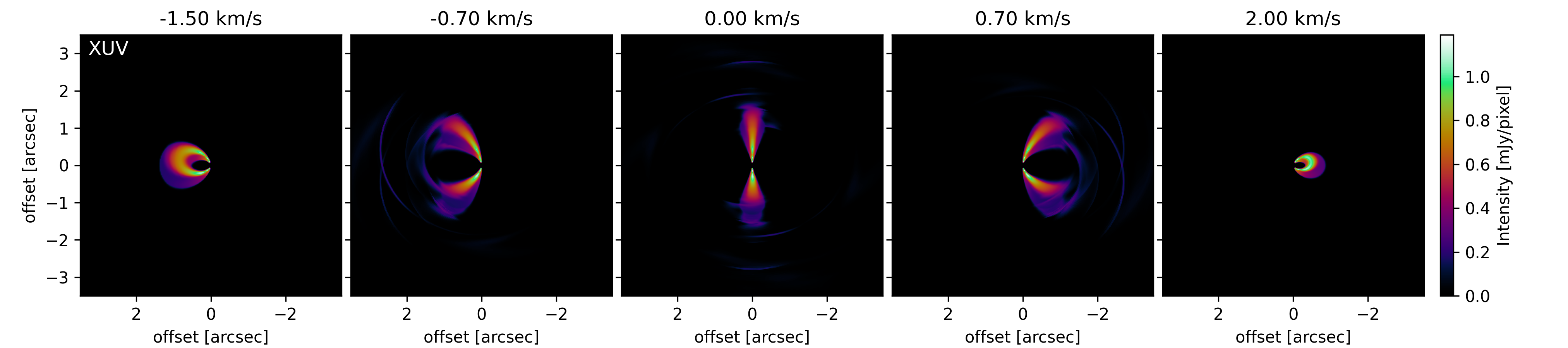}
    \caption{CO J=2-1 channel maps without beam convolution. From top to bottom: $\beta=10^4$, $\beta=10^4$ with 300 eV radiation, $\beta=10^5$, $\beta=10^5$ with 300 eV radiation, and PE model with 300 eV radiation. }
    \label{fig:peakchan_origin}
\end{figure*}

\begin{figure*}[htbp!]
\centering
    \includegraphics[width=1\textwidth]{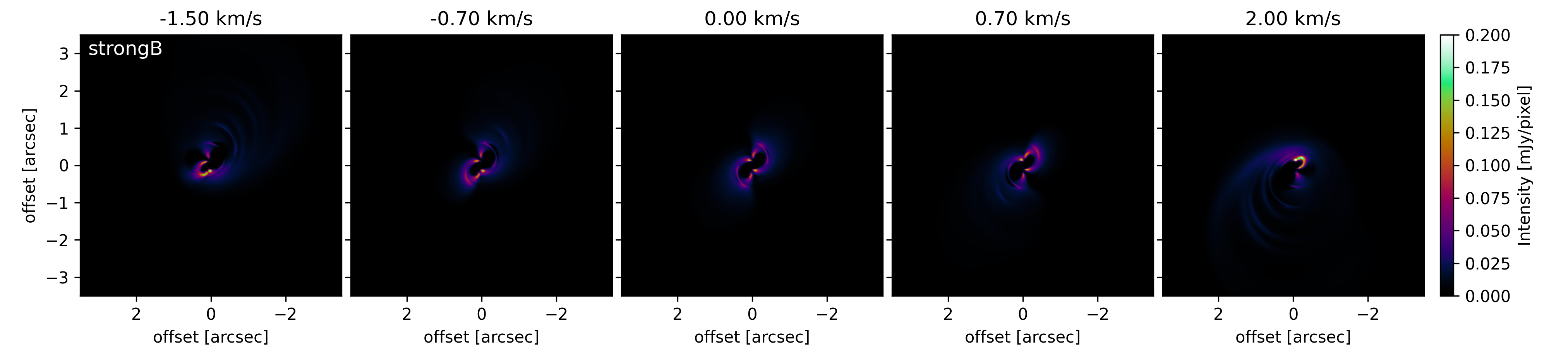}
    \includegraphics[width=1\textwidth]{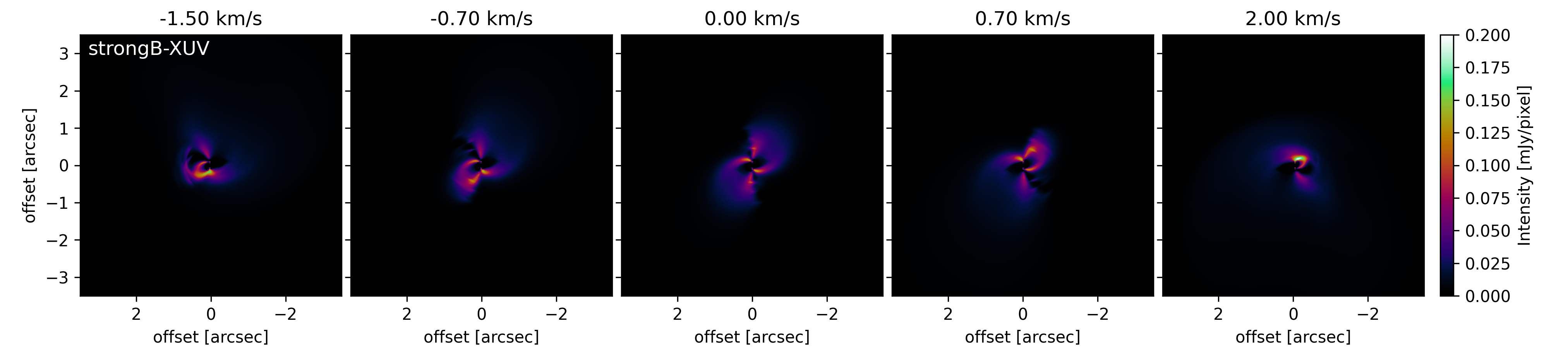}
    \includegraphics[width=1\textwidth]{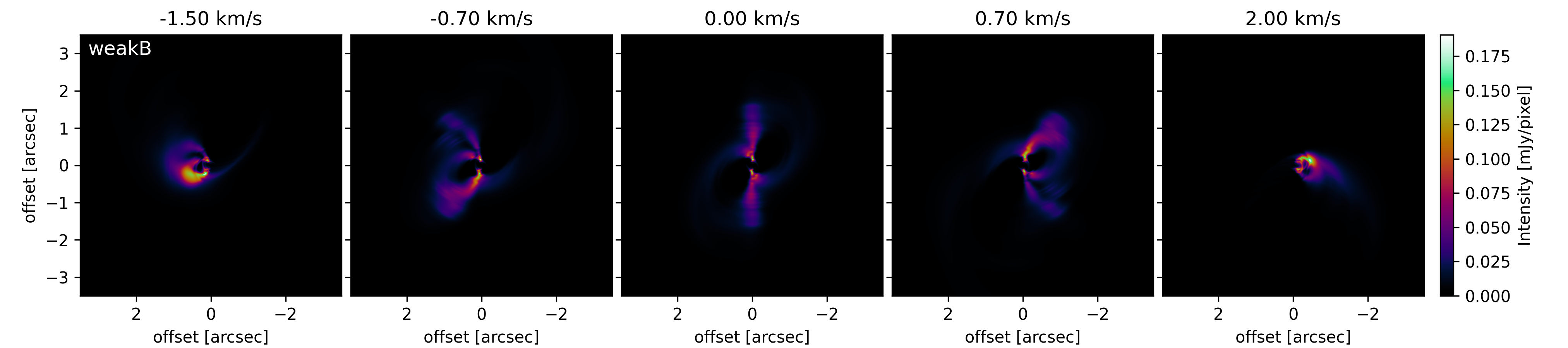}
    \includegraphics[width=1\textwidth]{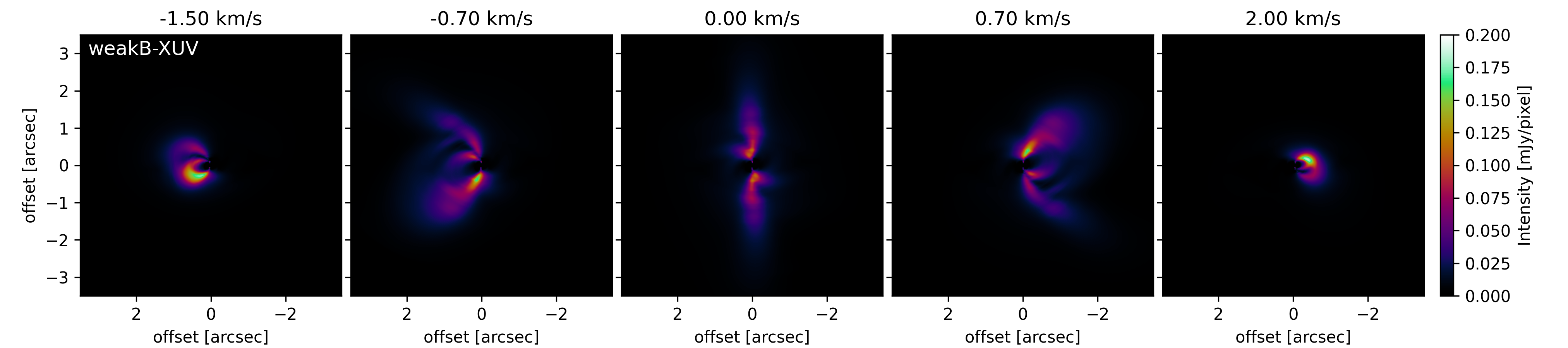}
    \includegraphics[width=1\textwidth]{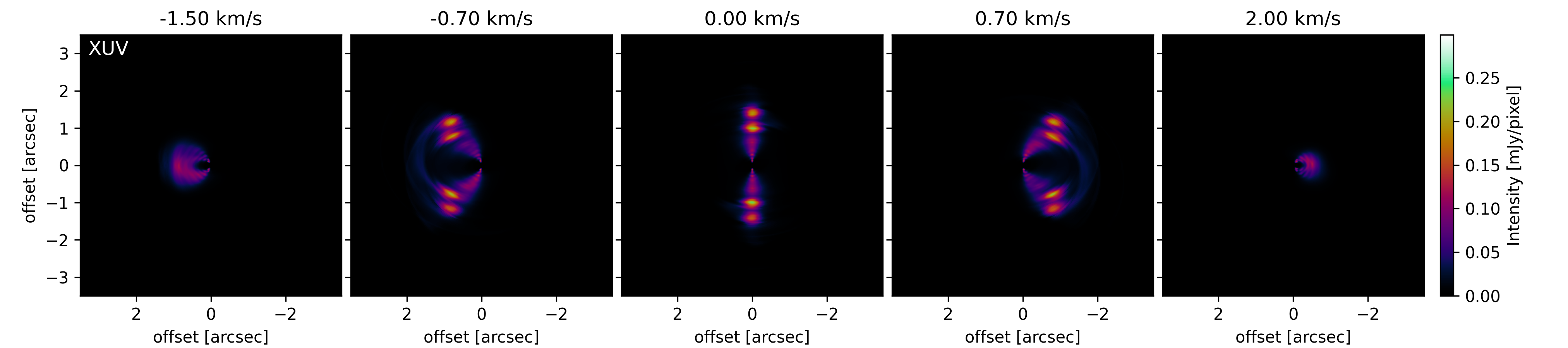}
    \caption{[C I] $^3P_1-^3P_0$ channel maps without beam convolution. From top to bottom: $\beta=10^4$, $\beta=10^4$ with 300 eV radiation, $\beta=10^5$, $\beta=10^5$ with 300 eV radiation, and PE model with 300 eV radiation.}
    \label{fig:peakchanC10_origin}
\end{figure*}

\begin{figure*}[htbp!]
\centering
    \includegraphics[width=1\textwidth]{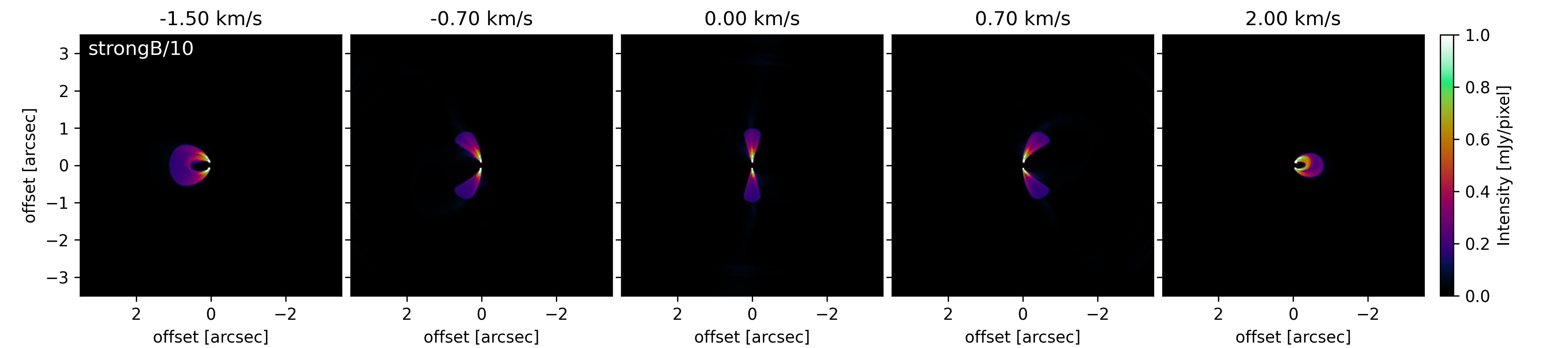}
    \includegraphics[width=1\textwidth]{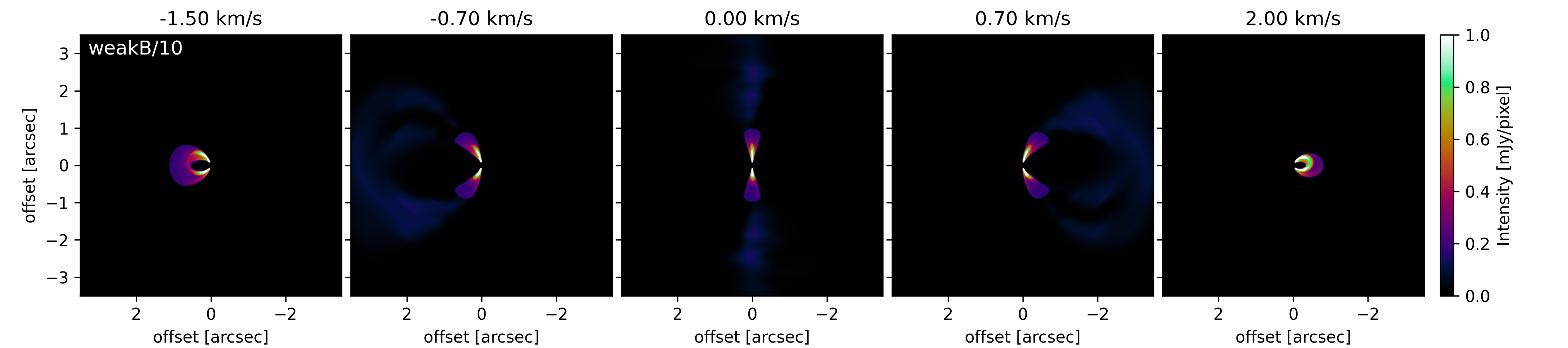}
    \includegraphics[width=1\textwidth]{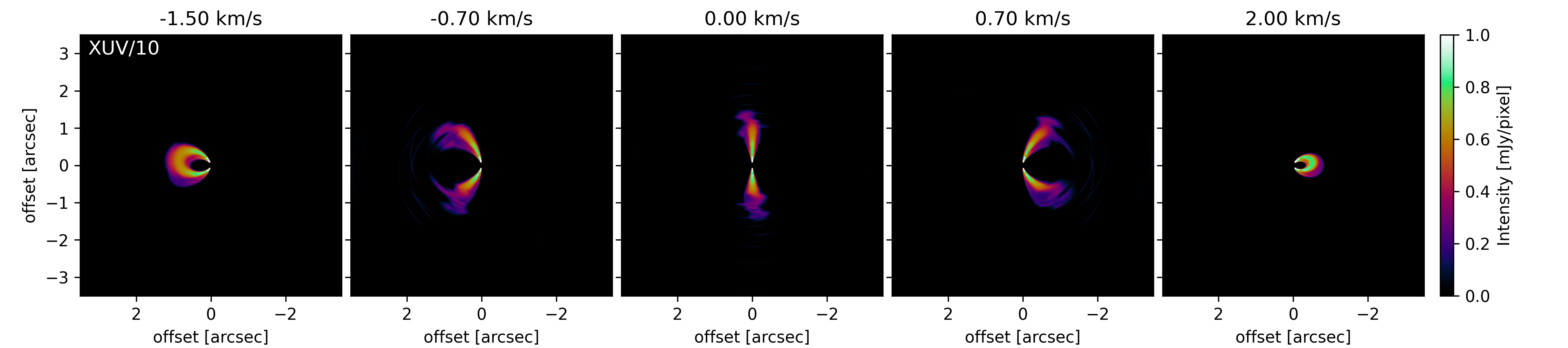}
    \caption{CO J=2-1 channel maps without beam convolution of a 0.002~$M_\odot$ mass disk, 1/10 of the fiducial disk mass.}
    \label{fig:peakchan_lowmass_origin}
\end{figure*}

\begin{figure*}[htbp!]
\centering
    \includegraphics[width=1\textwidth]{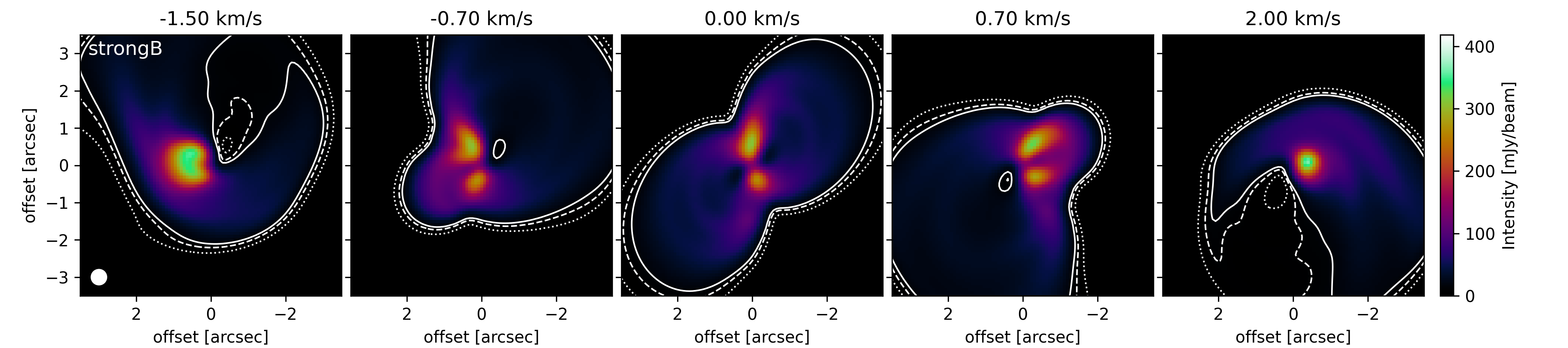}
    \includegraphics[width=1\textwidth]{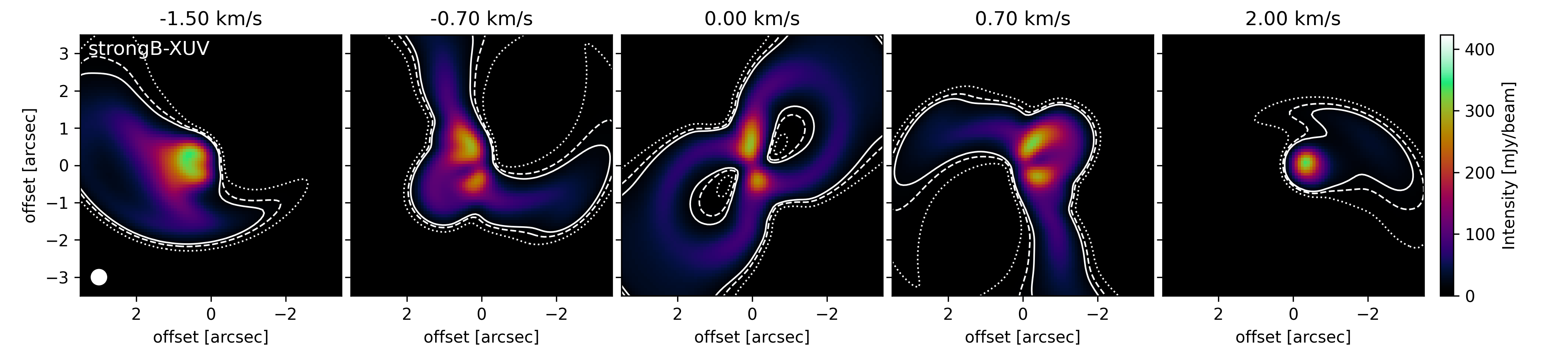}
    \includegraphics[width=1\textwidth]{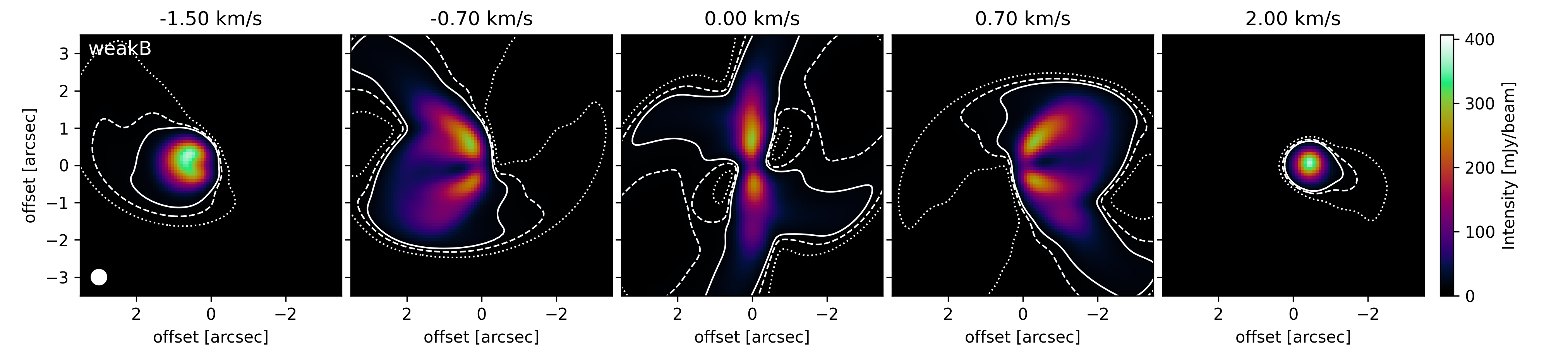}
    \includegraphics[width=1\textwidth]{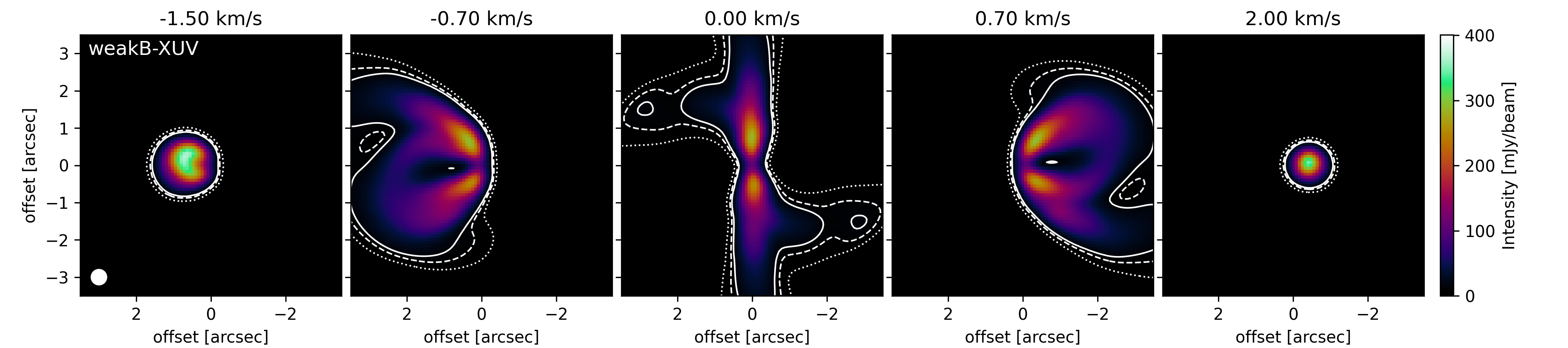}
    \includegraphics[width=1\textwidth]{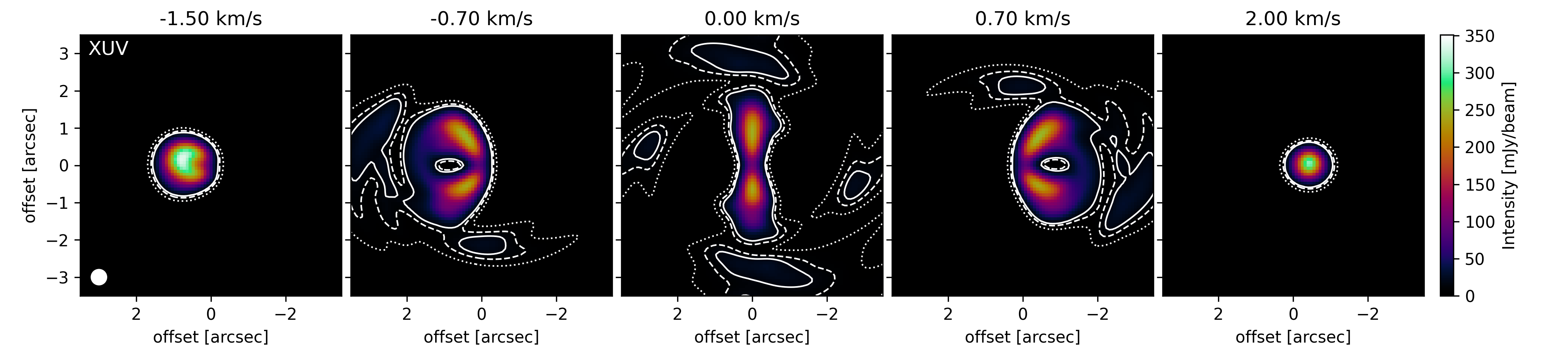}
    \caption{CO J=2-1 channel maps convolved with $0\farcs4\times0\farcs4$ beam (white spot at the bottom left corner). The white contours represent three detection limits: $5\sigma$ (solid), $3\sigma$ (dashed), and $1\sigma$ (dotted). }
    \label{fig:peakchanCO_0.4}
\end{figure*}

\begin{figure*}[htbp!]
\centering
    \includegraphics[width=1\textwidth]{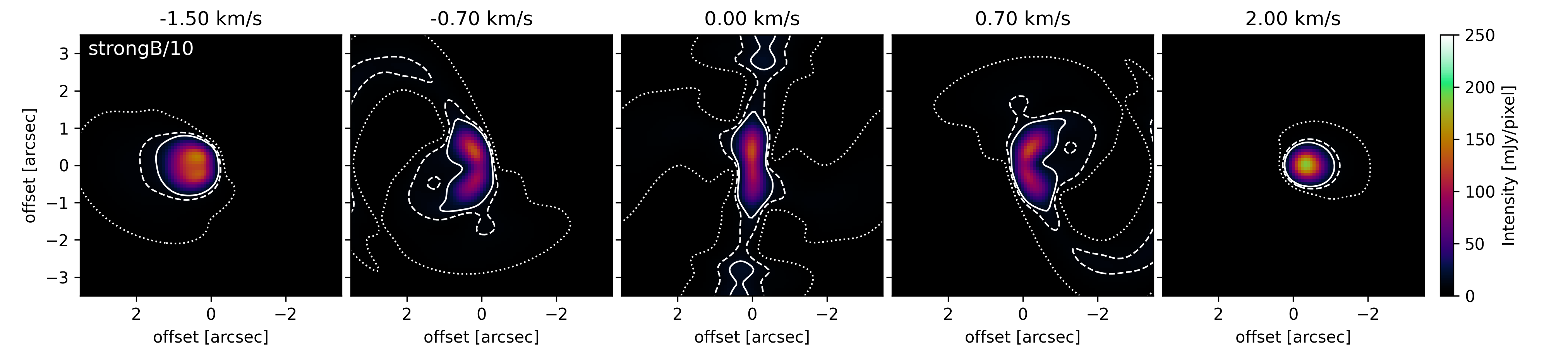}
    \includegraphics[width=1\textwidth]{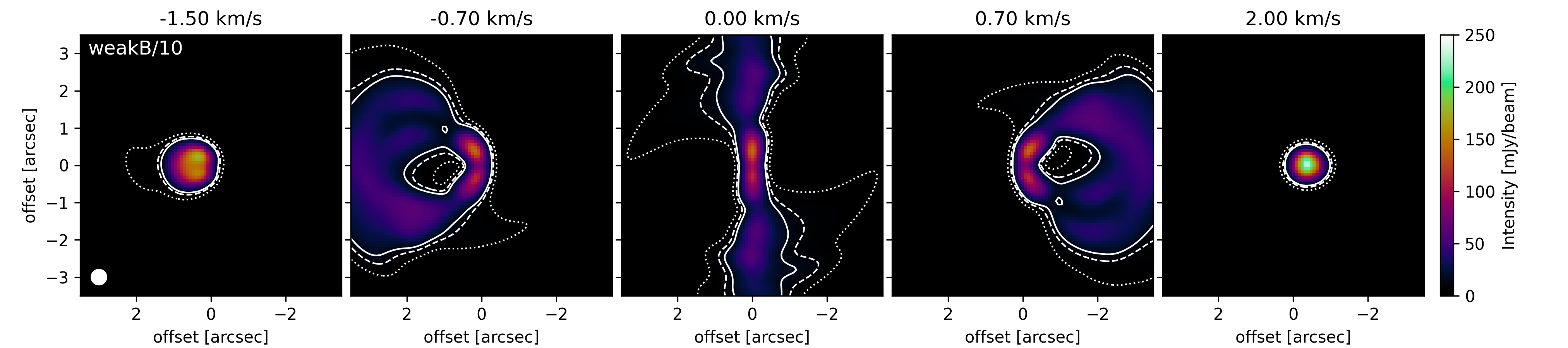}
    \includegraphics[width=1\textwidth]{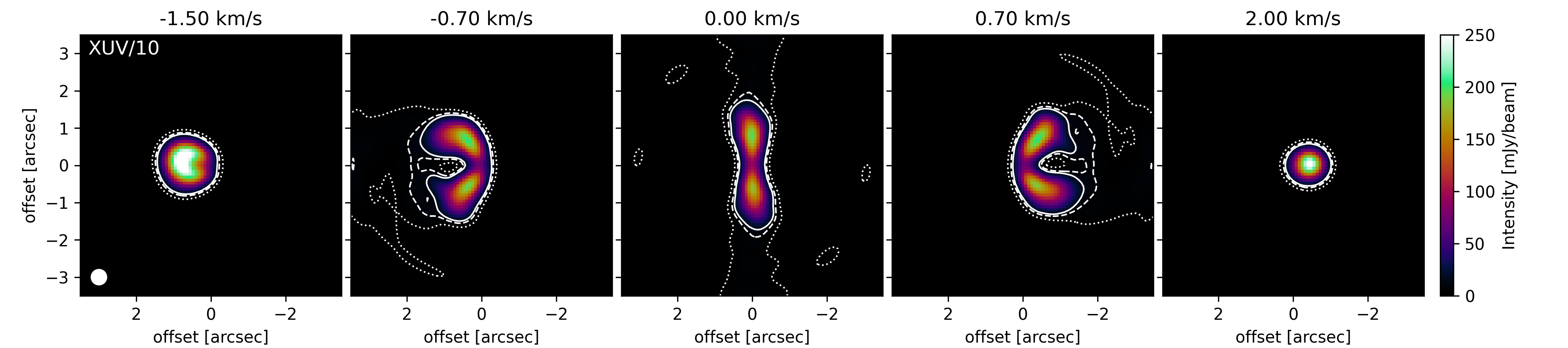}
    \caption{CO J=2-1 channel maps convolved with $0\farcs4\times0\farcs4$ beam, of a 0.002~$M_\odot$ mass disk, 1/10 of the fiducial disk mass. The white contours represent three detection limits: $5\sigma$ (solid), $3\sigma$ (dashed), and $1\sigma$ (dotted).}
    \label{fig:peakchan_lowmass_0.4}
\end{figure*}

The original channel maps, presented without beam convolution, are provided as reference materials. Specifically, Figure~\ref{fig:peakchan_origin} depicts CO J=2-1 maps for five models with fiducial disk masses (0.02$M_\odot$), Figure~\ref{fig:peakchanC10_origin} illustrates [C I] $^3P_1-^3P_0$ maps, and Figure~\ref{fig:peakchan_lowmass_origin} presents CO J=2-1 maps for the three low-mass disk configurations. Additionally, a comprehensive set of beam-convolved CO J=2-1 maps at $0\farcs\times0\farcs4$ resolution is included in Figure~\ref{fig:peakchanCO_0.4} and Figure~\ref{fig:peakchan_lowmass_0.4}. 

The front and back side emission surfaces are easily distinguishable in the original channel maps. For CO emission of the fiducial disks, we can even see a slight darker disk sandwiched between the two surfaces (e.g., \texttt{weakB}). For the more magnetized disks, they are not perfectly smooth from spontaneous substructure formation. The substructures from the redistributed magnetic flux could cause a 3\% deviation from the Keplerian rotation at the midplane and an even higher amplitude near the disk surface \citep{2022MNRAS.516.2006H}. This is most notable in the $0.7~{\rm km~s}^{-1}$ channel of model \texttt{strongB} and \texttt{strongB-XUV} by the ``spikes'' at the edge of the butterfly wings. The strong poloidal field drives faster accretion by exerting negative torque on the disk, leading to a sub-Keplerian gap with reduced disk material. A more detailed analysis of gas dynamics in similar gaps and rings with the effect on dust radial evolution is in \citet{2022MNRAS.516.2006H}. The stability of the substructures, especially under Rossby wave instability has been studied recently, both via numerical simulations \citep{2024arXiv240506026H,2024arXiv240802556C} and analytical methods \citep{2024arXiv240712722C}. Checking the $\pm0.7~{\rm km~s}^{-1}$ channels of model \texttt{weakB}, we can also find these substructures (``spikes'') at the outer edge with a lower amplitude. The same areas in \texttt{weakB-XUV} are smoother, which can be explained by the smoother $v_\phi$ profiles at $z/R=0.3$ and 0.45 in Figure~\ref{fig:3v3h}. The XUV driven outflow can nullify substructures carved by weak (midplane $\beta=10^5$) magnetic fields. 

We can also locate the emission surfaces more accurately in the non-convolved [C I] $^3P_1-^3P_0$ channel maps in Figure~
\ref{fig:peakchanC10_origin}. In the $-1.5~{\rm km~s}^{-1}$ channel, model \texttt{strongB} and \texttt{strongB-XUV} emit high up in the atmosphere with \texttt{strongB} carrying some bright rings in the outer half. The emission surfaces are lower in the weaker field setups, with a smaller but brighter back surface and a dimmer, slightly larger front surface. Both surfaces are on the left half and closely follow Keplerian patterns. In Figure~\ref{fig:2dcompC}, the lower boundary of atomic carbon distribution in model \texttt{XUV} has a ``step'': within 120 au, it is just below $z/R=0.45$, and drops down to $z/R=0.3$ before $\sim$180 au. Thus the [C I] stays within the Keplerian disk. The bright spots trace the local carbon enhancement around 150 au at $z/R=0.3$, and emissions from both surfaces overlap in $\pm0.7~{\rm km~s}^{-1}$ channels. The many pairs of rings and gaps are likely from the vertical shear instability at the disk surface \citep{2024ApJ...968...29Z}. In Figure~\ref{fig:peakchan_lowmass_origin}, without the CO surface that extends beyond the disk, the substructures caused by the magnetic field are visible in the magnetized disks with a low contrast.

With a beam size of $0\farcs4\times0\farcs4$, all magnetized disks in Figure~\ref{fig:peakchanCO_0.4} have strong ($>5\sigma$) wind features beyond 350 au. From Figure~\ref{fig:surf}, the $\tau=1$ surfaces in both \texttt{strongB} and \texttt{strongB-XUV} are constrained within 300 au, which means the CO wind can be detectable when it's optically thin. In \texttt{strongB}, three-quarters of the conic surface in $-1.5$ and $2~{\rm km~s}^{-1}$ channels are detectable. Compared to other morphological evidence, the conic shape is a direct reflection of the wind cone. Among the low mass disk setups, \texttt{XUV/10} does not have strong non-Keplerian signals even with a larger beam, but the $1\sigma$ signals outside the Keplerian disk might be noticeable thanks to their spatial scale.

\bibliography{ref_mhd_dust}{}
\bibliographystyle{aasjournal}

\end{document}

%% file: preamble.tex
\usepackage{float}
\usepackage{amsmath,amssymb}
\usepackage{natbib}    
\usepackage{todonotes}
\usepackage{graphicx} 
\usepackage{color}
\usepackage{verbatim}    
\usepackage{enumitem}
\usepackage{natbib}
\usepackage{CJK}
\usepackage{multirow}
\usepackage{comment}
\usepackage{bm}
\usepackage{physics}
\usepackage[version=4]{mhchem}
\usepackage{array}
\usepackage{orcidlink}  
\usepackage{soul}

\usepackage{hyperref} 

\newcommand{\cntextsc}[1]
{\begin{CJK*}{UTF8}{gbsn}#1\end{CJK*}}
\newcommand{\proptosim}{\mathrel{\vcenter{
 \offinterlineskip\halign{\hfil$##$\cr
 \propto\cr\noalign{\kern2pt}\sim\cr\noalign{\kern-2pt}}}}}
\renewcommand{\b}[1]{\boldsymbol{#1}}

\newcommand{\au}{\mathrm{AU}}

\newcommand{\m}{\mathrm{m}}

\newcommand{\g}{\mathrm{g}}
\newcommand{\G}{\mathrm{G}}
\newcommand{\K}{\mathrm{K}}

\newcommand{\eV}{\mathrm{eV}}
\newcommand{\keV}{\mathrm{keV}}
\newcommand{\s}{\mathrm{s}}

\newcommand{\B}{\bm B}
\renewcommand{\O}{\mathrm{O}}     
\newcommand{\A}{\mathrm{A}}     
\renewcommand{\H}{\mathrm{H}}     
\renewcommand{\P}{\mathrm{P}}     

\renewcommand{\d}{\mathrm{d}}
\newcommand{\e}{\mathrm{e}}

\newcommand{\Gr}{\ensuremath{\mathrm{Gr}}}

\newcommand{\p}{{\rm p}}
\newcommand{\J}{{\rm Jup}}

\newcommand*\chem[1]{\ensuremath{\mathrm{#1}}}

\newcommand{\E}{\bm E}     
\renewcommand{\v}{\bm v}   
\newcommand{\tot}{\mathrm{tot}}